\documentclass[a4paper,12pt]{article}
\pdfoutput=1
\usepackage{amsfonts}
\usepackage{graphicx}
\usepackage{subcaption}
\usepackage{pstricks}
\usepackage{pstricks-add}
\usepackage{tikz}
\usepackage{verbatim}
\usepackage{dsfont}
\usetikzlibrary{decorations.pathreplacing}
\usetikzlibrary{decorations.markings}
\usepackage{amsmath}
\usepackage{amssymb}
\usepackage[title]{appendix}
\usepackage[margin=2.3cm]{geometry}
\usepackage{cite} % To regroup citations: [1-3] instead of [1,2,3]

\DeclareMathOperator{\Tr}{Tr}

\newcommand{\beq}{\begin{equation}}
\newcommand{\dtt}{$D_2^2$ }
\newcommand{\eeq}{\end{equation}}
\newcommand{\no}{\noindent}

\newcommand{\StTL}[1]{\mathcal{W}_{#1}}
\newcommand{\q}{\mathfrak{q}}

\newcommand{\jesper}[1]{$\framebox{\tiny J}$\ \textbf{\texttt{{\color{cyan}\footnotesize#1}}}}

\begin{document}

\title{Lattice regularisation of a non-compact\\ boundary conformal field theory}
\date{}

\maketitle

\begin{center}

{Niall F.\ Robertson$^{1,2}$, Jesper Lykke Jacobsen$^{1,3,4,5}$, and Hubert Saleur$^{1,6}$}

\vspace{1.0cm}
{\sl\small $^1$ Universit\'e Paris Saclay, CNRS, CEA, Institut de Physique Th\'eorique, \\ F-91191 Gif-sur-Yvette, France\\}
{\sl\small $^2$ Departament de F\'isica Qu\`antica i Astrof\'isica and Institut de Ci\'encies del Cosmos (ICCUB), Universitat de Barcelona, Mart\'i i Franqu\`es 1, 08028 Barcelona, Catalonia, Spain \\}
{\sl\small $^3$ Laboratoire de Physique de l'\'Ecole Normale Sup\'erieure, ENS, Universit\'e PSL, \\
CNRS, Sorbonne Universit\'e, Universit\'e de Paris, F-75005 Paris, France\\}
{\sl\small $^4$ Sorbonne Universit\'e, \'Ecole Normale Sup\'erieure, CNRS, \\
Laboratoire de Physique (LPENS), F-75005 Paris, France\\}
{\sl\small $^5$ Institut des Hautes \'Etudes Scientifiques, Universit\'e Paris Saclay, CNRS, \\
Le Bois-Marie, 35 route de Chartres, F-91440 Bures-sur-Yvette, France\\}
{\sl\small $^6$ Department of Physics and Astronomy,
University of Southern California, \\
Los Angeles, CA 90089, USA\\}

\end{center}

\vskip 2cm

\begin{abstract}
Non-compact Conformal Field Theories (CFTs) are central to several aspects of string theory and condensed matter physics. They are characterised, in particular, by  the appearance of a continuum of conformal dimensions. Surprisingly, such CFTs have been identified as the continuum limits of lattice models with a finite number of degrees of freedom per site. However, results have so far been restricted to the case of periodic boundary conditions, precluding the exploration via lattice models of aspects of non-compact boundary CFTs and the corresponding D-brane constructions. 

The present paper follows a series of previous works on a $\mathbb{Z}_2$-staggered XXZ spin chain, whose continuum limit is known to be a non-compact CFT related with the Euclidian black hole sigma model.  By using the relationship of this spin chain with an integrable \dtt vertex model, we here identify integrable boundary conditions that lead to a continuous spectrum of boundary exponents, and thus correspond to non-compact branes. In the context of the Potts model on a square lattice, they correspond to wired boundary conditions at the physical antiferromagnetic critical point.
The relations with the boundary parafermion theories are discussed as well. We are also able to identify  a boundary renormalisation group flow from
the non-compact boundary conditions to the previously studied compact ones.

\end{abstract}

\newpage

\tableofcontents

\section{Introduction}\label{intro}

The study of the critical antiferromagnetic (AF) Potts model and the related $\mathbb{Z}_2$-staggered XXZ spin chain was initiated many years
ago \cite{B-AF,S-AF}. An extensive investigation---using a combination of integrability, conformal field theory (CFT) and
numerical techniques---paved the way to the understanding that this model has an unusual continuum limit: although the lattice model has a finite
number of degrees of freedom per site, its field-theory limit is a {\em non-compact} CFT, characterised in particular by the appearance of a continuous
spectrum of conformal dimensions \cite{JS-AF,IJS2008}.

This correspondence was subsequently made more precise by the identification \cite{IJS-AFlett}
of the continuum limit with the $sl(2,\mathbb{R})/u(1)$ Euclidian black hole CFT, a non-linear sigma model with a non-compact target space
that was first introduced in the string-theory literature \cite{Witten}. The consequences of this link go beyond the explanation of the observed continuous
spectrum and allows, in particular, for a lattice interpretation of the density of states---related in turn to the reflection amplitude for the scattering of a
closed string on the black hole---by showing that its continuum limit agrees with known results for the CFT \cite{MO01,HPT02}.
Further studies confirmed this identification in various ways \cite{CanduIkhlef,Frahm2014}. Finally, a very recent series of works
\cite{BKKL,BKKL20a,BKKL20b,BKKL20c} on the one hand confirms the coincidence of the partition function of the Euclidian black hole CFT with
{\em one half} of the partition function arising in the scaling limit of the lattice model with periodic boundary conditions, but on the other hand
refines the original identification by proposing that a part of the Hilbert space of the lattice model should coincide with the pseudo-Hilbert space
of the non-linear black hole sigma model with {\em Lorentzian} signature.

While all these studies were restricted to the case of periodic boundary conditions, the exploration of the AF Potts model in the presence of a
boundary is of undeniable interest. On general grounds, one would presume the existence of integrable boundary conditions in the lattice model,
which could in turn be identified with conformally invariant boundary conditions in the continuum limit. An attractive aspect of such investigations
would be to make precise contact with D-brane constructions in non-compact boundary CFTs of relevance in the context of string theory
\cite{RibaultSchomerus, Schomerus}.

We have initiated this work program in a series of two recent papers \cite{Robertson2019,Robertson2020}.
In \cite{Robertson2019}, a combination of numerical techniques was used to study the AF Potts model with open boundary conditions,
confirming again its close relationship with the Euclidean black hole CFT. A significant finding of \cite{Robertson2020} was to establish
the equivalence of the AF Potts model with an integrable model constructed from the twisted affine \dtt Lie algebra. This result paved
the way to studying the case of open boundary conditions with the Bethe Ansatz. Somewhat disappointingly, however, the boundary
conditions considered in both \cite{Robertson2019} and \cite{Robertson2020} led only to a {\em discrete} set of conformal weights in
the continuum-limit boundary CFT (BCFT), hence missing the more interesting non-compact aspects of the theory, and in particular 
the identification of non-compact branes \cite{RibaultSchomerus}. The motivation for the present work is to remedy this problem by, finally,
exhibiting integrable boundary conditions that produce a non-compact spectrum in the continuum limit.

\smallskip

In section \ref{review} we briefly review the critical antiferromagnetic Potts model with open boundary conditions. We shall pay particular attention
to the result from \cite{Robertson2020} showing that the integrable \dtt model is equivalent to the AF Potts model formulated
as a $\mathbb{Z}_2$-staggered six-vertex model. In section \ref{secnewK} the boundary conditions of interest will be presented. These
boundary conditions were found previously in \cite{Nepomechie2D22} by solving the boundary Yang-Baxter equation for the \dtt model.

%\jesper{Paragraph to be rewritten in light of final results.}
Our first key result is eq.~(\ref{hamtype3TL}), stating that the \dtt boundary conditions under study produce a remarkably simple
Hamiltonian that can be written entirely in terms of Temperley-Lieb (TL) algebra generators. In section \ref{sectmat}, we furthermore show
that---after a geometrical transformation discussed in section \ref{geomchange}---the transfer matrix also has a  simple interpretation in terms
of TL generators. In fact, it is given by the transfer matrix of the critical AF Potts model with {\em wired} boundary conditions (the dual
of free boundary conditions; see eq.~(\ref{tmataf})). We furthermore find that when we tune the spectral parameter $u$ to a different value and carry out the geometrical construction accordingly, the same \dtt boundary conditions can be interpreted as free boundary conditions in the AF Potts model. However, note that it is well established that the continuum limit of the AF Potts model with free boundary conditions is \textit{compact}.
Section \ref{secexactsoln} presents a complete Bethe Ansatz solution of the model with the \dtt boundary conditions considered in this paper,
based to a large extent on the work \cite{nepomechie2019towards} where a partial solution was already obtained. The complete Bethe
Ansatz equations are given in \eqref{baecorrect}.

In section \ref{seccontlim3} we consider the continuum limit of the model in different representations of the TL algebra. Most significantly, in
the loop representation the continuum limit is {\em non-compact} (see Figure \ref{noncompact2}), and the resulting continuous spectrum
of conformal dimensions can be understood in terms of those of the Euclidean black hole CFT---see eq.~(\ref{confw2}). This is in contrast
with the boundary condition studied previously in \cite{Robertson2020}, where we found a {\em compact} continuum limit.

To investigate the relation between these two boundary conditions we introduce, in section \ref{secboundaryrgflow}, a Hamiltonian \eqref{hgeneral}
with a boundary parameter
$\alpha$ that can interpolate between them. We show that $\alpha$ induces a boundary Renormalisation Group (RG) flow from
the non-compact BCFT studied here towards the compact BCFT of \cite{Robertson2020}. Moreover, we compute the boundary entropy
at either fixed point, confirming that it decreases under the boundary RG flow, in agreement with the ``$g$-conjecture'' by Affleck and Ludwig
\cite{affleck1991}.

\section{Conformally invariant boundary conditions in the AF Potts model}\label{review}

\begin{figure}
	\centering
\begin{tikzpicture}[scale=0.8]
\draw[black,line width = 1pt](1,1)--(9,9);
\draw[black,line width = 1pt](3,1)--(9,7);
\draw[black,line width = 1pt](5,1)--(9,5);
\draw[black,line width = 1pt](7,1)--(9,3);

\draw[black,line width = 1pt](1,3)--(7,9);
\draw[black,line width = 1pt](1,5)--(5,9);
\draw[black,line width = 1pt](1,7)--(3,9);

\draw[black,line width = 1pt](9,1)--(1,9);
\draw[black,line width = 1pt](7,1)--(1,7);
\draw[black,line width = 1pt](5,1)--(1,5);
\draw[black,line width = 1pt](3,1)--(1,3);

\draw[black,line width = 1pt](9,3)--(3,9);
\draw[black,line width = 1pt](9,5)--(5,9);
\draw[black,line width = 1pt](9,7)--(7,9);

\filldraw[black] (2,2) circle (4pt);
\filldraw[black] (4,2) circle (4pt);
\filldraw[black] (6,2) circle (4pt);
\filldraw[black] (8,2) circle (4pt);

\filldraw[black] (2,4) circle (4pt);
\filldraw[black] (4,4) circle (4pt);
\filldraw[black] (6,4) circle (4pt);
\filldraw[black] (8,4) circle (4pt);

\filldraw[black] (2,6) circle (4pt);
\filldraw[black] (4,6) circle (4pt);
\filldraw[black] (6,6) circle (4pt);
\filldraw[black] (8,6) circle (4pt);

\filldraw[black] (2,8) circle (4pt);
\filldraw[black] (4,8) circle (4pt);
\filldraw[black] (6,8) circle (4pt);
\filldraw[black] (8,8) circle (4pt);

\draw[black,line width = 1pt] (1.5,1.5) .. controls (1.25,2) .. (1.5,2.5);
\draw[black,line width = 1pt] (1.5,3.5) .. controls (1.25,4) .. (1.5,4.5);
\draw[black,line width = 1pt] (1.5,5.5) .. controls (1.25,6) .. (1.5,6.5);
\draw[black,line width = 1pt] (1.5,7.5) .. controls (1.25,8) .. (1.5,8.5);

\draw[black,line width = 1pt] (8.5,1.5) .. controls (8.75,2) .. (8.5,2.5);
\draw[black,line width = 1pt] (8.5,3.5) .. controls (8.75,4) .. (8.5,4.5);
\draw[black,line width = 1pt] (8.5,5.5) .. controls (8.75,6) .. (8.5,6.5);
\draw[black,line width = 1pt] (8.5,7.5) .. controls (8.75,8) .. (8.5,8.5);

\draw[black,line width = 2pt](2,2)--(2,4);
\draw[black,line width = 2pt](2,2)--(4,2);
\draw[black,line width = 2pt](2,4)--(6,4);
\draw[black,line width = 2pt](4,4)--(4,2);
\draw[black,line width = 2pt](6,4)--(6,2);
\draw[black,line width = 2pt](2,6)--(4,6);
\draw[black,line width = 2pt](2,8)--(4,8);
\draw[black,line width = 2pt](8,2)--(8,8);

\draw[black,line width = 1pt] (1.5,1.5) .. controls (2,1.25) .. (2.5,1.5);
\draw[black,line width = 1pt] (3.5,1.5) .. controls (4,1.25) .. (4.5,1.5);
\draw[black,line width = 1pt] (5.5,1.5) .. controls (6,1.25) .. (6.5,1.5);
\draw[black,line width = 1pt] (7.5,1.5) .. controls (8,1.25) .. (8.5,1.5);

\draw[black,line width = 1pt] (2.5,1.5) .. controls (3,1.75) .. (3.5,1.5);
\draw[black,line width = 1pt] (2.5,2.5) .. controls (3,2.25) .. (3.5,2.5);
\draw[black,line width = 1pt] (4.5,1.5) .. controls (4.75,2) .. (4.5,2.5);
\draw[black,line width = 1pt] (5.5,1.5) .. controls (5.25,2) .. (5.5,2.5);
\draw[black,line width = 1pt] (6.5,1.5) .. controls (6.75,2) .. (6.5,2.5);
\draw[black,line width = 1pt] (7.5,1.5) .. controls (7.25,2) .. (7.5,2.5);

\draw[black,line width = 1pt] (1.5,2.5) .. controls (1.75,3) .. (1.5,3.5);
\draw[black,line width = 1pt] (2.5,2.5) .. controls (2.25,3) .. (2.5,3.5);
\draw[black,line width = 1pt] (3.5,2.5) .. controls (3.75,3) .. (3.5,3.5);
\draw[black,line width = 1pt] (4.5,2.5) .. controls (4.25,3) .. (4.5,3.5);
\draw[black,line width = 1pt] (5.5,2.5) .. controls (5.75,3) .. (5.5,3.5);
\draw[black,line width = 1pt] (6.5,2.5) .. controls (6.25,3) .. (6.5,3.5);
\draw[black,line width = 1pt] (7.5,2.5) .. controls (7.75,3) .. (7.5,3.5);
\draw[black,line width = 1pt] (8.5,2.5) .. controls (8.25,3) .. (8.5,3.5);

\draw[black,line width = 1pt] (2.5,3.5) .. controls (3,3.75) .. (3.5,3.5);
\draw[black,line width = 1pt] (2.5,4.5) .. controls (3,4.25) .. (3.5,4.5);
\draw[black,line width = 1pt] (4.5,3.5) .. controls (5,3.75) .. (5.5,3.5);
\draw[black,line width = 1pt] (4.5,4.5) .. controls (5,4.25) .. (5.5,4.5);
\draw[black,line width = 1pt] (6.5,3.5) .. controls (6.75,4) .. (6.5,4.5);
\draw[black,line width = 1pt] (7.5,3.5) .. controls (7.25,4) .. (7.5,4.5);

\draw[black,line width = 1pt] (1.5,4.5) .. controls (2,4.75) .. (2.5,4.5);
\draw[black,line width = 1pt] (1.5,5.5) .. controls (2,5.25) .. (2.5,5.5);
\draw[black,line width = 1pt] (3.5,4.5) .. controls (4,4.75) .. (4.5,4.5);
\draw[black,line width = 1pt] (3.5,5.5) .. controls (4,5.25) .. (4.5,5.5);
\draw[black,line width = 1pt] (5.5,4.5) .. controls (6,4.75) .. (6.5,4.5);
\draw[black,line width = 1pt] (5.5,5.5) .. controls (6,5.25) .. (6.5,5.5);
\draw[black,line width = 1pt] (7.5,4.5) .. controls (7.75,5) .. (7.5,5.5);
\draw[black,line width = 1pt] (8.5,4.5) .. controls (8.25,5) .. (8.5,5.5);

\draw[black,line width = 1pt] (2.5,5.5) .. controls (3,5.75) .. (3.5,5.5);
\draw[black,line width = 1pt] (2.5,6.5) .. controls (3,6.25) .. (3.5,6.5);
\draw[black,line width = 1pt] (4.5,5.5) .. controls (4.75,6) .. (4.5,6.5);
\draw[black,line width = 1pt] (5.5,5.5) .. controls (5.25,6) .. (5.5,6.5);
\draw[black,line width = 1pt] (6.5,5.5) .. controls (6.75,6) .. (6.5,6.5);
\draw[black,line width = 1pt] (7.5,5.5) .. controls (7.25,6) .. (7.5,6.5);

\draw[black,line width = 1pt] (1.5,6.5) .. controls (2,6.75) .. (2.5,6.5);
\draw[black,line width = 1pt] (1.5,7.5) .. controls (2,7.25) .. (2.5,7.5);
\draw[black,line width = 1pt] (3.5,6.5) .. controls (4,6.75) .. (4.5,6.5);
\draw[black,line width = 1pt] (3.5,7.5) .. controls (4,7.25) .. (4.5,7.5);
\draw[black,line width = 1pt] (5.5,6.5) .. controls (6,6.75) .. (6.5,6.5);
\draw[black,line width = 1pt] (5.5,7.5) .. controls (6,7.25) .. (6.5,7.5);
\draw[black,line width = 1pt] (7.5,6.5) .. controls (7.75,7) .. (7.5,7.5);
\draw[black,line width = 1pt] (8.5,6.5) .. controls (8.25,7) .. (8.5,7.5);

\draw[black,line width = 1pt] (2.5,7.5) .. controls (3,7.75) .. (3.5,7.5);
\draw[black,line width = 1pt] (2.5,8.5) .. controls (3,8.25) .. (3.5,8.5);
\draw[black,line width = 1pt] (4.5,7.5) .. controls (4.75,8) .. (4.5,8.5);
\draw[black,line width = 1pt] (5.5,7.5) .. controls (5.25,8) .. (5.5,8.5);
\draw[black,line width = 1pt] (6.5,7.5) .. controls (6.75,8) .. (6.5,8.5);
\draw[black,line width = 1pt] (7.5,7.5) .. controls (7.25,8) .. (7.5,8.5);

\draw[black,line width = 1pt] (1.5,8.5) .. controls (2,8.75) .. (2.5,8.5);
\draw[black,line width = 1pt] (3.5,8.5) .. controls (4,8.75) .. (4.5,8.5);
\draw[black,line width = 1pt] (5.5,8.5) .. controls (6,8.75) .. (6.5,8.5);
\draw[black,line width = 1pt] (7.5,8.5) .. controls (8,8.75) .. (8.5,8.5);

\end{tikzpicture}

\caption{A configuration of the Potts model in the loop/cluster formulation on the square lattice.
The black dots are the loci of $Q$-component Potts spins $\sigma_i$.
The thick bars form clusters, whose total size is denoted $|G|$ in \eqref{looppartition}.
Cluster configurations are in one-to-one correspondence with their surrounding loop configurations, shown as thin curves in the figure.
Each pair of (horizontally or vertically) neighbouring black dots resides diagonally around a tile, in whose interior two pieces of loops split
in one of two possible ways.
}\label{fulllattice}
\end{figure}
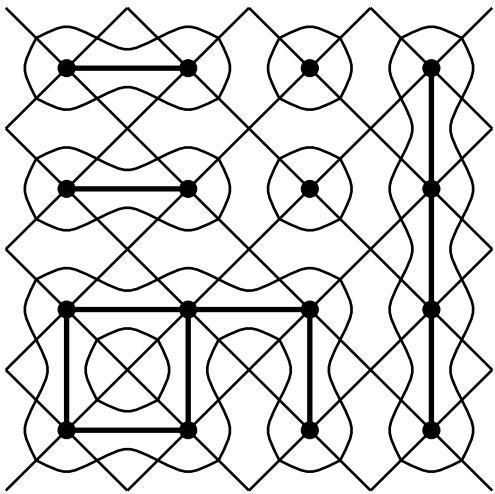

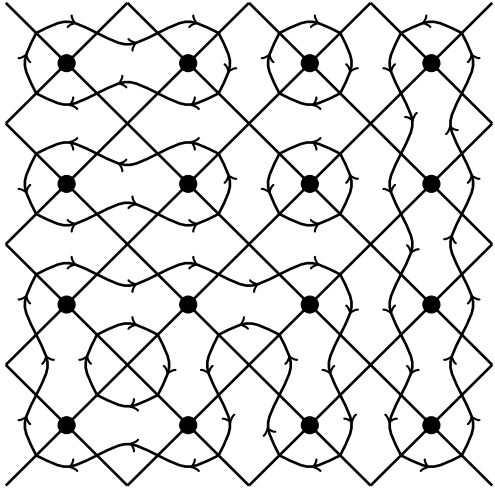
\begin{figure}
	\centering
\begin{tikzpicture}[scale=0.8]
\draw[black,line width = 1pt](1,1)--(9,9);
\draw[black,line width = 1pt](3,1)--(9,7);
\draw[black,line width = 1pt](5,1)--(9,5);
\draw[black,line width = 1pt](7,1)--(9,3);

\draw[black,line width = 1pt](1,3)--(7,9);
\draw[black,line width = 1pt](1,5)--(5,9);
\draw[black,line width = 1pt](1,7)--(3,9);

\draw[black,line width = 1pt](9,1)--(1,9);
\draw[black,line width = 1pt](7,1)--(1,7);
\draw[black,line width = 1pt](5,1)--(1,5);
\draw[black,line width = 1pt](3,1)--(1,3);

\draw[black,line width = 1pt](9,3)--(3,9);
\draw[black,line width = 1pt](9,5)--(5,9);
\draw[black,line width = 1pt](9,7)--(7,9);

\filldraw[black] (2,2) circle (4pt);
\filldraw[black] (4,2) circle (4pt);
\filldraw[black] (6,2) circle (4pt);
\filldraw[black] (8,2) circle (4pt);

\filldraw[black] (2,4) circle (4pt);
\filldraw[black] (4,4) circle (4pt);
\filldraw[black] (6,4) circle (4pt);
\filldraw[black] (8,4) circle (4pt);

\filldraw[black] (2,6) circle (4pt);
\filldraw[black] (4,6) circle (4pt);
\filldraw[black] (6,6) circle (4pt);
\filldraw[black] (8,6) circle (4pt);

\filldraw[black] (2,8) circle (4pt);
\filldraw[black] (4,8) circle (4pt);
\filldraw[black] (6,8) circle (4pt);
\filldraw[black] (8,8) circle (4pt);

			\begin{scope}[thick,decoration={
			    markings,
			    mark=at position 0.65 with {\arrow{>}}}
			    ] 

\draw[black,line width = 1pt,postaction={decorate}](1.5,1.5) .. controls (1.25,2) .. (1.5,2.5);
\draw[black,line width = 1pt,postaction={decorate}](1.5,3.5) .. controls (1.25,4) .. (1.5,4.5);
\draw[black,line width = 1pt,postaction={decorate}] (1.5,6.5).. controls (1.25,6) .. (1.5,5.5);
\draw[black,line width = 1pt,postaction={decorate}] (1.5,7.5) .. controls (1.25,8) .. (1.5,8.5);

			\end{scope}	

			\begin{scope}[thick,decoration={
			    markings,
			    mark=at position 0.65 with {\arrow{>}}}
			    ] 

\draw[black,line width = 1pt,postaction={decorate}] (8.5,1.5) .. controls (8.75,2) .. (8.5,2.5);
\draw[black,line width = 1pt,postaction={decorate}] (8.5,3.5) .. controls (8.75,4) .. (8.5,4.5);
\draw[black,line width = 1pt,postaction={decorate}] (8.5,5.5) .. controls (8.75,6) .. (8.5,6.5);
\draw[black,line width = 1pt,postaction={decorate}] (8.5,7.5) .. controls (8.75,8) .. (8.5,8.5);

			\end{scope}				

			\begin{scope}[thick,decoration={
			    markings,
			    mark=at position 0.65 with {\arrow{<}}}
			    ] 
				
			\draw[black,line width = 1pt,postaction={decorate}] (1.5,1.5) .. controls (2,1.25) .. (2.5,1.5);
			\draw[black,line width = 1pt,postaction={decorate}] (3.5,1.5) .. controls (4,1.25) .. (4.5,1.5);
			\draw[black,line width = 1pt,postaction={decorate}] (5.5,1.5) .. controls (6,1.25) .. (6.5,1.5);
			\draw[black,line width = 1pt,postaction={decorate}](8.5,1.5)  .. controls (8,1.25) .. (7.5,1.5);

			\end{scope}

			\begin{scope}[thick,decoration={
			    markings,
			    mark=at position 0.65 with {\arrow{<}}}
			    ] 
				
		\draw[black,line width = 1pt,postaction={decorate}] (2.5,1.5) .. controls (3,1.75) .. (3.5,1.5);
		\draw[black,line width = 1pt,postaction={decorate}] (2.5,2.5) .. controls (3,2.25) .. (3.5,2.5);
		\draw[black,line width = 1pt,postaction={decorate}] (4.5,1.5) .. controls (4.75,2) .. (4.5,2.5);
		\draw[black,line width = 1pt,postaction={decorate}] (5.5,2.5).. controls (5.25,2) .. (5.5,1.5) ;
		\draw[black,line width = 1pt,postaction={decorate}] (6.5,1.5) .. controls (6.75,2) .. (6.5,2.5);
		\draw[black,line width = 1pt,postaction={decorate}] (7.5,1.5) .. controls (7.25,2) .. (7.5,2.5);

			\end{scope}

			\begin{scope}[thick,decoration={
			    markings,
			    mark=at position 0.65 with {\arrow{>}}}
			    ] 
				
		\draw[black,line width = 1pt,postaction={decorate}] (1.5,2.5) .. controls (1.75,3) .. (1.5,3.5);
		\draw[black,line width = 1pt,postaction={decorate}] (2.5,2.5) .. controls (2.25,3) .. (2.5,3.5);
		\draw[black,line width = 1pt,postaction={decorate}] (3.5,3.5)  .. controls (3.75,3) .. (3.5,2.5);
		\draw[black,line width = 1pt,postaction={decorate}] (4.5,3.5).. controls (4.25,3) .. (4.5,2.5) ;
		\draw[black,line width = 1pt,postaction={decorate}] (5.5,2.5) .. controls (5.75,3) .. (5.5,3.5);
		\draw[black,line width = 1pt,postaction={decorate}] (6.5,3.5) .. controls (6.25,3) .. (6.5,2.5);
		\draw[black,line width = 1pt,postaction={decorate}] (7.5,3.5) .. controls (7.75,3) .. (7.5,2.5);
		\draw[black,line width = 1pt,postaction={decorate}] (8.5,2.5) .. controls (8.25,3) .. (8.5,3.5);

			\end{scope}

			\begin{scope}[thick,decoration={
			    markings,
			    mark=at position 0.65 with {\arrow{>}}}
			    ] 
				
				\draw[black,line width = 1pt,postaction={decorate}] (2.5,3.5) .. controls (3,3.75) .. (3.5,3.5);
				\draw[black,line width = 1pt,postaction={decorate}] (2.5,4.5) .. controls (3,4.25) .. (3.5,4.5);
				\draw[black,line width = 1pt,postaction={decorate}] (5.5,3.5) .. controls (5,3.75) .. (4.5,3.5);
				\draw[black,line width = 1pt,postaction={decorate}] (4.5,4.5) .. controls (5,4.25) .. (5.5,4.5);
				\draw[black,line width = 1pt,postaction={decorate}] (6.5,4.5) .. controls (6.75,4) .. (6.5,3.5);
				\draw[black,line width = 1pt,postaction={decorate}] (7.5,4.5) .. controls (7.25,4) .. (7.5,3.5);

			\end{scope}

			\begin{scope}[thick,decoration={
			    markings,
			    mark=at position 0.65 with {\arrow{>}}}
			    ] 
				
				\draw[black,line width = 1pt,postaction={decorate}] (1.5,4.5) .. controls (2,4.75) .. (2.5,4.5);
				\draw[black,line width = 1pt,postaction={decorate}] (1.5,5.5) .. controls (2,5.25) .. (2.5,5.5);
				\draw[black,line width = 1pt,postaction={decorate}] (3.5,4.5) .. controls (4,4.75) .. (4.5,4.5);
				\draw[black,line width = 1pt,postaction={decorate}] (3.5,5.5) .. controls (4,5.25) .. (4.5,5.5);
				\draw[black,line width = 1pt,postaction={decorate}] (5.5,4.5) .. controls (6,4.75) .. (6.5,4.5);
				\draw[black,line width = 1pt,postaction={decorate}] (5.5,5.5) .. controls (6,5.25) .. (6.5,5.5);
				\draw[black,line width = 1pt,postaction={decorate}] (7.5,5.5) .. controls (7.75,5) ..(7.5,4.5) ;
				\draw[black,line width = 1pt,postaction={decorate}] (8.5,4.5) .. controls (8.25,5) .. (8.5,5.5);

			\end{scope}

			\begin{scope}[thick,decoration={
			    markings,
			    mark=at position 0.65 with {\arrow{>}}}
			    ] 
				
				\draw[black,line width = 1pt,postaction={decorate}] (2.5,5.5) .. controls (3,5.75) .. (3.5,5.5);
				\draw[black,line width = 1pt,postaction={decorate}] (3.5,6.5).. controls (3,6.25) .. (2.5,6.5) ;
				\draw[black,line width = 1pt,postaction={decorate}] (4.5,5.5).. controls (4.75,6) .. (4.5,6.5) ;
				\draw[black,line width = 1pt,postaction={decorate}] (5.5,6.5) .. controls (5.25,6) .. (5.5,5.5);
				\draw[black,line width = 1pt,postaction={decorate}] (6.5,5.5) .. controls (6.75,6) .. (6.5,6.5);
				\draw[black,line width = 1pt,postaction={decorate}] (7.5,6.5) .. controls (7.25,6) .. (7.5,5.5);

			\end{scope}	

			\begin{scope}[thick,decoration={
			    markings,
			    mark=at position 0.65 with {\arrow{<}}}
			    ] 
				
			\draw[black,line width = 1pt,postaction={decorate}] (1.5,6.5) .. controls (2,6.75) .. (2.5,6.5);
			\draw[black,line width = 1pt,postaction={decorate}] (1.5,7.5) .. controls (2,7.25) .. (2.5,7.5);
			\draw[black,line width = 1pt,postaction={decorate}] (3.5,6.5) .. controls (4,6.75) .. (4.5,6.5);
			\draw[black,line width = 1pt,postaction={decorate}] (3.5,7.5) .. controls (4,7.25) .. (4.5,7.5);
			\draw[black,line width = 1pt,postaction={decorate}] (5.5,6.5) .. controls (6,6.75) .. (6.5,6.5);
			\draw[black,line width = 1pt,postaction={decorate}] (5.5,7.5) .. controls (6,7.25) .. (6.5,7.5);
			\draw[black,line width = 1pt,postaction={decorate}] (7.5,6.5) .. controls (7.75,7) .. (7.5,7.5) ;
			\draw[black,line width = 1pt,postaction={decorate}] (8.5,7.5) .. controls (8.25,7) .. (8.5,6.5);

			\end{scope}

			\begin{scope}[thick,decoration={
			    markings,
			    mark=at position 0.65 with {\arrow{>}}}
			    ] 
				
			\draw[black,line width = 1pt,postaction={decorate}] (3.5,7.5) .. controls (3,7.75) .. (2.5,7.5) ;
			\draw[black,line width = 1pt,postaction={decorate}] (2.5,8.5) .. controls (3,8.25) .. (3.5,8.5);
			\draw[black,line width = 1pt,postaction={decorate}] (4.5,8.5).. controls (4.75,8) .. (4.5,7.5) ;
			\draw[black,line width = 1pt,postaction={decorate}] (5.5,7.5) .. controls (5.25,8) .. (5.5,8.5);
			\draw[black,line width = 1pt,postaction={decorate}] (6.5,8.5) .. controls (6.75,8) .. (6.5,7.5);
			\draw[black,line width = 1pt,postaction={decorate}] (7.5,8.5) .. controls (7.25,8) .. (7.5,7.5);

			\end{scope}

			\begin{scope}[thick,decoration={
			    markings,
			    mark=at position 0.65 with {\arrow{>}}}
			    ]

			\draw[black,line width = 1pt,postaction={decorate}] (1.5,8.5) .. controls (2,8.75) .. (2.5,8.5);
			\draw[black,line width = 1pt,postaction={decorate}] (3.5,8.5) .. controls (4,8.75) .. (4.5,8.5);
			\draw[black,line width = 1pt,postaction={decorate}] (5.5,8.5) .. controls (6,8.75) .. (6.5,8.5);
			\draw[black,line width = 1pt,postaction={decorate}] (8.5,8.5) .. controls (8,8.75) .. (7.5,8.5);

			\end{scope}

\end{tikzpicture}

\caption{A configuration of oriented loops. The oriented loop model becomes a six-vertex model, after summing over the loop splittings.}\label{orientloop}
\end{figure}

\begin{figure}
	\centering
		\begin{tikzpicture}[scale=1]
		
		\node at (0.5,1.75) {\scriptsize{$x$}};
		\node at (2,1.75) {\scriptsize{$x$}};
		\node at (3.5,1.75) {\scriptsize{$1$}};
		\node at (5,1.75) {\scriptsize{$1$}};
		\node at (6.5,1.75) {\scriptsize{$x+e^{-i\gamma}$}};
		\node at (8,1.75) {\scriptsize{$x+e^{i\gamma}$}};

		\node at (0.5,0.25) {\scriptsize{$1$}};
		\node at (2,0.25) {\scriptsize{$1$}};
		\node at (3.5,0.25) {\scriptsize{$x$}};
		\node at (5,0.25) {\scriptsize{$x$}};
		\node at (6.5,0.25) {\scriptsize{$1+xe^{-i\gamma}$}};
		\node at (8,0.25) {\scriptsize{$1+xe^{i\gamma}$}};

			\begin{scope}[thick,decoration={
			    markings,
			    mark=at position 0.65 with {\arrow{>}}}
			    ] 
			\draw[black,line width = 1pt,postaction={decorate}] (0,0.5)--(0.5,1);
			\draw[black,line width = 1pt,postaction={decorate}] (1,0.5)--(0.5,1);
			\draw[black,line width = 1pt,postaction={decorate}] (0.5,1)--(1,1.5);
			 \draw[black,line width = 1pt,postaction={decorate}](0.5,1)--(0,1.5);
			\end{scope}	
			
			\begin{scope}[thick,decoration={
			    markings,
			    mark=at position 0.55 with {\arrow{<}}}
			    ] 
			\draw[black,line width = 1pt,postaction={decorate}] (1.5,0.5)--(2,1);
			\draw[black,line width = 1pt,postaction={decorate}] (2.5,0.5)--(2,1);
			\draw[black,line width = 1pt,postaction={decorate}] (2,1)--(2.5,1.5);
			 \draw[black,line width = 1pt,postaction={decorate}](2,1)--(1.5,1.5);
			\end{scope}

			\begin{scope}[thick,decoration={
			    markings,
			    mark=at position 0.65 with {\arrow{>}}}
			    ] 
			\draw[black,line width = 1pt,postaction={decorate}] (3,0.5)--(3.5,1);
			\draw[black,line width = 1pt,postaction={decorate}] (3.5,1)--(4,0.5);
			\draw[black,line width = 1pt,postaction={decorate}] (3.5,1)--(4,1.5);
			 \draw[black,line width = 1pt,postaction={decorate}](3,1.5)--(3.5,1);
			\end{scope}

			\begin{scope}[thick,decoration={
			    markings,
			    mark=at position 0.55 with {\arrow{<}}}
			    ] 
			\draw[black,line width = 1pt,postaction={decorate}] (4.5,0.5)--(5,1);
			\draw[black,line width = 1pt,postaction={decorate}] (5,1)--(5.5,0.5);
			\draw[black,line width = 1pt,postaction={decorate}] (5,1)--(5.5,1.5);
			 \draw[black,line width = 1pt,postaction={decorate}](4.5,1.5)--(5,1);
			\end{scope}

			\begin{scope}[thick,decoration={
			    markings,
			    mark=at position 0.55 with {\arrow{>}}}
			    ] 
			\draw[black,line width = 1pt,postaction={decorate}] (6,0.5)--(6.5,1);
			\draw[black,line width = 1pt,postaction={decorate}] (6.5,1)--(7,0.5);
			\draw[black,line width = 1pt,postaction={decorate}] (7,1.5)--(6.5,1);
			 \draw[black,line width = 1pt,postaction={decorate}](6.5,1)--(6,1.5);
			\end{scope}

			\begin{scope}[thick,decoration={
			    markings,
			    mark=at position 0.55 with {\arrow{<}}}
			    ] 
			\draw[black,line width = 1pt,postaction={decorate}] (7.5,0.5)--(8,1);
			\draw[black,line width = 1pt,postaction={decorate}] (8,1)--(8.5,0.5);
			\draw[black,line width = 1pt,postaction={decorate}] (8.5,1.5)--(8,1);
			 \draw[black,line width = 1pt,postaction={decorate}](8,1)--(7.5,1.5);
			\end{scope}

\end{tikzpicture}
\caption{The vertices and their Boltzmann weights, shown below (resp.\ above) the vertex for a tile corresponding to a horizontal (resp.\ vertical)
pair of Potts spins.}\label{vertexmodel}
\end{figure}
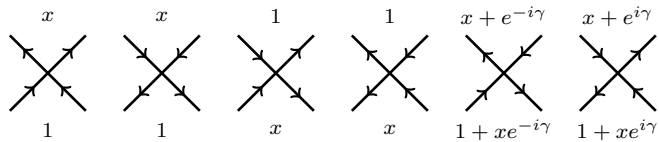
The two-dimensional $Q$-state Potts model is defined by the classical Hamiltonian
\beq\label{classicalpotts}
\mathcal{H} = - K\sum\limits_{\langle ij \rangle }\delta_{\sigma_i \sigma_j} \,,
\eeq
where $\sigma_i = 1,2,\ldots,Q$ and $\langle ij \rangle$ denotes the set of nearest neighbours on the square lattice. This model has been reviewed and discussed extensively in our previous works \cite{Robertson2019,Robertson2020} so only the most important points will be repeated here. It is well known that the partition function 
\beq\label{Pottspart}
\mathcal{Z}=\sum\limits_{\{\sigma\}}\exp(-\mathcal{H})=\sum\limits_{\{\sigma\}}\prod\limits_{\langle ij \rangle}\exp(K\delta_{\sigma_i,\sigma_j})
\eeq
arising from \eqref{classicalpotts} can be reformulated, first as a sum over configurations of clusters of spins \cite{FK72}, and subsequently as a sum over configurations of loops \cite{BKW76} (see Figure \ref{fulllattice}):
\beq\label{looppartition}
\mathcal{Z}=Q^{\frac{|V|}{2}}\sum\limits_{\text{loops}}x^{|G|}Q^{\frac{\ell}{2}} \,,
\eeq
with $x=\frac{e^K-1}{\sqrt{Q}}$, whereas $|G|$ denotes the number of edges in the clusters, and $\ell$ is the number of loops in each configuration.

One can further define an oriented loop model by assigning an orientation to each of the loops, and summing
over both orientations with appropriate weights \cite{BKW76} to recover the unoriented loop weight $Q^{1/2}$.
Keeping then these orientations but summing instead over the possible loop splittings at each vertex, for any
fixed choice of orientations of the adjacent edges, results finally
in a six-vertex model where the vertices live on the tiles in Figure \ref{orientloop} and take the configurations shown in Figure \ref{vertexmodel}.
This vertex model is in general {\em staggered}, meaning that the Boltzmann weights of the vertices take two different values depending on
whether they live on a tile that corresponds to a vertical or horizontal
coupling $(ij)$ between two neighbouring Potts spins, $\sigma_i$ and $\sigma_j$. The vertex weights are shown in Figure \ref{vertexmodel},
with the case of a horizontal (resp.\ vertical) coupling shown below (resp.\ above) the vertices.

The case $x_{\rm FM}=1$ is the ferromagnetic critical point of the Potts model, for which the staggering disappears. But in this paper we focus instead on the
AF critical point \cite{B-AF}
\beq \label{xAF}
 x_{\rm AF}^\pm = \frac{-2 \pm \sqrt{4-Q}}{\sqrt{Q}} \,,
\eeq
and we refer to this vertex model as the $\mathbb{Z}_2$-staggered six-vertex model \cite{IJS2008}. 
The two possible signs in $x_{\rm AF}^\pm$ correspond to a pair of mutually dual models---we shall come back to the issue
of duality at the end of our analysis.

Previous work on the AF Potts model imposed
periodic boundary conditions on the vertex orientations, or on the corresponding loops. If one moreover imposes the same weight for
a non-contractible loop as for a contractible one---namely $Q^{1/2}$, by \eqref{looppartition}---the vertex model must be appropriately twisted.

We consider instead {\em open boundary conditions}, corresponding to the loops being reflected back at the boundaries. A first study of such an
open AF Potts model appeared in \cite{Robertson2019}, where we constructed conformally invariant boundary conditions by assigning particular
Boltzmann weights to the loops touching the boundaries. These boundary conditions were shown to produce the discrete characters of
the $sl(2,\mathbb{R})/u(1)$  CFT in the continuum limit \cite{RibaultSchomerus}. However, no sign of a continuous spectrum of conformal
weights was found for that choice of boundary weights.

The open AF Potts model was further studied in our second paper \cite{Robertson2020}, this time using the tools of integrability.
Recall first that in the case of {\em periodic} boundary conditions the staggered six-vertex model can be described by an integrable row-to-row
transfer matrix
\beq\label{tmatrixperiodic}
t(u)=\Tr_a(R_{a,1}(u) R_{a,2}(u) \cdots R_{a,L}(u)) \,,
\eeq
where the subscripts of the $R$-matrix refer to the quantum spaces $i=1,2,\ldots,L$ and the auxiliary space $a$.
Since $R$ satisfies the Yang-Baxter equation, the theory is solvable by Bethe Ansatz \cite{B-AF,IJS2008}. A main result
of \cite{Robertson2020} was to show that the $R$-matrix appearing in (\ref{tmatrixperiodic}) and the $R$-matrix of a particular
integrable model constructed from the twisted affine \dtt Lie algebra are equivalent, in the following sense.
First recall from \cite{IJS2008} that the staggered six-vertex model $R$-matrix is written in terms of two parameters,
$\gamma$ (crossing parameter) and $u$ (spectral parameter). The former is related to the $Q$ appearing in the $Q$-state Potts model by
\beq\label{qgamma}
 \sqrt{Q}=e^{i\gamma}+e^{-i\gamma} \,.
\eeq
In the standard work \cite{Jimbo1986}, the integrable \dtt $R$-matrix is written instead in terms of two other parameters, $\tilde{k}$ and $x$.%
\footnote{Note that in \cite{Jimbo1986} the parameter $\tilde{k}$ is just written as $k$, but to be consistent with our previous work we need to reserve the label $k$ for a different parameter appearing in eq.~(\ref{cpf2}).}
If we relate the parameters by \cite{Robertson2020}
\begin{subequations}\label{d22notation}
\begin{eqnarray}
\tilde{k} &=& e^{2i\gamma} \,, \\
x &=& e^{2iu} \,,
\end{eqnarray}
\end{subequations}
the transfer matrix in (\ref{tmatrixperiodic}), built from the \dtt $R$-matrix, is equivalent---via a change of basis and a gauge transformation
\cite{Robertson2020}---to the transfer matrix built from the staggered six-vertex model $R$-matrix.

This equivalence between $R$-matrices allows one to study as well the staggered six-vertex model with {\em open} boundary conditions,
by exploiting the known integrable boundary conditions in the \dtt model. In the open case we need the double-row transfer matrix
\beq\label{tmatrixopen}
t(u)=\Tr_a K_a^+(u)R_{a,L}(u) \cdots R_{a,1}(u)K_a^-(u)R_{1,a}(u)...R_{L,a}(u) \,,
\eeq
where the $K^{\pm}$-matrices encode the Boltzmann weights of the left ($K^-$) and right ($K^+$) boundary vertices respectively.
$K^-$ satisfies the \textit{boundary} Yang-Baxter equation
\beq\label{reflection}
R_{1,2}(u-v)K_1^-(u)R_{2,1}(u+v)K_2^-(v)=K_2^-(u)R_{1,2}(u+v)K_1^-(u)R_{2,1}(u-v) \,.
\eeq
To ensure that $K^+$ satisfies the analogue of (\ref{reflection}) for the right boundary we take
\beq\label{Kright}
K^+(\lambda)=K^{-{\rm t}}(-\rho-\lambda)M \,,
\eeq
where ${\rm t}$ denotes the transpose. The model-dependent quantities $\rho$ and $M$ are given by $\rho = -\log \tilde{k}$ and
$M = \text{diag}(\tilde{k},1,1,\tilde{k}^{-1})$ for the \dtt model.

%\jesper{To be rewritten in light of our final conclusions. Stress also that we now have free bcs in {\em finite} size.}
A number of solutions to (\ref{reflection}) for the \dtt model were found in \cite{Nepomechie2D22}. Our previous work \cite{Robertson2020}
focussed on one of these, showing that its continuum limit coincides with that of the AF Potts model with free boundary conditions for the
Potts spins. By contrast with that work, we here find a nice interpretation of the boundary conditions (namely {\em wired}, dual to free) already in finite size.
Relating the double-row ``integrable geometry'' of the transfer matrix \eqref{tmatrixopen} to the ``diagonal geometry'' of the
lattice model (see Figures \ref{fulllattice}--\ref{orientloop}) requires a few tricks, which will be exposed in section \ref{geomchange}.

The present work considers another solution to (\ref{reflection}) that will turn out to admit a remarkably simple lattice interpretation.
But most importantly, the corresponding boundary conditions will be shown to produce an interesting continuum limit, given by a non-compact boundary CFT. 

\section{New boundary conditions}\label{secnewK}

The new boundary conditions to be considered here correspond to the integrable $K$-matrix \cite{Nepomechie2D22}
\beq\label{Kmatrixtype3}
K^-(\lambda)=
\frac{1}{\sinh(\lambda+\eta)}
\begin{pmatrix}
	-\sinh(\lambda-\eta) & 0 & 0 & 0 \\
	0 & \!\!\!\!\!\! \cosh\lambda\sinh\eta & -\sinh\lambda\cosh\eta & 0\\
	0 & \!\!\!\!\!\! -\sinh\lambda\cosh\eta & \cosh\lambda\sinh\eta & 0\\
	0 & 0 & 0 & \!\!\!\!\!\! -\sinh(\lambda-\eta) 
\end{pmatrix} \,,
\eeq
where the correspondence with our previous notation \eqref{d22notation} is
\begin{subequations}\label{lambdaurelation}
\begin{eqnarray}
	\lambda&=&2iu \,, \\
	\eta&=& i\gamma \,.
\end{eqnarray}
\end{subequations}
The edges in the \dtt vertex model can be in four possible states
\beq\label{d22basis}
\{|1\rangle,|2\rangle,|3\rangle,|4\rangle \} \,.
\eeq
In the $\mathbb{Z}_2$-staggered six-vertex model, the spectral parameter alternates on every second quantum space \cite{IJS2008},
so to it is natural to regroup these spaces two by two and choose again a basis of four states
\beq\label{s6vbasis}
\{|\!\uparrow\uparrow\rangle, |\!\uparrow\downarrow\rangle, |\!\downarrow\uparrow\rangle, |\!\downarrow\downarrow\rangle \} \,,
\eeq
corresponding to the arrow orientations in Figures \ref{orientloop}--\ref{vertexmodel}. There is a similar alternation on the auxiliary spaces,
so the integrable $R$-matrix is formed by regrouping an array of $2 \times 2$ elementary $R$-matrices of the usual six-vertex model.
This leads to a 38-vertex model (an $R$-matrix of size $16 \times 16$ with 38 non-zero entries), as first noticed in \cite{IJS2008} and
further reviewed in \cite{Robertson2020}. One central result of \cite{Robertson2020} was to relate this to the integrable $R$-matrix of the
\dtt model.

\subsection{Change of basis} \label{sec-cob}

To interpret the $K$-matrix \eqref{Kmatrixtype3} within the staggered six-vertex model, we briefly review the corresponding change
of basis from \eqref{d22basis} to \eqref{s6vbasis}, first given in \cite{Robertson2020}.
The main idea is to first write the $K$-matrix in an intermediate basis $\{|1\rangle,|\tilde{2}\rangle,|\tilde{3}\rangle,|4\rangle \}$, where
\begin{subequations}
\begin{eqnarray}
|\tilde{2}\rangle &=& \frac{1}{\sqrt{2}}(|2\rangle+|3\rangle) \,, \\
|\tilde{3}\rangle &=& \frac{1}{\sqrt{2}}(|2\rangle-|3\rangle) \,,
\end{eqnarray}
\end{subequations}
and then make one more basis change to \eqref{s6vbasis} as follows:
\begin{subequations}
\begin{eqnarray}
|\tilde{2} \rangle &=&\frac{1}{\sqrt{2\cos\gamma}} \left( e^{\frac{i\gamma}{2}}| \! \uparrow\downarrow\rangle-e^{-\frac{i\gamma}{2}}| \! \downarrow\uparrow\rangle \right) \,. \\
|\tilde{3}\rangle &=& \frac{1}{\sqrt{2\cos\gamma}} \left( e^{-\frac{i\gamma}{2}}| \! \uparrow\downarrow\rangle+e^{\frac{i\gamma}{2}}| \! \downarrow\uparrow\rangle \right) \,.
\end{eqnarray}
\end{subequations}

\subsection{Hamiltonian limit}

The fully anisotropic limit is obtained by sending the spectral parameter $u \to 0$. A quantum Hamiltonian $\mathcal{H}$ is defined as the
logarithmic derivative of the transfer matrix $t(u)$ in that limit. Due to the integrability, we have $[t(u),\mathcal{H}] = 0$, so the two operators
share the same eigenvectors. Therefore, studying the eigenvalue problem of $\mathcal{H}$ defines a quantum spin chain problem which
is equivalent to the original two-dimensional statistical problem, up to an anisotropic rescaling.

In the periodic case, the precise relation between the quantum Hamiltonian and the integrable $R$-matrix is thus
\beq\label{hperiodic}
\mathcal{H}_{\rm periodic}  \equiv \sum\limits_{n=1}^{L} \left. P_{n,n+1}\frac{{\rm d}}{{\rm d}u}R_{n,n+1}(u) \right|_{u=0} \,,
\eeq
where $P_{n,n+1}$ denotes the permutation operator. It was found in \cite{IJS2008} that for the $\mathbb{Z}_2$-staggered six-vertex model
\beq\label{hperiodictl}
\mathcal{H}_{\rm periodic} = \sum\limits_{m=1}^{2L}\left(2\cos\gamma\  e_m-(e_m e_{m+1}+e_{m+1} e_m)\right) \,,
\eeq
where $e_i$ are certain operators acting on sites $i,i+1$ of the chain. The site labels here pertain to the case before the two-by-two regrouping
of spaces---hence to a chain of length $2L$---and the periodic boundary conditions imply that labels are considered modulo $2L$.
More precisely, the $e_i$ are the generators of the (periodic) Temperley-Lieb (TL) algebra, satisfying the abstract algebraic relations%
\footnote{The proper definition of the periodic TL algebra requires a few further elements, but since our aim is to move to the open case,
we shall not discuss them further here.}
\begin{subequations}
\label{TLrelations}
\begin{eqnarray}
e_i^2 &=&  \sqrt{Q} e_i \,, \\
e_i e_{i\pm1}e_i &=& e_i \,, \\
e_i e_j &=& e_je_i \text{ for } |i-j|\geq2 \,.
\end{eqnarray}
\end{subequations}
where $Q$ is related to $\gamma$ by (\ref{qgamma}).

The TL algebra admits several interesting representations. Of particular interest to us is the so-called vertex representation, where the $e_i$ are
written in terms of Pauli matrices $\sigma_i$ as
\beq\label{esigma}
e_i = \frac{1}{2}\left[\sigma_i^x\sigma_{i+1}^x+\sigma_i^y\sigma_{i+1}^y-\cos\gamma\ \sigma_i^z\sigma_{i+1}^z+\cos\gamma-i\sin\gamma\ (\sigma_i^z-\sigma_{i+1}^z) \right] \,,
\eeq
or, in an explicit tensor-product notation,
\beq\label{ei}
e_i=
\mathds{1}^{\otimes i-1}
\otimes
\begin{pmatrix}
	0 & 0 & 0 & 0 \\
	0 & e^{-i\gamma} & 1 & 0\\
	0 & 1 & e^{i\gamma} & 0\\
	0 & 0 & 0 & 0 
\end{pmatrix}
\otimes
\mathds{1}^{\otimes 2L-i-1} \,,
\eeq
where $\mathds{1}$ denotes the $2 \times 2$ identity matrix.
The corresponding Hamiltonian---i.e., \eqref{esigma} inserted into \eqref{hperiodictl}---defines the periodic $\mathbb{Z}_2$-staggered XXZ spin chain,
which was studied extensively in \cite{IJS2008} and subsequent works (see the Introduction).

Turning now to the open case, the relationship between the double-row transfer matrix \eqref{tmatrixopen} and the
 Hamiltonian is (usually---but see the discussion below) given by  \cite{sklyanin1988}
\beq\label{tmatham}
\left. \frac{{\rm d}}{{\rm d}u}t(u) \right|_{u=0}=2H \Tr K^+(0)+\Tr K^{+'}(0) \,,
\eeq
leading to the following expression for the Hamiltonian $H$:
\beq\label{hamintopen}
H=\sum\limits_{n=1}^{L-1}H_{n,n+1}+\frac{1}{2}K^{-'}_1(0)+\frac{\Tr_a(K^+_a(0)H_{L,a})}{\Tr K^+_a(0)} \,,
\eeq
where $a$ is an auxiliary space and where $H_{n,n+1}=P_{n,n+1} \left. \frac{{\rm d}}{{\rm d}u}R_{n,n+1}(u) \right|_{u=0}$.
However, for the $K$-matrix (\ref{Kmatrixtype3}) the quantity $\Tr_a K^+(0)$ vanishes, and hence the $H$ defined by \eqref{tmatham} is ill-defined.
To construct a meaningful Hamiltonian for the $K$-matrix under study, one must take instead the \textit{second} derivative of the transfer matrix,
yielding \cite{Cuerno1993}
\begin{eqnarray}
\label{hamdef}
\mathcal{H} = \frac{t''(0)}{4(T+2A)} &=& \sum\limits_{n=1}^{L-1}H_{n,n+1}+\frac{1}{2}K^{-'}(0)\\
&+&\frac{1}{2(T+2A)} \left( \Tr_a(K^+_a(0)G_{L,a}) + 2\Tr_a(K^{+'}_a(0)H_{L,a}) + \Tr_a(K^+_a(0)H_{L,a}^2) \right) \,, \nonumber
\end{eqnarray}
where the quantities $A,T$ and $G$ are defined by 
\begin{subequations}
\begin{eqnarray}
\Tr_a (K^+_a(0)H_{L,a}) &=& A \mathds{1} \,, \\
T &=& \Tr_a K^{+'}(0) \,, \\
G_{j,j+1} &=& \mathcal{P}_{j,j+1} \left. \frac{{\rm d}^2 R_{j,j+1}}{{\rm d}u^2} \right|_{u=0} \,,
\end{eqnarray}
\end{subequations}
where $\mathds{1}$ is now the $4\times 4$ identity matrix. The Hamiltonian (\ref{hamdef}) reads
\beq\label{hopengen}
\mathcal{H}=A_{\rm left}+A_{\rm right}+\cos\gamma(e_1+e_{2L-1})+2\cos\gamma\sum\limits_{m=2}^{2L-2}e_m-\sum\limits_{m=1}^{2L-2}(e_me_{m+1}+e_{m+1}e_m) \,,
\eeq
where $A_{\rm left}$ and $A_{\rm right}$ denote the second and third terms in (\ref{hamdef}), built from $K^-$ and $K^+$ respectively.
They can be computed from (\ref{Kmatrixtype3}) via the change of basis discussed in section \ref{sec-cob}.
Omitting additive constants proportional to the identity, the results are
\beq\label{lbdry}
A_{\rm left}=\cos\gamma
\begin{pmatrix}
	0 & 0 & 0 & 0 \\
	0 & e^{-i\gamma} & 1 & 0\\
	0 & 1 & e^{i\gamma} & 0\\
	0 & 0 & 0 & 0
\end{pmatrix}
\otimes
\mathbb{I}^{\otimes 2L-2}
\eeq
and
\beq\label{rbdry}
A_{\rm right}=
\mathbb{I}^{\otimes 2L-2}
\otimes
\cos\gamma
\begin{pmatrix}
		0 & 0 & 0 & 0 \\
		0 & e^{-i\gamma} & 1 & 0\\
		0 & 1 & e^{i\gamma} & 0\\
		0 & 0 & 0 & 0
\end{pmatrix} \,.
\eeq
Comparing with \eqref{ei}, the total Hamiltonian (\ref{hopengen}) therefore becomes
\beq\label{hamtype3TL}
\mathcal{H} = 2\cos\gamma\sum\limits_{j=1}^{2L-1}e_j -\sum\limits_{j=1}^{2L-2}(e_j e_{j+1} + e_{j+1}e_j) \,.
\eeq
The simple expression \eqref{hamtype3TL} is a key result of this paper.
When compared to the open Hamiltonian studied in our paper \cite{Robertson2020},
\beq\label{hamtype2TL}
\widetilde{\mathcal{H}}=-\frac{1}{\cos\gamma}(e_1+e_{2L-1})+2\cos\gamma\sum\limits_{m=1}^{2L-1}e_m-\sum\limits_{m=1}^{2L-2}(e_me_{m+1}+e_{m+1}e_m) \,,
\eeq
we notice that the difference is a boundary term, $-\frac{1}{\cos\gamma}(e_1+e_{2L-1})$. 
The new Hamiltonian \eqref{hamtype3TL} appears in some sense very natural, since it is related to the periodic one \eqref{hperiodictl} by a simple
change of the summation ranges. This simplicity, however, does certainly not mean that \eqref{hamtype3TL} is a trivial result. Indeed, suppose
one defined an open model in a naive fashion, by simply replacing $K^\pm(u)$ by the identity matrix. This would lead to
\beq
\mathcal{H}_{\rm naive} \equiv \sum\limits_{n=1}^{L-1}\frac{{\rm d}R_{n,n+1}}{{\rm d}u}\propto \cos\gamma \left(e_1+e_{2L-1}+2\sum\limits_{m=2}^{2L-2}e_m\right)-\sum\limits_{m=1}^{2L-2}(e_me_{m+1}+e_{m+1}e_m)\,,
\eeq
hence again boundary terms, yet different from those of \eqref{hamtype2TL}. But the identity matrix does not satisfy the boundary
Yang-Baxter equation \eqref{reflection}, so $\mathcal{H}_{\rm naive}$ does not define an integrable model! From this perspective,
it is a remarkable fact that the $K$-matrix in (\ref{Kmatrixtype3}) ensures that all the $e_i$ terms in \eqref{hamtype3TL} get the same coefficient.

\subsection{Change of geometry}\label{geomchange}

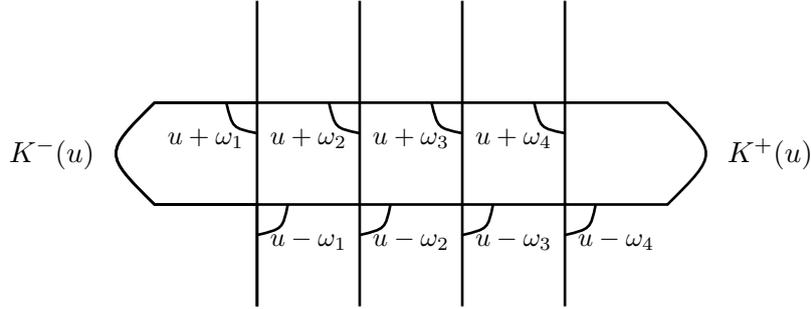
\begin{figure}
	\centering
	\begin{tikzpicture}[scale=1.35]

\draw[black,line width = 1pt] (1,0)--(1,1);

\draw[black,line width = 1pt] (1,1)-- (0,1) .. controls (-0.5,1.5) .. (0,2)--(1,2);
\draw[black,line width = 1pt] (1,1)-- (0,1) .. controls (-0.5,1.5) .. (0,2)--(1,2);

\draw[black,line width = 1pt](4,1)--(5,1) .. controls (5.5,1.5) .. (5,2)--(4,2);

\draw[black,line width = 1pt](1,0)--(1,3);
\draw[black,line width = 1pt](2,0)--(2,3);
\draw[black,line width = 1pt](3,0)--(3,3);
\draw[black,line width = 1pt](4,0)--(4,3);

\draw[black,line width = 1pt](1,1)--(4,1);
\draw[black,line width = 1pt](1,2)--(4,2);

\draw[black,line width = 1pt] (4,0.7) .. controls (4.25,0.76) .. (4.3,1);
\draw[black,line width = 1pt] (3,0.7) .. controls (3.25,0.76) .. (3.3,1);
\draw[black,line width = 1pt] (2,0.7) .. controls (2.25,0.76) .. (2.3,1);
\draw[black,line width = 1pt] (1,0.7) .. controls (1.25,0.76) .. (1.3,1);

\draw[black,line width = 1pt] (0.7,2) .. controls (0.76,1.75) .. (1,1.7);
\draw[black,line width = 1pt] (1.7,2) .. controls (1.76,1.75) .. (2,1.7);
\draw[black,line width = 1pt] (2.7,2) .. controls (2.76,1.75) .. (3,1.7);
\draw[black,line width = 1pt] (3.7,2) .. controls (3.76,1.75) .. (4,1.7);

\node at (4.5,0.65) {\footnotesize{$u-\omega_4$}};
\node at (3.5,0.65) {\footnotesize{$u-\omega_3$}};
\node at (2.5,0.65) {\footnotesize{$u-\omega_2$}};
\node at (1.5,0.65) {\footnotesize{$u-\omega_1$}};

\node at (0.5,1.65) {\footnotesize{$u+\omega_1$}};
\node at (1.5,1.65) {\footnotesize{$u+\omega_2$}};
\node at (2.5,1.65) {\footnotesize{$u+\omega_3$}};
\node at (3.5,1.65) {\footnotesize{$u+\omega_4$}};

\node at (6,1.5) {\small{$K^+(u)$}};
\node at (-1,1.5) {\small{$K^-(u)$}};

\end{tikzpicture}
\caption{Geometrical interpretation of the transfer matrix \eqref{tmatrixopendefects}. When the parameters $\omega_i$ are set to zero we recover the transfer matrix \eqref{tmatrixopen}.}\label{tmatgeneral}
\end{figure}

The double-row transfer matrix (\ref{tmatrixopen}) has the diagrammatic interpretation shown in Figure \ref{tmatgeneral}.
This ``integrable geometry'' is different from the ``diagonal geometry'' of Figures \ref{orientloop}--\ref{vertexmodel}, in which the Potts spins reside on
an axially oriented square lattice, while the vertices of the corresponding six-vertex model are oriented diagonally (cf.\ Figure \ref{vertexmodel}).
Fortunately the two geometries can be related via a simple construction \cite{Destri1992,BatchelorYung1994} that we now review.
The resulting change of geometry will ultimately allow us to interpret
the integrable $K$-matrix (\ref{Kmatrixtype3}) in terms of the diagonal geometry, which is more convenient for
describing the AF Potts model.

\medskip

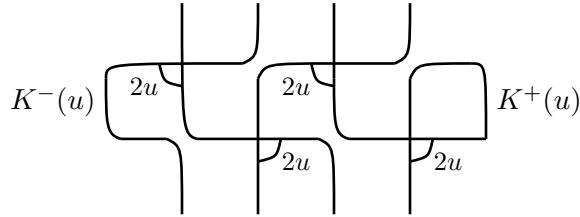
\begin{figure}
	\centering
	\begin{tikzpicture}

\draw[black,line width = 1pt] (4,0)--(4,1.8);

\draw[black,line width = 1pt] (4,1.8) .. controls (4.1,2) .. (4.8,2);
\draw[black,line width = 1pt] (4.8,2) .. controls (5,2) .. (5,1);

\draw[black,line width = 1pt] (3.2,1)--(5,1);

\draw[black,line width = 1pt] (3.2,1) .. controls (3,1.05) and (3,1.2) .. (3,1.8);

\draw[black,line width = 1pt] (3,0) .. controls (3,0.9) .. (2.8,1);

\draw[black,line width = 1pt] (2.8,1)--(1.2,1);
\draw[black,line width = 1pt] (1.2,1) .. controls (1,1.05) and (1,1.2) .. (1,2.8);

\draw[black,line width = 1pt] (2,0)--(2,1.8);

\draw[black,line width = 1pt] (1,0) .. controls (1,0.9) .. (0.8,1);

\draw[black,line width = 1pt] (0.8,1)--(0.2,1);

\draw[black,line width = 1pt] (0.2,1) .. controls (0,1.05) and (0,1.2) .. (0,1.8);

\draw[black,line width = 1pt] (0,1.8) .. controls (0,2) .. (1.8,2);

\draw[black,line width = 1pt] (1.8,2) .. controls (2,2.1) .. (2,2.8);

\draw[black,line width = 1pt] (2,1.8) .. controls (2.1,2) .. (3,2);

\draw[black,line width = 1pt] (3,2)--(3.8,2);

\draw[black,line width = 1pt] (3,1.8)--(3,2.8);

\draw[black,line width = 1pt] (3.8,2) .. controls (4,2.1) .. (4,2.8);

\draw[black,line width = 1pt] (4,0.7) .. controls (4.25,0.76) .. (4.3,1);
\draw[black,line width = 1pt] (2,0.7) .. controls (2.25,0.76) .. (2.3,1);

\draw[black,line width = 1pt] (0.7,2) .. controls (0.76,1.75) .. (1,1.7);
\draw[black,line width = 1pt] (2.7,2) .. controls (2.76,1.75) .. (3,1.7);

\node at (4.5,0.7) {\footnotesize{$2u$}};
\node at (2.5,0.7) {\footnotesize{$2u$}};

\node at (0.5,1.7) {\footnotesize{$2u$}};
\node at (2.5,1.7) {\footnotesize{$2u$}};

\node at (5.7,1.5) {\small{$K^+(u)$}};
\node at (-0.7,1.5) {\small{$K^-(u)$}};

\end{tikzpicture}
\caption{The geometrical interpretation of the transfer matrix in (\ref{tmatrixopendefects}) when the parameters $\omega_j$ are given by equation (\ref{defectchoice}).}\label{tmatdefects}
\end{figure}

\begin{figure}
	\centering
	\begin{tikzpicture}

\draw[black,line width = 1pt] (1.5,1.5)--(3.5,3.5);
\draw[black,line width = 1pt] (0,2)--(1.5,3.5);
\draw[black,line width = 1pt] (3.5,1.5)--(4,2);

\draw[black,line width = 1pt] (0.5,1.5)--(0,2);
\draw[black,line width = 1pt] (2.5,1.5)--(0.5,3.5);
\draw[black,line width = 1pt] (4,2)--(2.5,3.5);

\draw[black,line width = 1pt] (4,2).. controls (4.8,2.8) .. (5.6,2);
\draw[black,line width = 1pt] (5.6,2).. controls (4.8,1.2) .. (4,2);

\draw[black,line width = 1pt] (0.7,2.7).. controls (1,2.5) .. (1.3,2.7);
\node at (1,2.3) {$2u$};

\draw[black,line width = 1pt] (2.7,2.7).. controls (3,2.5) .. (3.3,2.7);
\node at (3,2.3) {$2u$};

\draw[black,line width = 1pt] (1.7,1.7).. controls (2,1.5) .. (2.3,1.7);
\node at (2,1.3) {$2u$};

\draw[black,line width = 1pt] (3.7,1.7).. controls (4,1.5) .. (4.3,1.7);
\node at (4,1.3) {$2u$};

\node at (-0.6,2) {$K^-(u)$};

\node at (7.5,2) {$\Tr_0(K_0(u)^+R_{N0}(2u))$};

\end{tikzpicture}
\caption{A rearrangement of the lattice in Figure \ref{tmatdefects}---the two lattices are topologically equivalent and correspond to the same transfer matrix.}\label{tmatdiag}
\end{figure}
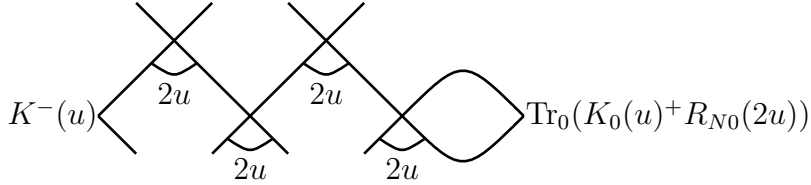
The transfer matrix \eqref{tmatrixopen} can be modified to include inhomogeneous spectral parameters $\omega_n$ on each
quantum space $n=1,2,\ldots,L$, while conserving the integrability of the model. We keep the spectral parameter $u$ on the auxiliary spaces.
The resulting inhomogeneous transfer matrix reads
\beq\label{tmatrixopendefects}
t(u)=\Tr_a K_a^+(u)R_{a,L}(u+\omega_L) \cdots R_{a,1}(u+\omega_1)K_a^-(u)R_{1,a}(u-\omega_1) \cdots R_{L,a}(u-\omega_L)
\eeq
and is depicted graphically in Figure \ref{tmatgeneral}. The arguments of each $R$-matrix are given by the usual difference property.
The choice \cite{Destri1992,BatchelorYung1994}
\beq\label{defectchoice}
\omega_n=(-1)^{n+1}u
\eeq
implies that the arguments of the $R$-matrices alternate between $0$ and $2u$, in a checkerboard pattern. Since $R_{i,j}(0) \propto P_{i,j}$,
the permutation operator, the choice \eqref{defectchoice} can be graphically depicted as in Figure \ref{tmatdefects}, which can in turn be redrawn
as Figure \ref{tmatdiag}. This is precisely the diagonal geometry required by Figures \ref{fulllattice}--\ref{orientloop}. In this geometry, the Boltzmann weights of
bulk vertices are given by $R(2u)$, while the left and right boundary weights are $K^-(u)$ and $\Tr_a(K_a(u)^+R_{L,a}(2u))$ respectively.

\subsection{Potts transfer matrix}\label{sectmat}

We now work out the corresponding transfer matrix for the AF Potts model, by which we mean the transfer matrix describing the Potts spins in the
geometry of Figures \ref{fulllattice}--\ref{orientloop}. First recall that each space in the integrable \dtt model carries the four different states \eqref{d22basis}, which can be related to a tensor product of two spin-$\frac{1}{2}$ spaces \eqref{s6vbasis} by means of the basis change of section \ref{sec-cob}. To revert to the six-vertex model we must thus replace each edge in Figure \ref{tmatdiag} by a pair of edges, each of which can be in two possible states, $|\!\! \uparrow\rangle$ and $|\!\!\downarrow\rangle$. Under this doubling, the geometry in Figure \ref{tmatdiag} becomes the lattice shown in Figure \ref{tmatdiagboundaries1}, where the dotted lines
represent the edges carrying the six-vertex arrows, while the full lines are the surrounding tiles (like in Figure \ref{orientloop}).

The staggered six-vertex model $R$-matrix can be written in terms of Temperley-Lieb operators as \cite{IJS2008}
\beq\label{rxei2}
R_{i,i+1}=(x+e_{2i})(1+xe_{2i-1})(1+xe_{2i+1})(x+e_{2i}) \,.
\eeq
In the doubled lattice (Figure \ref{tmatdiagboundaries1}) each of the four factors then corresponds to the interaction between
a pair of dotted lines within one tile. Making this identification for each $R$-matrix yields the interactions of the full lattice model, as shown
in Figure \ref{tmatdiagboundaries2}.
\begin{figure}
	\centering
	\begin{tikzpicture}[scale=1.5]
		
\draw[black,line width = 1pt] (2,0)--(0,2)--(0,6)--(2,8)--(4,6)--(2,4)--(4,2)--(2,0);
\draw[black,line width = 1pt] (0,6)--(2,4)--(0,2);

\node at (4.5,4) {$\cdots$};

\draw[black,line width = 1pt] (7,0)--(5,2)--(7,4)--(5,6)--(7,8)--(9,6)--(7,4)--(9,2)--(7,0);

\draw[black,line width = 1pt] (9,2)--(9,6);

\draw[black,line width = 1pt] (1,1)--(3,3);
\draw[black,line width = 1pt] (3,1)--(1,3);
\draw[black,line width = 1pt] (1,5)--(3,7);
\draw[black,line width = 1pt] (3,5)--(1,7);

\draw[black,line width = 1pt] (6,1)--(8,3);
\draw[black,line width = 1pt] (8,1)--(6,3);
\draw[black,line width = 1pt] (6,5)--(8,7);
\draw[black,line width = 1pt] (8,5)--(6,7);

\draw[black, dotted, line width = 1pt] (1.5,0.5)--(3.5,2.5);
\draw[black, dotted, line width = 1pt] (0.5,1.5)--(2.5,3.5);
\draw[black, dotted, line width = 1pt] (0.5,2.5)--(2.5,0.5);
\draw[black, dotted, line width = 1pt] (1.5,3.5)--(3.5,1.5);

\draw[black, dotted, line width = 1pt] (1.5,4.5)--(3.5,6.5);
\draw[black, dotted, line width = 1pt] (0.5,5.5)--(2.5,7.5);
\draw[black, dotted, line width = 1pt] (0.5,6.5)--(2.5,4.5);
\draw[black, dotted, line width = 1pt] (1.5,7.5)--(3.5,5.5);

\draw[black, dotted, line width = 1pt] (6.5,0.5)--(8.5,2.5);
\draw[black, dotted, line width = 1pt] (5.5,1.5)--(7.5,3.5);
\draw[black, dotted, line width = 1pt] (5.5,2.5)--(7.5,0.5);
\draw[black, dotted, line width = 1pt] (6.5,3.5)--(8.5,1.5);

\draw[black, dotted, line width = 1pt] (6.5,4.5)--(8.5,6.5);
\draw[black, dotted, line width = 1pt] (5.5,5.5)--(7.5,7.5);
\draw[black, dotted, line width = 1pt] (5.5,6.5)--(7.5,4.5);
\draw[black, dotted, line width = 1pt] (6.5,7.5)--(8.5,5.5);

\end{tikzpicture}
\caption{The lattice from Figure \ref{tmatdiag} but with each edge replaced by two spin-$\frac{1}{2}$ edges, shown as dotted lines. The full lines represent the tiles that surround each vertex.}\label{tmatdiagboundaries1}
\end{figure}
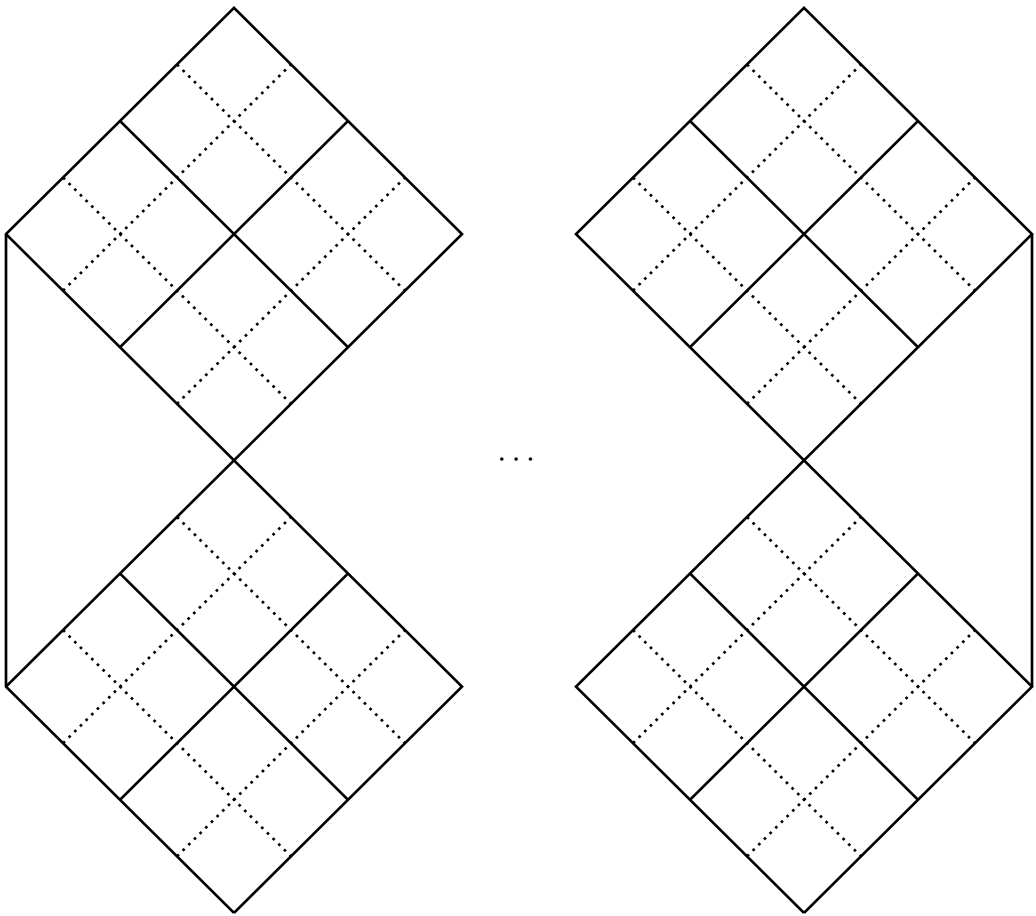
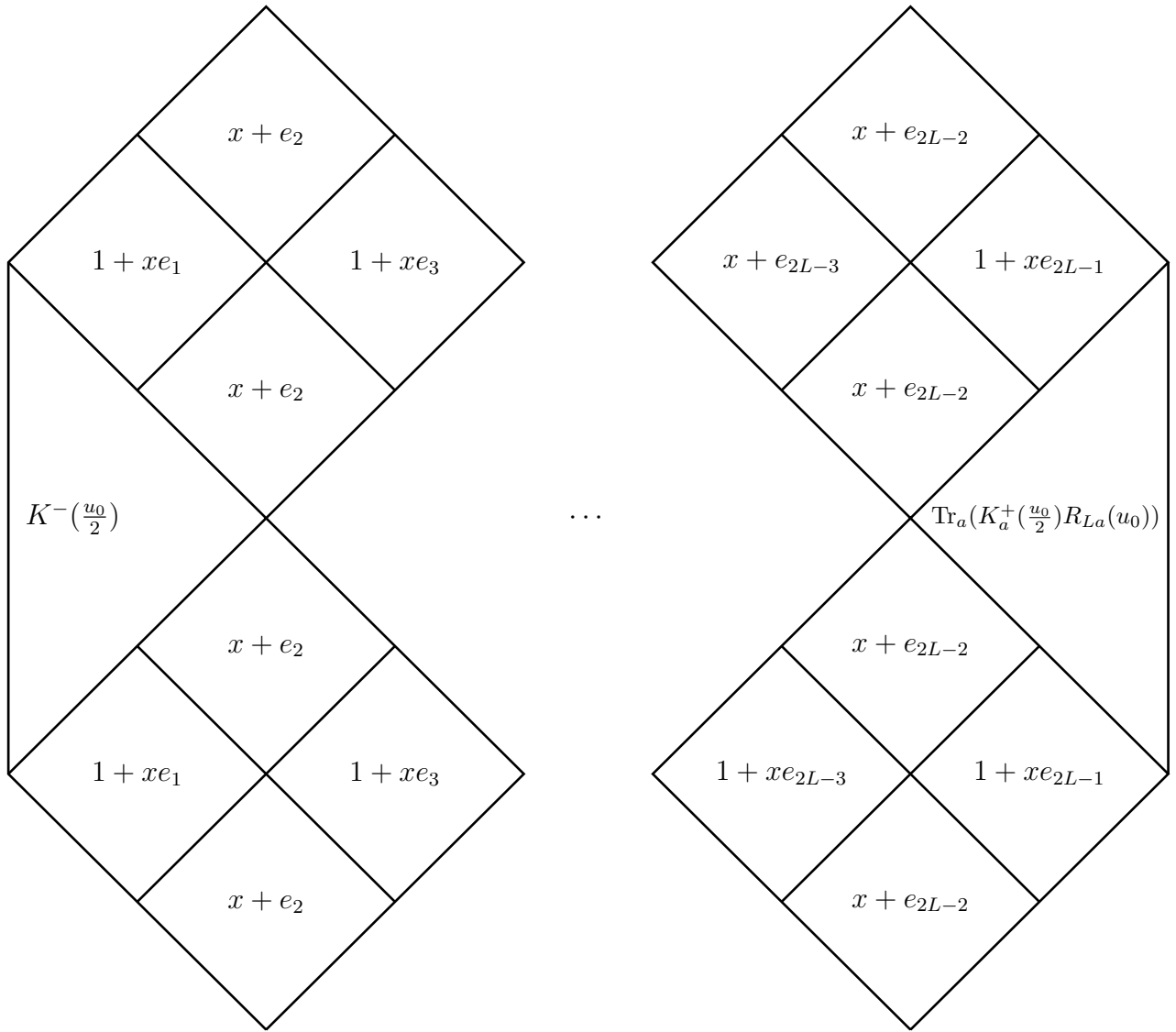
\begin{figure}
	\centering
	\begin{tikzpicture}[scale=1.85]
		
\draw[black,line width = 1pt] (2,0)--(0,2)--(0,6)--(2,8)--(4,6)--(2,4)--(4,2)--(2,0);
\draw[black,line width = 1pt] (0,6)--(2,4)--(0,2);

\node at (4.5,4) {$\cdots$};

\draw[black,line width = 1pt] (7,0)--(5,2)--(7,4)--(5,6)--(7,8)--(9,6)--(7,4)--(9,2)--(7,0);

\draw[black,line width = 1pt] (9,2)--(9,6);

\draw[black,line width = 1pt] (1,1)--(3,3);
\draw[black,line width = 1pt] (3,1)--(1,3);
\draw[black,line width = 1pt] (1,5)--(3,7);
\draw[black,line width = 1pt] (3,5)--(1,7);

\draw[black,line width = 1pt] (6,1)--(8,3);
\draw[black,line width = 1pt] (8,1)--(6,3);
\draw[black,line width = 1pt] (6,5)--(8,7);
\draw[black,line width = 1pt] (8,5)--(6,7);

\node at (2,1) {$x+e_2$};
\node at (2,3) {$x+e_2$};
\node at (2,5) {$x+e_2$};
\node at (2,7) {$x+e_2$};

\node at (1,2) {$1+xe_1$};
\node at (1,6) {$1+xe_1$};
\node at (3,2) {$1+xe_3$};
\node at (3,6) {$1+xe_3$};

\node at (7,1) {$x+e_{2L-2}$};
\node at (7,3) {$x+e_{2L-2}$};
\node at (7,5) {$x+e_{2L-2}$};
\node at (7,7) {$x+e_{2L-2}$};

\node at (6,2) {$1+xe_{2L-3}$};
\node at (6,6) {$x+e_{2L-3}$};

\node at (8,2) {$1+xe_{2L-1}$};
\node at (8,6) {$1+xe_{2L-1}$};

\node at (0.5,4) {$K^-(\frac{u_0}{2})$};

\node at (8.05,4) {\footnotesize{$\Tr_a(K_a^+(\frac{u_0}{2})R_{La}(u_0))$}};

\end{tikzpicture}
\caption{Same as Figure \ref{tmatdiagboundaries1}, but with the action of the $R$-matrix and $K$-matrix written at the place of each vertex, in terms of Temperley-Lieb generators.}\label{tmatdiagboundaries2}
\end{figure}
In general, the para\-meter $x$ appearing in \eqref{rxei2} corresponds to the spectral-parameter dependent function \cite{JS-AF}
\beq\label{xspectparam}
x(u)=\frac{\sin u}{\sin(\gamma-u)} \,,
\eeq
but in order for the {\em same} parameter $x$ to appear on even and odd tiles, as in \eqref{rxei2}, it is crucial to evaluate $x(u)$ at a value of $u$ that corresponds to
an {\em isotropic point} of the integrable vertex model \cite{JS-AF}. The isotropic point of relevance to us is
\beq\label{u0def}
u_0=\frac{\gamma}{2}-\frac{\pi}{4} \,,
\eeq
for which we have
\beq \label{xu0trig}
x(u_0)=\frac{\sin\frac{\gamma}{2} -\cos\frac{\gamma}{2}}{\sin\frac{\gamma}{2} +\cos\frac{\gamma}{2}} = x_{\rm AF}^+ \,,
\eeq
according to eqs.\ \eqref{xAF} and \eqref{qgamma}. Indeed this $u_0$ belongs to the so-called Regime I of the spectral parameter $u$,
\beq\label{regime1}
{\rm Regime\ I}: \quad \gamma-\frac{\pi}{2} < u < 0 \,,
\eeq
for which the continuum limit of the model with periodic boundary conditions is described by a non-compact CFT \cite{IJS2008, ikhlef2010}. We remark that the range of $u$ in \eqref{regime1} that we use to define Regime I is different from the range used in \cite{ikhlef2010}. In that case, Regime I was defined as
\beq\label{regime1b}
\gamma < u < \frac{\pi}{2} \,.
\eeq
The two ranges of $u$, i.e., \eqref{regime1} and \eqref{regime1b}, are related by the {\em duality transformation} (to be discussed in more detail below) $u \rightarrow \gamma - u$, under which $x(\gamma -u) = \frac{1}{x(u)}$ from \eqref{xspectparam}. When periodic boundary conditions are imposed, both ranges \eqref{regime1} and \eqref{regime1b} correspond to a non-compact CFT in the continuum limit. In addition to the isotropic point in \eqref{u0def} which lies in the range \eqref{regime1}, there is an additional isotropic point,
\beq\label{u0primedef}
u_0' = \frac{\gamma}{2}+\frac{\pi}{4} \,,
\eeq
which lies in the range \eqref{regime1b} and is related to $u_0$ by the duality transformation $u \rightarrow \gamma - u$. We have that
\begin{equation}
 \label{xAFminus}
 x(u_0') = x_{\rm AF}^- \,.
\end{equation}
We shall for now focus on the point $u=u_0$ and remark on $u=u_0'$ later. We set
\begin{equation}\label{xAFu0}
 x \equiv x(u_0) = x_{\rm AF}^+ \,.
\end{equation}

It remains to determine the boundary interactions in Figure \ref{tmatdiagboundaries2}. According to 
Figures \ref{tmatdefects}--\ref{tmatdiag}, if the $R$ matrix is evaluated at $u$, then $K$ must be evaluated at $\frac{u}{2}$.
At the left boundary we therefore have, using \eqref{Kmatrixtype3},
\beq\label{Kbyung}
\begingroup
K^-\left(\frac{u_0}{2}\right)=
\renewcommand*{\arraystretch}{2}
\begin{pmatrix}
	\frac{\sin\frac{\gamma}{2}+\cos\frac{\gamma}{2}}{\sin\frac{3\gamma}{2}-\cos\frac{3\gamma}{2}} & 0 & 0 & 0 \\
	0 & \frac{\sin\gamma(\cos\frac{\gamma}{2}+\sin\frac{\gamma}{2})}{\sin\frac{3\gamma}{2}-\cos\frac{3\gamma}{2}} & \frac{\cos\gamma(\cos\frac{\gamma}{2}-\sin\frac{\gamma}{2})}{\sin\frac{3\gamma}{2}-\cos\frac{3\gamma}{2}} & 0\\
	0 & \frac{\cos\gamma(\cos\frac{\gamma}{2}-\sin\frac{\gamma}{2})}{\sin\frac{3\gamma}{2}-\cos\frac{3\gamma}{2}} & \frac{\sin\gamma(\cos\frac{\gamma}{2}+\sin\frac{\gamma}{2})}{\sin\frac{3\gamma}{2}-\cos\frac{3\gamma}{2}} & 0\\
	0 & 0 & 0 & \frac{\sin\frac{\gamma}{2}+\cos\frac{\gamma}{2}}{\sin\frac{3\gamma}{2}-\cos\frac{3\gamma}{2}} 
\end{pmatrix}
\endgroup
\eeq
in the basis \eqref{d22basis} of the \dtt model. Rewriting this in the six-vertex basis \eqref{s6vbasis}, using the basis change of section \ref{sec-cob},
we obtain
\beq
K^-\left(\frac{u_0}{2}\right)\propto \mathds{1} + \left(\frac{\sin\frac{\gamma}{2} -\cos\frac{\gamma}{2}}{\sin\frac{\gamma}{2} +\cos\frac{\gamma}{2}}\right) e_1 \,.
\eeq
Recalling \eqref{xu0trig} and \eqref{xAFu0} we finally arrive at
\beq\label{kminusu0}
K^-\left(\frac{u_0}{2}\right)\propto \mathds{1}+x e_1 \,.
\eeq
A similar calculation for the right boundary leads to%
\footnote{To arrive at the result in (\ref{tracekplus}) one must be careful with the form of the $R$-matrix used. As discussed in \cite{Robertson2020},
the $R$-matrix of the \dtt model differs from that of the staggered six-vertex model by some minus signs of certain matrix components
(although the transfer matrices of both models are identical). The result in (\ref{tracekplus}) is valid only for the \dtt $R$-matrix.}
\beq\label{tracekplus}
\Tr_a \left( K_a^+\left(\frac{u_0}{2}\right)R_{L,a}(u_0) \right) \propto \mathds{1} +x e_{2L-1} \,.
\eeq
\begin{figure}
	\centering
	\begin{tikzpicture}[scale=0.65]
		
\draw[black,line width = 1pt] (0,4)--(2,6)--(4,4)--(8,8)--(12,4)--(16,8)--(20,4)--(22,6)--(24,4);
\draw[black,line width = 1pt] (0,4)--(2,2)--(4,4)--(8,0)--(12,4)--(16,0)--(20,4)--(22,2)--(24,4);

\draw[black,line width = 1pt] (2,6)--(0,8)--(4,12)--(8,8)--(12,12)--(16,8)--(20,12)--(24,8)--(22,6);

\draw[black, dotted, line width = 1pt] (2,6)--(6,10);
\draw[black, dotted, line width = 1pt] (2,10)--(10,2);

\draw[black, dotted, line width = 1pt] (6,2)--(14,10);
\draw[black, dotted, line width = 1pt] (10,10)--(18,2);

\draw[black, dotted, line width = 1pt] (14,2)--(22,10);
\draw[black, dotted, line width = 1pt] (18,10)--(22,6);

\node at (2,4) {$1+xe_1$};
\node at (6,4) {$1+xe_3$};
\node at (10,4) {$1+xe_5$};
\node at (14,4) {$1+xe_7$};
\node at (18,4) {$1+xe_9$};
\node at (22,4) {$1+xe_{11}$};

\node at (2,8) {$1+xe_1$};
\node at (6,8) {$1+xe_3$};
\node at (10,8) {$1+xe_5$};
\node at (14,8) {$1+xe_7$};
\node at (18,8) {$1+xe_9$};
\node at (22,8) {$1+xe_{11}$};

\node at (4,6) {$x+e_2$};
\node at (8,6) {$x+e_4$};
\node at (12,6) {$x+e_6$};
\node at (16,6) {$x+e_8$};
\node at (20,6) {$x+e_{10}$};

\node at (4,10) {$x+e_2$};
\node at (12,10) {$x+e_6$};
\node at (20,10) {$x+e_{10}$};

\node at (8,2) {$x+e_4$};
\node at (16,2) {$x+e_8$};

\end{tikzpicture}
\caption{The transfer matrix $T$ defined in (\ref{bigt}) with $L=6$, corresponding to the integrable transfer matrix in (\ref{tmatdefects}) evaluated at $\frac{u_0}{2}$ and with defects $\omega_i$ defined in (\ref{defectchoice}).}\label{tmatgeometry1}
\end{figure}
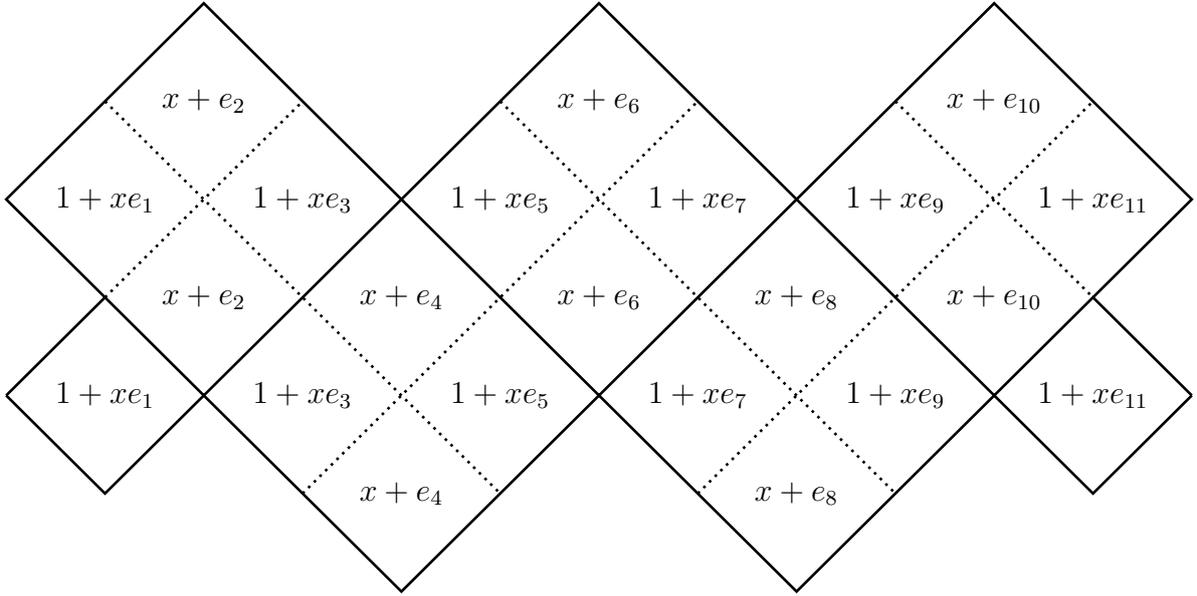
\begin{figure}
	\centering
	\begin{tikzpicture}[scale=0.65]
		
\draw[black,line width = 1pt] (0,4)--(4,0)--(8,4)--(12,0)--(16,4)--(20,0)--(24,4);
\draw[black,line width = 1pt] (2,2)--(4,4)--(6,2)--(8,4)--(10,2)--(12,4)--(14,2)--(16,4)--(18,2)--(20,4)--(22,2)--(24,4);
\draw[black,line width = 1pt] (6,2)--(8,0)--(10,2);
\draw[black,line width = 1pt] (14,2)--(16,0)--(18,2);

\draw[black,line width = 1pt] (0,4)--(2,6)--(4,4)--(6,6)--(8,4)--(10,6)--(12,4)--(14,6)--(16,4)--(18,6)--(20,4)--(22,6)--(24,4);

\node at (2,4) {$1+xe_1$};
\node at (6,4) {$1+xe_3$};
\node at (10,4) {$1+xe_5$};
\node at (14,4) {$1+xe_7$};
\node at (18,4) {$1+xe_9$};
\node at (22,4) {$1+xe_{11}$};

\node at (4,2) {$x+e_2$};
\node at (8,2) {$x+e_4$};
\node at (12,2) {$x+e_6$};
\node at (16,2) {$x+e_8$};
\node at (20,2) {$x+e_{10}$};

\end{tikzpicture}
\caption{The simpler transfer matrix $T_1$ defined in (\ref{tmatafdual}) with $L=6$. This transfer matrix is not integrable but describes the same model as the transfer matrix $T_1$, shown in Figure \ref{tmatgeometry1}, which is integrable.}\label{tmatgeometry2}
\end{figure}
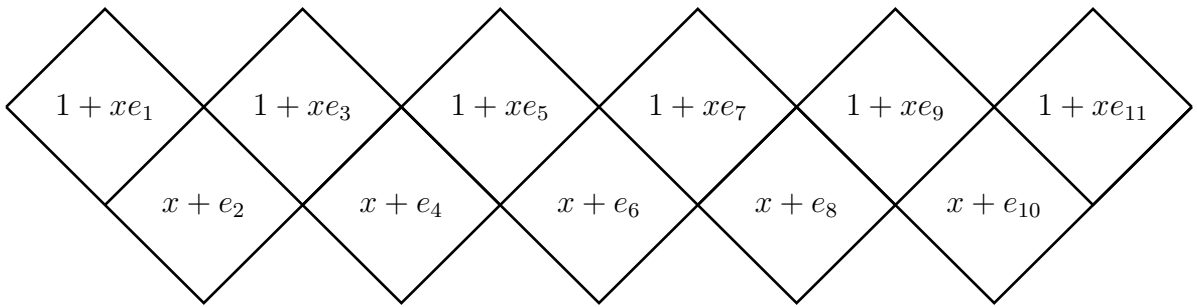

The two results \eqref{kminusu0}--\eqref{tracekplus} allow us to complete Figure \ref{tmatdiagboundaries2}: the interactions within the boundary
triangles simply take (up to proportionality coefficients) the same form as that of the tiles within the corresponding columns. The full transfer matrix defined by \eqref{tmatrixopendefects} and \eqref{defectchoice} is written as a product of transfer matrices $t_1$, $t_2$, $t_3$, $t_4$:
\beq\label{bigt}
T = t_4 t_2 t_3 t_2 t_1 \,,
\eeq
where
\begin{subequations}
\label{t1t2t3t4}
\begin{eqnarray}
t_1 &=& (x+e_4)(x+e_8) \cdots (x+e_{2L-4}) \,, \\
t_2 &=& (1+xe_1)(1+xe_3) \cdots (1+xe_{2L-1}) \,, \\
t_3 &=& (x+e_2)(x+e_4) \cdots (x+e_{2L-2}) \,, \\
t_4 &=& (x+e_2)(x+e_6) \cdots (x+e_{2L-2}) \,.
\end{eqnarray}
\end{subequations}
This is depicted graphically for $L=6$ in Figure \ref{tmatgeometry1}. By cyclicity of the factors, it is equivalent to consider
\begin{equation} \label{tmatafdual}
 \widetilde{T} = t_2 t_3 \times t_2 t_1 t_4 = (t_2 t_3)^2 \equiv (T_1)^2 \,,
\end{equation}
where the simper transfer matrix $T_1$, corresponding to one row of even tiles (with generators $e_{2i}$) and one row of odd tiles (with $e_{2i-1}$), is shown graphically in Figure \ref{tmatgeometry2}.

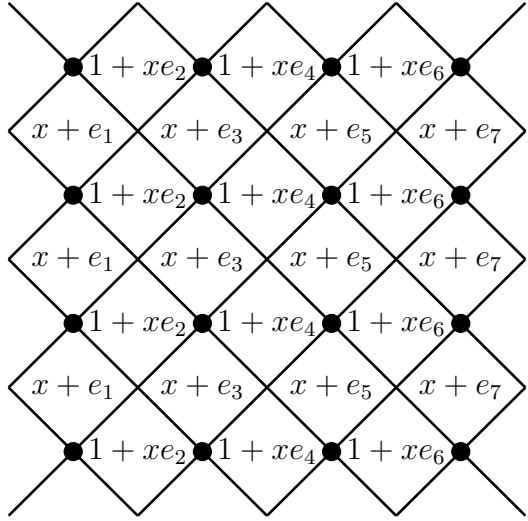
\begin{figure}

\centering
\begin{tikzpicture}[scale=0.85]

\draw[black,line width = 1pt](1,1)--(9,9);
\draw[black,line width = 1pt](3,1)--(9,7);
\draw[black,line width = 1pt](5,1)--(9,5);
\draw[black,line width = 1pt](7,1)--(9,3);

\draw[black,line width = 1pt](1,3)--(7,9);
\draw[black,line width = 1pt](1,5)--(5,9);
\draw[black,line width = 1pt](1,7)--(3,9);

\draw[black,line width = 1pt](9,1)--(1,9);
\draw[black,line width = 1pt](7,1)--(1,7);
\draw[black,line width = 1pt](5,1)--(1,5);
\draw[black,line width = 1pt](3,1)--(1,3);

\draw[black,line width = 1pt](9,3)--(3,9);
\draw[black,line width = 1pt](9,5)--(5,9);
\draw[black,line width = 1pt](9,7)--(7,9);

\filldraw[black] (2,2) circle (4pt);
\filldraw[black] (4,2) circle (4pt);
\filldraw[black] (6,2) circle (4pt);
\filldraw[black] (8,2) circle (4pt);

\filldraw[black] (2,4) circle (4pt);
\filldraw[black] (4,4) circle (4pt);
\filldraw[black] (6,4) circle (4pt);
\filldraw[black] (8,4) circle (4pt);

\filldraw[black] (2,6) circle (4pt);
\filldraw[black] (4,6) circle (4pt);
\filldraw[black] (6,6) circle (4pt);
\filldraw[black] (8,6) circle (4pt);

\filldraw[black] (2,8) circle (4pt);
\filldraw[black] (4,8) circle (4pt);
\filldraw[black] (6,8) circle (4pt);
\filldraw[black] (8,8) circle (4pt);

\node at (2,3) {$x+e_1$};
\node at (2,5) {$x+e_1$};
\node at (2,7) {$x+e_1$};

\node at (4,3) {$x+e_3$};
\node at (4,5) {$x+e_3$};
\node at (4,7) {$x+e_3$};

\node at (6,3) {$x+e_5$};
\node at (6,5) {$x+e_5$};
\node at (6,7) {$x+e_5$};

\node at (8,3) {$x+e_7$};
\node at (8,5) {$x+e_7$};
\node at (8,7) {$x+e_7$};

\node at (3,2) {$1+xe_2$};
\node at (3,4) {$1+xe_2$};
\node at (3,6) {$1+xe_2$};
\node at (3,8) {$1+xe_2$};

\node at (5,2) {$1+xe_4$};
\node at (5,4) {$1+xe_4$};
\node at (5,6) {$1+xe_4$};
\node at (5,8) {$1+xe_4$};

\node at (7,2) {$1+xe_6$};
\node at (7,4) {$1+xe_6$};
\node at (7,6) {$1+xe_6$};
\node at (7,8) {$1+xe_6$};

\end{tikzpicture}

 \captionof{figure}{The AF Potts model with free boundary conditions, described by the graphical expansion of the transfer matrix (\ref{tmataf}).}
 \label{l1}
\end{figure}

The analysis at the other isotropic point $u_0'$ can be carried out similarly. The result is that the expressions \eqref{rxei2} for the $R$-matrix,
\eqref{kminusu0} and \eqref{tracekplus} for the $K$-matrices, and \eqref{t1t2t3t4} for the transfer matrices are only modified by replacing each
factor of type $(1+x e_i)$ by $(x+e_i)$, and vice versa. In particular, the final one-row transfer matrix $T_1$ given by \eqref{tmatafdual} reads
now, at $u = u_0'$,
\begin{equation}
\label{tmataf}
 T_1' =(x+e_1)(x+e_3)\cdots(x+e_{2L-1})(1+xe_2)(1+xe_4)\cdots(1+xe_{2L-2}) \,.
\end{equation}
Comparing the graphical expansion of $T_1'$ (Figure~\ref{l1}) with the lattice on which the Potts model was originally defined
(Figure~\ref{fulllattice}), it is seen that $T_1'$ is precisely the transfer matrix of a Potts model with free boundary conditions
and temperature parameter $x$. Section \ref{secexactsoln} will present a complete Bethe Ansatz solution of the transfer matrix in \eqref{tmatrixopendefects} with the $K$-matrix in \eqref{Kmatrixtype3} for all values of $u$ and $\omega_i$. The transfer matrices in \eqref{tmatafdual} and \eqref{tmataf} are special cases to which our exact solution of the general case can be specialised.

The exchange of the factors of types $(1+x e_i)$ and $(x+e_i)$ is obviously equivalent to applying the transformation ${\cal D}: x \mapsto \frac{1}{x}$,
up to an irrelevant overall factor. But ${\cal D}$ is nothing but the familiar duality transformation of the Potts model
(see \cite{Robertson2019, JS-AF}). A horizontal (resp.\ vertical) coupling between a pair of Potts spins
situated diagonally across a tile is mapped, under duality, to a vertical (resp.\ horizontal) coupling between the
corresponding pair of dual Potts spins, situated across the other diagonal of the same tile. It follows that
when ${\cal D}$ is applied to a lattice with free boundary conditions, the result is a lattice where all the spins along
the boundary are coupled to a single exterior (dual) spin by a layer of horizontal couplings. Such boundary conditions are
called {\em wired}. One can imagine the single exterior spin to be the contraction of one spin per row, all constrained to take the same value. This argument
shows that wired boundary conditions can be identified with fixed boundary conditions, but where the ``fixed'' spin is being summed over all $Q$ values.

We can now summarise the findings of this section.
The transfer matrix (\ref{tmatrixopendefects}) with the boundary condition defined by the $K$-matrix \eqref{Kmatrixtype3},
evaluated at the isotropic point $\frac{u_0}{2}$, and with the parameters $\omega_i$ defined in \eqref{defectchoice}, is equivalent to either of the following two situations:
\begin{itemize}
\item An AF Potts model at temperature parameter $x_{\rm AF}^-$ with {\em free} boundary conditions along both sides of the strip.
\item An AF Potts model at temperature parameter $x_{\rm AF}^+$ with {\em wired} boundary conditions along both sides of the strip.
\end{itemize}

Similarly, the transfer matrix (\ref{tmatrixopendefects}), with the boundary condition defined by the $K$-matrix \eqref{Kmatrixtype3}, and with the parameters $\omega_i$ defined in \eqref{defectchoice},
evaluated at the isotropic point $\frac{u_0'}{2}$, is equivalent to either of the following two situations:
\begin{itemize}
\item An AF Potts model at temperature parameter $x_{\rm AF}^+$ with {\em free} boundary conditions along both sides of the strip.
\item An AF Potts model at temperature parameter $x_{\rm AF}^-$ with {\em wired} boundary conditions along both sides of the strip.
\end{itemize}

\begin{comment}
\jesper{Added:}
Before concluding this section, we briefly remark on the choice of regime for the integrable model. The spectral-parameter
dependent coupling constant $x(u)$ in \eqref{xspectparam} is obviously $\pi$-periodic. It also satisfies
%
\begin{equation}
 x\left(u+\frac{\pi}{2}\right) = \frac{1}{x(u)} \,,
\end{equation}
%
so a shift $u \mapsto u + \frac{\pi}{2}$ is just equivalent to a duality transformation. The Regime I, as written in \eqref{regime1}, is thus related
to the range $\gamma < u < \frac{\pi}{2}$ used in \cite{ikhlef2010} by  a duality transformation. The two isotropic points are similarly
related, $u_0' = u_0 + \frac{\pi}{2}$. Another Regime II, defined by $\frac{\pi}{2} < u < \gamma + \frac{\pi}{2}$ (or equivalently
$0 < u < \gamma$, by duality), was analysed in \cite{ikhlef2010} and found to give rise to a {\em compact} CFT,
different from the Euclidean black hole CFT. We stress that our treatment of boundary conditions in the present paper only apply to
the first, non-compact Regime I.
\end{comment}

\section{Finding an exact solution}\label{secexactsoln}

The preceding sections have shown that the $K$-matrix \eqref{Kmatrixtype3} has a very clean and convenient interpretation in both the Hamiltonian
and transfer matrix formulations---see eqs.~\eqref{hamtype3TL}, \eqref{bigt} and \eqref{tmataf}. % \eqref{bigtildet}.
However an exact solution of the model with these boundary
conditions has, until now, been lacking. In \cite{nepomechie2019towards} the authors presented Bethe Ansatz equations that successfully accounted
for only a part of the spectrum of the model with boundary conditions \eqref{Kmatrixtype3}.
We present here a slight modification to the Bethe Ansatz solution of \cite{nepomechie2019towards} which accounts for all of the states in the spectrum.

\medskip

We start with a brief review of the partial solution presented in \cite{nepomechie2019towards}. Let $\Lambda(\lambda)$ denote an eigenvalue 
of the transfer matrix \eqref{tmatrixopendefects} with the $K$-matrices \eqref{Kmatrixtype3}. The parameter $u$ used in \cite{nepomechie2019towards} is what we call $\lambda$, defined in (\ref{lambdaurelation}). Using this notation, we write
\beq\label{eigval}
\Lambda(\lambda)=\phi(\lambda)\kappa(\lambda) \,,
\eeq
where
\small{
\begin{subequations}
\begin{eqnarray}
\phi(\lambda) &=& \frac{\sinh(\lambda)\sinh(\lambda-2\eta)}{\sinh(\lambda+\eta)\sinh(\lambda-3\eta)} \,, \\
 \kappa(\lambda) &=& Z_1(\lambda)+Z_2(\lambda)+Z_3(\lambda)+Z_4(\lambda) \,.
\end{eqnarray}
\end{subequations}
The functions $Z_k(\lambda)$ are defined as
\begin{subequations}\no
\begin{eqnarray}
 Z_1(\lambda) &=& a(\lambda)\frac{Q(\lambda+\eta)Q(\lambda+\eta+i\pi)}{Q(\lambda-\eta)Q(\lambda-\eta+i\pi)}\prod\limits_{k=1}^{L}16\sinh^{2}(\lambda-2i\omega_k-2\eta)\sinh^{2}(\lambda+2i\omega_k-2\eta) \, \\
 Z_2(\lambda) &=& b(\lambda)\frac{Q(\lambda-3\eta)Q(\lambda+\eta+i\pi)}{Q(\lambda-\eta)Q(\lambda-\eta+i\pi)} \\
&\times& \prod\limits_{k=1}^{L}16\sinh(\lambda-2i\omega_k-2\eta)\sinh(\lambda+2i\omega_k-2\eta)\sinh(\lambda-2i\omega_k)\sinh(\lambda+2i\omega_k)  \, \\
 Z_3(\lambda) &=& b(-\lambda+2\eta)\frac{Q(\lambda+\eta)Q(\lambda-3\eta+i\pi)}{Q(\lambda-\eta)Q(\lambda-\eta+i\pi)}\\
&\times&
\prod\limits_{k=1}^{L}16\sinh(\lambda-2i\omega_k-2\eta)\sinh(\lambda+2i\omega_k-2\eta)\sinh(\lambda-2i\omega_k)\sinh(\lambda+2i\omega_k)\, \\
 Z_4(\lambda) &=& a(-\lambda+2\eta)\frac{Q(\lambda-3\eta)Q(\lambda-3\eta+i\pi)}{Q(\lambda-\eta)Q(\lambda-\eta+i\pi)}\prod\limits_{k=1}^{L}16\sinh^{2}(\lambda-2i\omega_k)\sinh^{2}(\lambda+2i\omega_k) \,
\end{eqnarray}
\end{subequations}
}
where we have
\begin{subequations}
\label{abdef}
\begin{eqnarray}
 a(\lambda) &=& \frac{\cosh^2(\lambda-2\eta)}{\cosh^2(\lambda-\eta)} \,, \label{adef} \\
 b(\lambda) &=& \frac{\cosh(\lambda)\cosh(\lambda-2\eta)}{\cosh^2(\lambda-\eta)} \label{bdef}
\end{eqnarray}
\end{subequations}
and
\beq\label{qold}
Q(\lambda)=\prod\limits_{j=1}^{m}\sinh \left( \frac{1}{2}(\lambda-u_j) \right) \sinh \left( \frac{1}{2}(\lambda+u_j) \right) \,.
\eeq
The $u_j$ are the Bethe roots that can be found in the usual way by requiring that the residues of the poles of $\Lambda(u)$ all cancel, leading to the
following Bethe Ansatz equations
\begin{eqnarray} \label{baeoriginal}
\left(\frac{\sinh(u_j+\eta)}{\sinh(u_j-\eta)}\right)^{2L} &=& \frac{\sinh(u_j+\eta)}{\sinh(u_j-\eta)}\frac{\cosh(u_j-\eta)}{\cosh(u_j+\eta)} \\
&\times& \prod\limits_{k\neq j}^{m}\frac{\sinh(\frac{1}{2}(u_j-u_k)+\eta)}{\sinh(\frac{1}{2}(u_j-u_k)-\eta)}\frac{\sinh(\frac{1}{2}(u_j+u_k)+\eta)}{\sinh(\frac{1}{2}(u_j+u_k)-\eta)} \,. \nonumber
\end{eqnarray}
As already discussed in \cite{nepomechie2019towards}, this Bethe Ansatz solution is \textit{not} complete. In fact it is found that all eigenvalues with
even degeneracy are not accounted for, but all of the eigenvalues with odd degeneracy are. Furthermore, these states with odd degeneracies correspond to an even number $m$ of Bethe roots $u_j$ that come in pairs $\{u_j,u_j+i\pi\}$, with $j=1,\ldots,\frac{m}{2}$. The Bethe Ansatz equations for roots of this form
simplify to
\beq\label{baetype3subset}
\left(\frac{\sinh(u_j+\eta)}{\sinh(u_j-\eta)}\right)^{2L}=\frac{\sinh(u_j+\eta)}{\sinh(u_j-\eta)}\prod\limits_{k\neq j}^{\frac{m}{2}}\frac{\sinh(u_j-u_k+2\eta)}{\sinh(u_j-u_k-2\eta)}\frac{\sinh(u_j+u_k+2\eta)}{\sinh(u_j+u_k-2\eta)} \,.
\eeq
We shall present in section \ref{newsolution} a new solution that accounts for the missing states, but whose Bethe Ansatz equations also reduce to (\ref{baetype3subset}) for states that come in pairs $\{u_j,u_j+i\pi\}$.

\subsection{The complete solution}\label{newsolution}

To identify the complete solution we have employed the McCoy method \cite{fabricius2001bethe}, which is a procedure for obtaining
the Bethe roots corresponding to a particular eigenvalue. Although we do not possess advance knowledge of the general form of the eigenvalues,
we shall see that this method will nevertheless provide fruitful in the case at hand.

The first step of the McCoy method is to calculate numerically the eigenvectors of
the transfer matrix $t(\lambda)$ at a particular value $\lambda_0$ of the spectral parameter.
As the integrable transfer matrix satisfies $[t(\lambda_1),t(\lambda_2)]=0$, its eigenvectors are independent of $\lambda$.
One can therefore now find a polynomial expression in the variable $x \equiv e^{\lambda}$ for any eigenvalue $\Lambda(\lambda)$,
by acting with the transfer matrix $t(\lambda)$ on the corresponding eigenvector. Next one returns to the general expression
of the eigenvalue \eqref{eigval} in terms of the $Q$-function \eqref{qold}. Using the polynomial $\Lambda(x)$ found in the first step,
one can solve \eqref{eigval}--\eqref{qold} to obtain a polynomial expression $Q(x)$. Finally, one finds the roots $x_j$ of the $Q(x)$,
which gives the Bethe roots $u_j$, via $x_j \equiv e^{u_j}$.

In terms of the exponentiated variables the $Q$-function \eqref{qold} reads
\beq
Q(x)=\prod\limits_{k=1}^{m}\frac{1}{4} \left( x+x^{-1}-x_k-x_k^{-1} \right)=\sum\limits_{k=-m}^{m}a_kx^k
\eeq 
for some coefficients $a_k$ that depend on the Bethe roots $u_j$. Eq.~\eqref{qold} clearly satisfies $Q(\lambda)=Q(-\lambda)$, corresponding
to $Q(x)=Q(x^{-1})$, or $a_k=a_{-k}$.

We have first checked that the McCoy method applied to one of the states with odd degeneracy---those that
can be obtained by the previous incomplete solution---indeed produces a $Q$-polynomial whose coefficients satisfy $a_k=a_{-k}$.
However, for any of the states \textit{not} obtained by the partial solution \eqref{eigval}--\eqref{qold}, we find 
instead $a_k=(-1)^ka_{-k}$. Eq.~\eqref{qold} does not have this symmetry and must be modified to
\beq\label{qnew}
Q(\lambda)=\prod\limits_{j=1}^{m}\sinh \left( \frac{1}{2}(\lambda-u_j) \right) \cosh \left( \frac{1}{2}(\lambda+u_j) \right) \,.
\eeq
We find that changing the form of $Q(\lambda)$ by replacing \eqref{qold} with \eqref{qnew}, while keeping \eqref{eigval}--\eqref{abdef},
leads to a complete Bethe Ansatz solution that accounts for all of the eigenvalues. The new Bethe Ansatz equations, obtained
by requiring that the residues of the poles of $\Lambda(\lambda)$ all cancel, are now given by replacing \eqref{baetype3subset} with
\beq\label{baecorrect}
\left(\frac{\sinh(u_j+\eta)}{\sinh(u_j-\eta)}\right)^{2L}=\prod\limits_{k\neq j}^{m}\frac{\sinh(\frac{1}{2}(u_j-u_k)+\eta)}{\sinh(\frac{1}{2}(u_j-u_k)-\eta)}\frac{\cosh(\frac{1}{2}(u_j+u_k)+\eta)}{\cosh(\frac{1}{2}(u_j+u_k)-\eta)} \,.
\eeq

There are a few important remarks to be made on this complete solution. Firstly, note that the subset of solutions with paired Bethe roots of the form
$\{u_j,u_j+i\pi\}$ still satisfy (\ref{baetype3subset}), since \eqref{baecorrect} reduces to \eqref{baetype3subset} in this case. Our redefinition of $Q(\lambda)$
thus leaves unaffected the states that were already accounted for by the original partial solution in \cite{nepomechie2019towards}.

Secondly, observe that the new set of Bethe Ansatz equations do not obey the symmetry $u_j\rightarrow -u_j$ but instead $u_j\rightarrow\pi-u_j$.
This appears to be related to the fact that, as we shall see in section \ref{secxxxcorr}, the model with these boundary conditions maps to the
XXX model with \textit{periodic} boundary conditions in the $\gamma\rightarrow 0$ limit; the ``admissible solutions'' \cite{NepomechieWang2013,Avdeev1987,Marboe2016,Granet2019,Bajnok2019} to the Bethe Ansatz equations of
the periodic XXX model allow for vanishing Bethe roots $u_j$ \cite{Granet2019}, whereas the open XXX model does not \cite{Bajnok2019}.

Finally, recall from \eqref{hamdef} that the Hamiltonian ${\mathcal H}$ corresponding to the transfer matrix $t(\lambda)$ is, in this case, given by
the {\em second} derivative of the transfer matrix. From \eqref{eigval} we observe then that the energy eigenvalues are given by
\beq
E\propto \Lambda''(0)=\phi'(0)\lambda'(0) \,,
\eeq
leading to
\beq\label{d22eigs3}
E_{D_2^2}=\sum\limits_{j=1}^{m}\frac{2\sin^2(2\gamma)}{\cosh2u_j-\cos2\gamma} \,,
\eeq
which is exactly the same form as the eigenvalues of the Hamiltonian with the boundary conditions considered in our paper \cite{Robertson2020}.

\medskip

Summarising, the eigenvalues of the Hamiltonian defined in \eqref{hamtype3TL} are given by \eqref{d22eigs3}, where the $u_j$ satisfy the
Bethe Ansatz equations \eqref{baecorrect}.

\subsection{Correspondence with the XXX model}\label{secxxxcorr}

It is enlightening to study the present model in the limit $\gamma\rightarrow 0$. We will show that in this limit the Hamiltonian \eqref{hamtype3TL}
becomes equivalent to that of a \textit{periodic} XXX chain.

\begin{figure}
\centering
\begin{tikzpicture}

		\node at (0,-0.4) {1};
		\node at (1,1.4) {2};
		\node at (2,-0.4) {3};
		\node at (3,1.4) {4};
		\node at (4,-0.4) {5};
		\node at (5,1.4) {6};

		\filldraw[black] (0,0) circle (1.5pt);
		\filldraw[black] (1,1) circle (1.5pt);
		\filldraw[black] (2,0) circle (1.5pt);
		\filldraw[black] (3,1) circle (1.5pt);
		\filldraw[black] (4,0) circle (1.5pt);
		\filldraw[black] (5,1) circle (1.5pt);
		
		\draw[black,line width = 1pt,dotted] (0,0)--(1,1);
		\draw[black,line width = 1pt,dotted] (2,0)--(3,1);
		\draw[black,line width = 1pt,dotted] (4,0)--(5,1);
		
		\draw[black,line width = 1pt,dotted] (1,1)--(2,0);
		\draw[black,line width = 1pt,dotted] (3,1)--(4,0);
		
		\draw[blue,line width = 1pt] (0,0)--(2,0);
		\draw[blue,line width = 1pt] (2,0)--(4,0);
		
		\draw[blue,line width = 1pt] (1,1)--(5,1);
		
		\draw[red,line width = 1pt,dotted] (0,0) .. controls (0.2,0.8) .. (1,1);
		\draw[red,line width = 1pt,dotted] (4,0) .. controls (4.8,0.2) .. (5,1);

		\node at (10,-0.4) {1};
		\node at (11,1.4) {2};
		\node at (12,-0.4) {3};
		\node at (13,1.4) {4};
		\node at (14,-0.4) {5};
		\node at (15,1.4) {6};

		\filldraw[black] (10,0) circle (1.5pt);
		\filldraw[black] (11,1) circle (1.5pt);
		\filldraw[black] (12,0) circle (1.5pt);
		\filldraw[black] (13,1) circle (1.5pt);
		\filldraw[black] (14,0) circle (1.5pt);
		\filldraw[black] (15,1) circle (1.5pt);

		\draw[blue,line width = 1pt] (10,0)--(12,0);
		\draw[blue,line width = 1pt] (12,0)--(14,0);
		
		\draw[blue,line width = 1pt] (11,1)--(15,1);
		
		\draw[blue,line width = 1pt] (10,0) .. controls (10.2,0.8) .. (11,1);
		\draw[blue,line width = 1pt] (14,0) .. controls (14.8,0.2) .. (15,1);
		
		\begin{scope}[thick,decoration={
		    markings,
		    mark=at position 1 with {\arrow{>}}}
		    ] 
		\draw[black,line width = 1pt,postaction={decorate}] (6,0.5)--(9,0.5);
		\end{scope}
	
	\node at (7.5,0.8) {$\gamma\rightarrow 0$};
	
\end{tikzpicture}
\caption{The graphical interpretation of the terms in the Hamiltonian (\ref{hsigmaboundary}) for a chain of length $2L=6$ are shown in the left part of the figure.
The nearest-neighbour interaction is represented by dotted black lines, the XXX next-to-nearest neighbour interaction by solid blue lines,
and the boundary interactions by the dotted red lines between sites $1$ and $2$ and between sites $5$ and $6$. The three-site
interaction is not represented here. The right part of the figure shows the interactions in the limit $\gamma\rightarrow 0$, where one observes that
the Hamiltonian becomes equivalent to a periodic XXX Hamiltonian.}\label{figgammazero}
\end{figure}
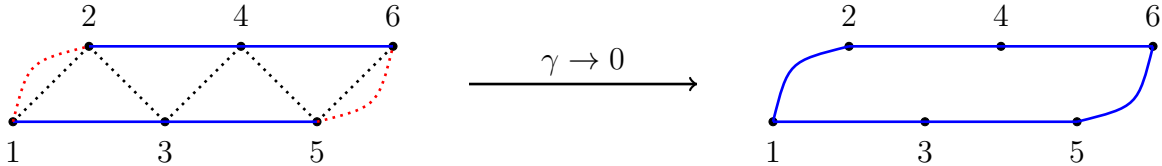

To prove this statement we first consider the Hamiltonian \eqref{hamtype3TL} written in terms of Pauli matrices:
\begin{eqnarray} 
\mathcal{H} &=& -\frac{1}{2}\sum\limits_{i=1}^{L-1}(\sigma_{2i-1}^x\sigma_{2i+1}^x+\sigma_{2i-1}^y\sigma_{2i+1}^y+\sigma_{2i-1}^z\sigma_{2i+1}^z)-\frac{1}{2}\sum\limits_{i=1}^{L-1}(\sigma_{2i}^x\sigma_{2i+2}^x+\sigma_{2i}^y\sigma_{2i+2}^y+\sigma_{2i}^z\sigma_{2i+2}^z) \nonumber \\
&-&\frac{1}{2}i\sin\gamma\sum\limits_{i=3}^{2L}(\sigma_i^z(\sigma_{i-1}^x\sigma_{i-2}^x+\sigma_{i-1}^y\sigma_{i-2}^y)-\sigma_{i-2}^z(\sigma_{i-1}^x\sigma_{i}^x+\sigma_{i-1}^y\sigma_{i}^y)) \nonumber \\
&+&\frac{1}{2}\sin^2\gamma\sum\limits_{i=1}^{2L-2}(\sigma_i^z\sigma_{i+1}^z+\sigma_{i+1}^z\sigma_{i+2}^z) \nonumber \\
&-&\frac{1}{2}i\sin\gamma\cos\gamma(\sigma_1^z+\sigma_2^z-\sigma_{2L-1}^z-\sigma_{2L}^z)+\frac{1}{2}\cos\gamma(\sigma_{1}^x\sigma_{2}^x+\sigma_{1}^y\sigma_{2}^y+\sigma_{2L-1}^x\sigma_{2L}^x+\sigma_{2L-1}^y\sigma_{2L}^y) \nonumber \\
&-&\frac{1}{2}\cos^2\gamma(\sigma_{1}^z\sigma_{2}^z+\sigma_{2L-1}^z\sigma_{2L}^z)-i\sin\gamma\cos\gamma(\sigma_1^z-\sigma_{2L}^z) \,,
\label{hsigmaboundary}
\end{eqnarray}
where we have used the identity \eqref{esigma}. We notice that there are four types of terms in (\ref{hsigmaboundary}):
the first line is an XXX interaction between next-to-nearest neighbours, the second line is a three-site interaction,
the third line is a nearest-neighbour interaction, and the last two lines are the boundary terms. All of these interactions,
except for the three-site interaction, are represented graphically for a chain of length $2L=6$ in the diagram on the
left-hand-side of Figure \ref{figgammazero}.
In the limit $\gamma\rightarrow 0$, the only non-vanishing terms are the next-to-nearest neighbour XXX interaction (the first line of (\ref{hsigmaboundary})) and a contribution from the boundary terms (the last two lines of (\ref{hsigmaboundary})), so that the Hamiltonian becomes
\beq\label{hsigma}
\begin{aligned}
\mathcal{H}=&-\frac{1}{2}\sum\limits_{i=1}^{L-1}(\sigma^x_{2i}\sigma^x_{2i+2}+\sigma^y_{2i}\sigma^y_{2i+2}+\sigma^z_{2i}\sigma^z_{2i+2}+\sigma^x_{2i-1}\sigma^x_{2i+1}+\sigma^y_{2i-1}\sigma^y_{2i+1}+\sigma^z_{2i-1}\sigma^z_{2i+1})\\
&+\frac{1}{2}(\sigma_1^x\sigma_2^x+\sigma_1^y\sigma_2^y-\sigma_1^z\sigma_2^z)\\
&+\frac{1}{2}(\sigma_{2L-1}^x\sigma_{2L}^x+\sigma_{2L-1}^y\sigma_{2L}^y-\sigma_{2L-1}^z\sigma_{2L}^z) \,.
\end{aligned}
\eeq 
This Hamiltonian is represented graphically in the diagram on the right-hand side of Figure \ref{figgammazero}. The term on the first line
of (\ref{hsigma}) corresponds to two decoupled open XXX chains, while the second and third terms introduce an interaction between
the two chains whose effect is to produce a periodic chain. We observe that the signs of the $\sigma^x\sigma^x+\sigma^y\sigma^y$ terms on the second and
third lines of (\ref{hsigma}) are opposite to the signs of the corresponding terms in the first line. However, it is easily verified that one can always
change the sign of an {\em even} number of $\sigma^x\sigma^x+\sigma^y\sigma^y$ terms without changing the spectrum of the XXX Hamiltonian.
Doing so for the two boundary terms precisely results in the XXX Hamiltonian. We have therefore shown that the spectrum of (\ref{hamtype3TL}),
in the limit $\gamma\rightarrow 0$, is identical to that of the periodic XXX chain, as is illustrated in Figure \ref{figgammazero}.

We may discuss this equivalence as well on the level of the Bethe Ansatz solution presented in section \ref{newsolution}. We first rewrite the periodic XXX Hamiltonian \eqref{hsigma} in the customary form
\beq\label{hxxxperiodic}
H=-\frac{1}{2}\sum\limits_{i=1}^{2L}(\sigma_i^x\sigma_{i+1}^x+\sigma_i^y\sigma_{i+1}^y+\sigma_i^z\sigma_{i+1}^z) \,.
\eeq
The corresponding eigenvalues are
\beq\label{xxxeigs1}
E=\sum\limits_{k=1}^{m}\frac{4}{\lambda_k^2+1} \,,
\eeq
where the Bethe roots $\lambda_k$ satisfy the Bethe Ansatz equations (BAE)
\beq\label{xxxbaesimp}
\left(\frac{\lambda_j+i}{\lambda_j-i}\right)^{2L}=\prod\limits_{k\neq j}^{m}\left(\frac{\lambda_j-\lambda_k+2i}{\lambda_j-\lambda_k-2i} \right) \,,
\eeq
with $m$ denoting the number of Bethe roots in any given solution. We now compare this to the Bethe Ansatz solution of the \dtt chain
(see section \ref{newsolution}). Upon rescaling the Bethe roots, $u_j\rightarrow\gamma u_j$,
the eigenvalues (\ref{d22eigs3}) of the \dtt chain become, in the limit $\gamma\rightarrow 0$,
\beq
E=\sum\limits_{k=1}^{m}\frac{4}{u_k^2+1} \,,
\eeq
which we see is identical to (\ref{xxxeigs1}) for $u_k=\lambda_k$.

We next consider the \dtt BAE in the limit $\gamma\rightarrow 0$. Let us first consider the simplest case of roots that
come in pairs $\{u_j,u_j+i\pi\}$, which we recall satisfy the simplified BAE \eqref{baetype3subset}.
Recalling that $\eta\equiv i \gamma$, eq.~\eqref{baetype3subset} can be rewritten as
\beq\label{d22subsetlimit}
\left(\frac{u_j+i}{u_j-i}\right)^{2L}=\left(\frac{u_j+i}{u_j-i}\right)\prod\limits_{k\neq j}^{\frac{m}{2}}\left( \frac{(u_j-u_k+2i)(u_j+u_k+2i)}{(u_j-u_k-2i)(u_j+u_k-2i)} \right) \,,
\eeq
when $\gamma\rightarrow 0$. Consider a solution to the XXX BAE \eqref{xxxbaesimp} with Bethe roots $\lambda_j$ that are symmetric about $0$,
i.e., roots that come in pairs $\{\lambda_j, -\lambda_j\}$. For roots of this form, \eqref{xxxbaesimp} is indeed identical to \eqref{d22subsetlimit}
for $u_k=\lambda_k$.

We may also examine the original partial solution, with BAE \eqref{baetype3subset}, in view of establishing how it breaks down for states that do
not come in pairs $\{u_j,u_j+i\pi\}$. The simplest example of roots that are not of this form are a set with an odd number of roots, including one vanishing root which we take as $u_1=0$, while all of the other roots come in pairs $\{u_j,u_j+i\pi\}$ as before. The BAE \eqref{baeoriginal} for roots of this form become
\beq\label{subsetrootzero}
\begin{aligned}
\left(\frac{\sinh(u_j+\eta)}{\sinh(u_j-\eta)}\right)^{2L}=\frac{\sinh(u_j+\eta)}{\sinh(u_j-\eta)}&\left(\prod\limits_{k\neq j}^{\frac{m}{2}}\frac{\sinh(u_j-u_k+2\eta)}{\sinh(u_j-u_k-2\eta)}\frac{\sinh(u_j+u_k+2\eta)}{\sinh(u_j+u_k-2\eta)}\right)\\
& \times \left(\frac{\sinh(\frac{1}{2}u_j+\eta)}{\sinh(\frac{1}{2}u_j-\eta)} \right)^2 \,,
\end{aligned}
\eeq
which in the limit $\gamma\rightarrow 0$ simplify to
\beq\label{oddbaelimit}
\left(\frac{u_j+i}{u_j-i}\right)^{2L}=\left(\frac{u_j+i}{u_j-i}\right)\left(\frac{u_j+2i}{u_j-2i} \right)^2\prod\limits_{k\neq j}^{\frac{m}{2}}\left( \frac{(u_j-u_k+2i)(u_j+u_k+2i)}{(u_j-u_k-2i)(u_j+u_k-2i)} \right) \,.
\eeq
We would like to compare these BAE with the BAE \eqref{xxxbaesimp} for the periodic XXX model. In analogy with the previous case, we should
consider solutions to (\ref{xxxbaesimp}) with one root $\lambda_1=0$ and all other roots coming in pairs $\{\lambda_j,-\lambda_j\}$. For roots of
this form, (\ref{xxxbaesimp}) becomes
\beq\label{oddxxx0}
\left(\frac{\lambda_j+i}{\lambda_j-i}\right)^{2L}=\left(\frac{\lambda_j+i}{\lambda_j-i}\right)\left(\frac{\lambda_j+2i}{\lambda_j-2i} \right)\prod\limits_{k\neq j}^{\frac{m}{2}}\left( \frac{(\lambda_j-\lambda_k+2i)(\lambda_j+\lambda_k+2i)}{(\lambda_j-\lambda_k-2i)(\lambda_j+\lambda_k-2i)} \right) \,.
\eeq
Comparing (\ref{oddxxx0}) with (\ref{oddbaelimit}) one observes that when $\lambda_j=u_j$ the two sets of equations differ as a result of the square
on the second factor on the right-hand side of (\ref{oddbaelimit}).

One can use this result to guess the correct BAE---i.e., to provide an alternative
argument for the BAE (\ref{baecorrect}) that were obtained in section \ref{newsolution} by the McCoy method. One starts with the observation that the square in the second factor in (\ref{oddbaelimit}) comes from the two $\sinh$ terms within the product in (\ref{baeoriginal}). To get rid of this square one changes one of these $\sinh$ terms to a $\cosh$, whose leading-order term vanishes in the $\gamma\rightarrow 0$ limit. To ensure that the new equations still reduce to (\ref{baetype3subset}) for roots that come in pairs $\{u_j,u_j+i\pi\}$, one must remove the terms outside of the product in (\ref{baeoriginal}), and one is left with the new BAE (\ref{baecorrect}) indeed.

\section{Continuum limit}\label{seccontlim3}
The results presented so far relate to the boundary conditions (\ref{Kmatrixtype3}) on a finite lattice. We will now consider the description of the model in
terms of boundary CFT (BCFT) when we take the continuum limit. In particular, we will be interested in calculating the generating function of scaling levels
corresponding to the Hamiltonian (\ref{hamtype3TL}). The continuum limit is studied by finite-size scaling of the energy eigenvalues (\ref{d22eigs3}).
We have \cite{Blote1986}
\beq\label{flscaling}
E_i=f_0L+f_{\rm s}-\frac{\pi v_{\rm F}(\frac{c}{24}-h_i)}{L}+o\left( \frac{1}{L} \right) \,,
\eeq
where $L$ is the system size, $c$ the central charge, $h_i$ the conformal weight of the primary field corresponding to the $E_i$,
$f_0$ the bulk energy density, and $f_{\rm s}$ the surface energy. The Fermi velocity $v_{\rm F}$ was calculated in \cite{IJS2008}
\beq\label{vfd22}
v_{\rm F}=\frac{2\pi\sin(\pi-2\gamma)}{\pi-2\gamma} \,.
\eeq
The Hamiltonian (\ref{hamtype3TL}) and transfer matrix (\ref{tmatafdual}) are written entirely in terms of Temperley-Lieb (TL) algebra generators $e_i$.
We have so far considered only the vertex representation \eqref{esigma} of the TL algebra. In this section we consider as well
other representations of $e_i$ that satisfy (\ref{TLrelations}): the loop representation (section \ref{secloopreptype3}) and the RSOS representation
(section \ref{secrsostype3}). In the loop and vertex representations we find a continuous spectrum which is the identifying property of a non-compact
BCFT. In the RSOS restriction for $\gamma={\pi\over k}$, with $k$ an integer, the states in the loop or vertex model that form the continuum disappear, and the generating function of the remaining
scaling levels is given by a combination of string functions.

\subsection{The loop model}\label{secloopreptype3}

\begin{figure}
	\centering
\begin{tikzpicture}[scale=0.5]
	
	\node at (-1,1) {$e_1=$};
	
	\draw[black,line width = 1pt] (0,2) .. controls (0.25,1.4) and (0.75,1.4) .. (1,2);
	\draw[black,line width = 1pt] (0,0) .. controls (0.25,0.6) and (0.75,0.6) .. (1,0);
	\draw[black,line width = 1pt] (2,0) -- (2,2);
	\draw[black,line width = 1pt] (3,0) -- (3,2);

	\node at (3.8,0.2) {,};
	
	\node at (5.5,1) {$e_2=$};
	
	\draw[black,line width = 1pt] (7,0) -- (7,2);
	\draw[black,line width = 1pt] (8,0) .. controls (8.25,0.6) and (8.75,0.6) .. (9,0);
	\draw[black,line width = 1pt] (8,2) .. controls (8.25,1.4) and (8.75,1.4) .. (9,2);
	\draw[black,line width = 1pt] (10,0) -- (10,2);
	
	\node at (11.3,0.2) {,};
	
	\node at (13,1) {$e_3=$};
	
	\draw[black,line width = 1pt] (14,0) -- (14,2);
	\draw[black,line width = 1pt] (15,0) -- (15,2);
	\draw[black,line width = 1pt] (16,0) .. controls (16.25,0.6) and (16.75,0.6) .. (17,0);
	\draw[black,line width = 1pt] (16,2) .. controls (16.25,1.4) and (16.75,1.4) .. (17,2);
	
	\node at (18.3,0.2) {,};
	
	\node at (20,1) {$\mathcal{I}=$};
	\draw[black,line width = 1pt] (21,0) -- (21,2);
	\draw[black,line width = 1pt] (22,0) -- (22,2);
	\draw[black,line width = 1pt] (23,0) -- (23,2);
	\draw[black,line width = 1pt] (24,0) -- (24,2);

\end{tikzpicture}
\caption{The graphical interpretation of the Temperley-Lieb loop representation.}\label{TL}
\end{figure}

\begin{figure}
	\centering
\begin{tikzpicture}[scale=0.5]
	
	\node at (-1,2) {$e_1^2=$};
	
	\draw[black,line width = 1pt] (0,2) .. controls (0.25,1.4) and (0.75,1.4) .. (1,2);
	\draw[black,line width = 1pt] (0,0) .. controls (0.25,0.6) and (0.75,0.6) .. (1,0);
	\draw[black,line width = 1pt] (2,0) -- (2,2);
	\draw[black,line width = 1pt] (3,0) -- (3,2);
	
	\draw[black,line width = 1pt] (0,4) .. controls (0.25,3.4) and (0.75,3.4) .. (1,4);
	\draw[black,line width = 1pt] (0,2) .. controls (0.25,2.6) and (0.75,2.6) .. (1,2);
	\draw[black,line width = 1pt] (2,2) -- (2,4);
	\draw[black,line width = 1pt] (3,2) -- (3,4);
	
	\node at (5,2) {$=\sqrt{Q}$};
	
	\draw[black,line width = 1pt] (6.5,3) .. controls (6.75,2.4) and (7.25,2.4) .. (7.5,3);
	\draw[black,line width = 1pt] (6.5,1) .. controls (6.75, 1.6) and (7.25, 1.6) .. (7.5,1);
	\draw[black,line width = 1pt] (8.5,1) -- (8.5,3);
	\draw[black,line width = 1pt] (9.5,1) -- (9.5,3);
	
	\node at (12,2) {$=\sqrt{Q}e_1$};

\end{tikzpicture}
\caption{Graphical interpretation of the relation $e_i^2=\sqrt{Q}e_i$.}\label{e1squared}
\end{figure}

\begin{figure}
	\centering
\begin{tikzpicture}[scale=0.5]
	
	\node at (-2,3) {$e_1e_2e_1=$};
	
	\draw[black,line width = 1pt] (0,2) .. controls (0.25,1.4) and (0.75,1.4) .. (1,2);
	\draw[black,line width = 1pt] (0,0) .. controls (0.25,0.6) and (0.75, 0.6).. (1,0);
	\draw[black,line width = 1pt] (2,0) -- (2,2);
	\draw[black,line width = 1pt] (3,0) -- (3,4);
	
	\draw[black,line width = 1pt] (0,2) -- (0,4);
	\draw[black,line width = 1pt] (1,4) .. controls (1.25,3.4) and (1.75,3.4) .. (2,4);
	\draw[black,line width = 1pt] (1,2) .. controls (1.25,2.6) and (1.75,2.6) .. (2,2);
	
	\draw[black,line width = 1pt] (0,6) .. controls (0.25,5.4) and (0.75,5.4) .. (1,6);
	\draw[black,line width = 1pt] (0,4) .. controls (0.25,4.6) and (0.75,4.6) .. (1,4);
	\draw[black,line width = 1pt] (2,4) -- (2,6);
	\draw[black,line width = 1pt] (3,4) -- (3,6);

	\node at (5,3) {$=$};

	\draw[black,line width = 1pt] (6,4) .. controls (6.25,3.4) and (6.75,3.4) .. (7,4);
	\draw[black,line width = 1pt] (6,2) .. controls (6.25,2.6) and (6.75,2.6) .. (7,2);
	\draw[black,line width = 1pt] (8,2) -- (8,4);
	\draw[black,line width = 1pt] (9,2) -- (9,4);
	
	\node at (11,3) {$=e_1$};
	
\end{tikzpicture}
\caption{Graphical interpretation of $e_1e_2e_1=e_1$.}\label{e1e2e1}
\end{figure}

\begin{figure}
	\centering
\begin{tikzpicture}[scale=0.5]
	
	\node at (-2,0.8) {$\mathcal{W}_0=$};
	
	\node at (-0.5,0.9) {$\{$};
	
	\draw[black,line width = 1pt] (0,1) .. controls (0.25,0.5) and (0.75,0.5) .. (1,1);
	\draw[black,line width = 1pt] (2,1) .. controls (2.25,0.5) and (2.75,0.5) .. (3,1);
	
	\node at (3.5,0.7) {,};
	
	\draw[black,line width = 1pt] (4,1) .. controls (5,0.2) and (6, 0.2) .. (7,1);
	\draw[black,line width = 1pt] (5,1) .. controls (5.25,0.5) and (5.75,0.5) .. (6,1);
	
	\node at (7.4,0.9) {$\}$};

	\node at (-2,-1.2) {$\mathcal{W}_1=$};
	
	\node at (-0.5,-1.1) {$\{$};
	
	\draw[black,line width = 1pt] (0,-1) .. controls (0.25,-1.5) and (0.75,-1.5) .. (1,-1);
	\draw[black,line width = 1pt] (2,-1.5) -- (2,-1);
	\draw[black,line width = 1pt] (3,-1.5) -- (3,-1);
	
	\node at (3.5,-1.4) {,};
	
	\draw[black,line width = 1pt] (4.5,-1.5) -- (4.5,-1);
	\draw[black,line width = 1pt] (5.5,-1) .. controls (5.75,-1.5) and (6.25,-1.5) .. (6.5,-1);
	\draw[black,line width = 1pt] (7.5,-1.5) -- (7.5,-1);
	
	\node at (8,-1.4) {,};
	
	\draw[black,line width = 1pt] (9,-1.5) -- (9,-1);
	\draw[black,line width = 1pt] (10,-1.5) -- (10,-1);
	\draw[black,line width = 1pt] (11,-1) .. controls (11.25,-1.5) and (11.75,-1.5) .. (12,-1);
	
	\node at (12.5,-1.4) {$\}$};

	\node at (-2,-3.2) {$\mathcal{W}_2=$};
	
	\node at (-0.5,-3.1) {$\{$};
	
	\draw[black,line width = 1pt] (0,-3.5) -- (0,-2.8);
	\draw[black,line width = 1pt] (1,-3.5) -- (1,-2.8);
	\draw[black,line width = 1pt] (2,-3.5) -- (2,-2.8);
	\draw[black,line width = 1pt] (3,-3.5) -- (3,-2.8);

	\node at (3.5,-3.1) {$\}$};

\end{tikzpicture}
\caption{The representation spaces of the Temperley-Lieb algebra acting on $N=4$ strands.}\label{reps}
\end{figure}
One obtains the loop representation of the TL algebra by assigning a diagrammatic representation to each generator $e_i$. Figure \ref{TL} shows these
diagrams for the case $N=4$. In the loop representation, multiplication corresponds to stacking diagrams vertically. The first relation in (\ref{TLrelations})
implies that a loop can be replaced by its corresponding weight $\sqrt{Q}$ (Figure \ref{e1squared}), while the second relation shows
that the curves can be deformed by topological equivalence (Figure \ref{e1e2e1}). The representation spaces upon which the $e_i$ act are called
{\em standard modules}\/; they are shown in Figure \ref{reps}. We see that in the case $N=4$ these states are divided into three sectors:
$\mathcal{W}_0$, $\mathcal{W}_1$ and $\mathcal{W}_2$. A {\em through-line} is defined as a connection between the top and the bottom of the
diagram, so $\mathcal{W}_j$ is the sector with precisely $2j$ through-lines. Within the standard modules, the action of $e_i$ on two adjacent
through-lines is defined to be zero. For a system of size $N$ there can be at most $N$ through-lines, and hence the maximum value of $j$ is $\frac{N}{2}$
in the general case.

We consider now the Hamiltonian (\ref{hamtype3TL}) with the $e_i$ in the loop representation. A principal result of this paper is that the BCFT
describing the continuum limit in the loop representation contains a continuum of critical exponents---a phenomenon that had until now only been
observed in the bulk spectrum, and which was absent in our earlier papers on the boundary model \cite{Robertson2019,Robertson2020}.
In addition to this continuous part of the spectrum, there is also a discrete set of exponents, such that the full partition function of the model is given by
\beq\label{contdis}
\mathcal{Z}=\mathcal{Z}^{\text{disc.}}+\mathcal{Z}^{\text{cont.}} \,.
\eeq
Here and in the following, the continuum-limit partition function is understood to mean the generating function of conformal weights
\begin{equation}
\mathcal{Z}=\hbox{Tr } q^{L_0-{c\over 24}} \,,
\end{equation}
where the trace is taken over  the Hilbert space (or some subspaces) of the BCFT. This ${\mathcal Z}$ is known to coincide with the
continuum limit of the lattice-model partition function on the annulus, with $q = e^{-\pi \tau}$ being the modular parameter and $\tau$  the aspect ratio. We also note that the spectrum of the  transfer matrix can be split into sectors with a fixed number $2j$ of through-lines, with $j$ an integer (whose quantum-group meaning is discussed below). This will allow us to split (\ref{contdis}) further into $\mathcal{Z}^{\text{disc.}}_{2j}$ and 
$\mathcal{Z}^{\text{cont.}}_{2j}$. 

We will consider the continuous part of the spectrum in section \ref{seccontspec} and the discrete part in section \ref{secdiscretestates}. We calculate the eigenvalues of (\ref{hamtype3TL}) using the Bethe Ansatz solution outlined in section \ref{newsolution}, and use equation (\ref{flscaling}) to extract $c$ and the various values of $h_i$. There is a simple correspondence between the number of Bethe roots $m$ appearing in any given solution to the BAE in equation \eqref{baecorrect} and the sector $\mathcal{W}_j$ that such a solution corresponds to in the loop model. We have
\beq
m = L - j \,.
\eeq
We can therefore study each $\mathcal{Z}^{\text{disc.}}_{2j}$ and 
$\mathcal{Z}^{\text{cont.}}_{2j}$ separately by fixing the number of Bethe roots in the solutions to the Bethe Ansatz equations in \eqref{baecorrect}. Let us start with the observation that the central charge is given by \cite{S-AF}
\beq\label{cpf2}
c=2-\frac{6}{k} \,,
\eeq
where we recall that $k$ is related to $\gamma$ and $Q$ by
\beq
 \gamma = \frac{\pi}{k}
\eeq
as well as \eqref{qgamma}.
This central charge coincides with the one of the $su(2)_{k-2}/u(1)$, $Z_{k-2}$ parafermions \cite{FateevZamolodchikov,GepnerQiu},
although (\ref{cpf2}) holds for $k\in \mathbb{R}$, while of course parafermions are a priori well defined only for $k>2$ integer. The relationship with ``genuine'' parafermions when $k$ is an integer will be discussed in more detail below. We shall see below that, for certain values of $k$, the leading exponent $h$ is \textit{negative}, leading to an effective central charge $c_{\text{eff}}=c-24h$ that is \textit{greater} than $c$.

\subsubsection{Continuous spectrum}\label{seccontspec}

We examine separately each of the sectors with $2j$ through-lines. In each sector we observe both a discrete and a continuous set of critical exponents. The continuum is found to begin at the following threshold value (the ``bottom'' of the spectrum)
\beq\label{hparaf0}
h = \frac{l(l+2)}{4k} \,,
\eeq
with $l$ given by
\beq\label{lcont}
l=k-2j-2 \,.
\eeq
Note that (\ref{lcont}) means that for certain values of $k$ the exponent in (\ref{hparaf0}) can be negative. In particular, when $k$ is an odd integer we can have $l=-1$, corresponding to $2j=k-1$.

\begin{figure}
	\centering
\includegraphics[scale=0.5]{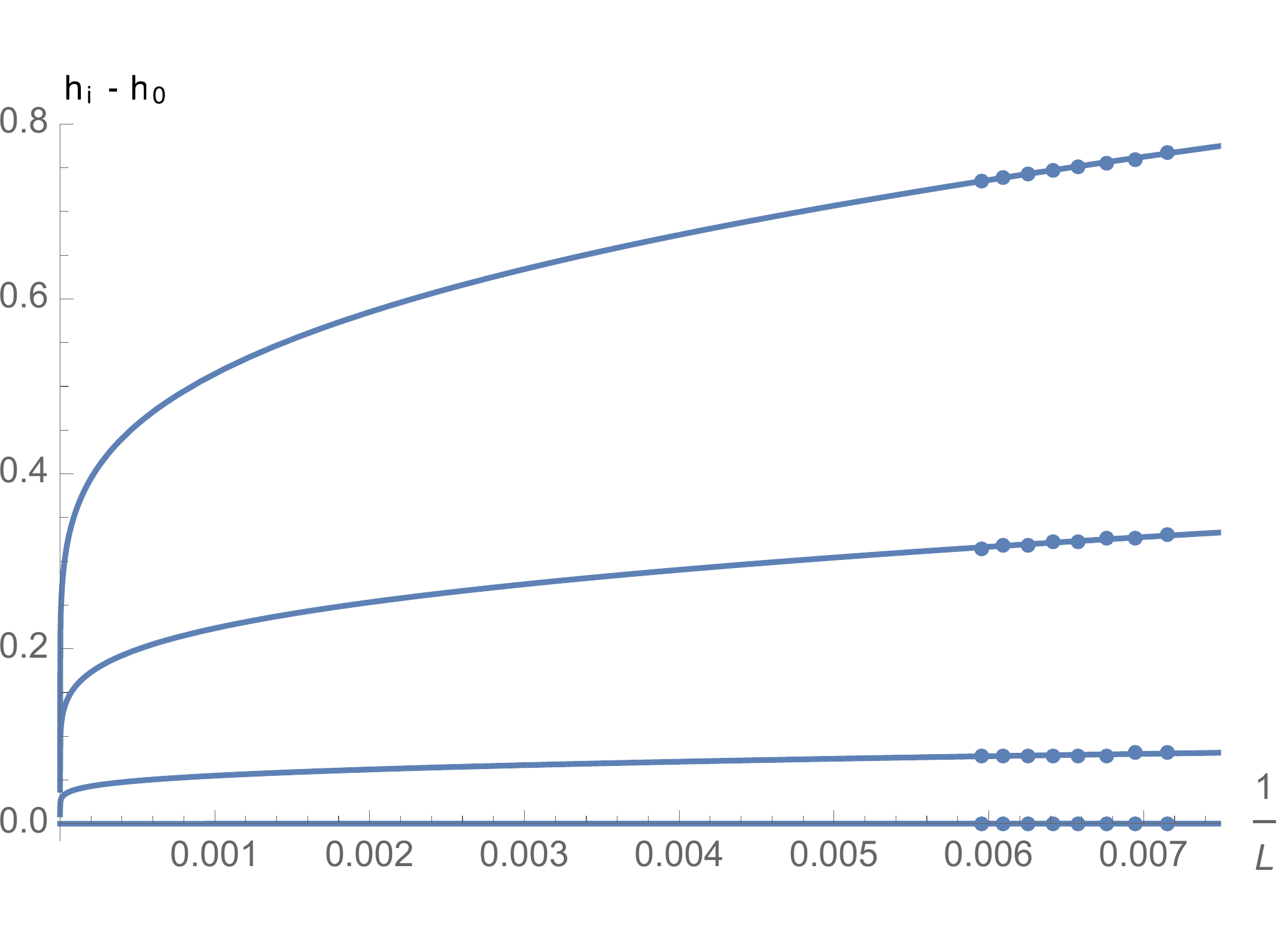}
	\caption{The scaling behaviour of the gaps generated from the Hamiltonian (\ref{hamtype3TL}) with $\gamma=\frac{\pi}{4}$, in the sector $2j=4$. The gaps $h_i-h_0$ are calculated numerically using (\ref{flscaling}). We observe that they converge logarithmically to $0$ as we expect from (\ref{hparaf0})--(\ref{lcont}), and that a large (presumably infinite) number of such states appear in the limit $L\rightarrow \infty$---the identifying properties of a non-compact scaling limit. The extrapolations are made by finding the best fit to (\ref{flscaling}) where the $o(1/L)$ term is taken to be of order $O(1/(L \log L))$.
	}\label{noncompact2}
\end{figure}

The appearance of a continuum is apparent from the logarithmic convergence of exponents extracted from the lattice model towards the value (\ref{hparaf0}), as $L\rightarrow \infty$, and from the fact that a large (supposedly infinite in the limit) number of lattice states converge to the same value of $h$, so as to create a finite density at and above this value. More details about the identification of a non-compact CFT emerging as the scaling limit of a lattice model were discussed in \cite{IJS2008}. Figure \ref{noncompact2} gives an example of some numerical results of states converging logarithmically for the case under study. In particular, this means that the $o(1/L)$ term in (\ref{flscaling}) is in fact of order $O(1/(L \log L))$. This form of the correction term was then used to produce the extrapolations shown in the figure.

\medskip

We now wish to interpret these results in the context of the Euclidean black hole CFT.
Recall that this CFT has central charge (see, e.g., \cite{RibaultSchomerus} and references therein)
 \begin{equation}\label{cbh}
 c_{\rm BH}=2+{6\over k-2}
 \end{equation}
and conformal weights
 \begin{equation}\label{confw2} 
 h_{\rm BH}=-{J(J-1)\over k-2}+{(n \pm wk)^2\over 4k} \,,
 \end{equation}
where $n$ and $w$ are integers. There is a continuous series of conformal weights with
\beq\label{eqJdef}
J=\frac{1}{2}+is \,, \quad \mbox{with } s \in \mathbb{R}_+
\eeq
and a discrete set with ${1\over 2}<J<{k-1\over 2}$, where $2J\in\mathbb{N}$. According to (\ref{flscaling}) we only observe the central charge and
conformal dimensions in the combination $\frac{c}{24}-h_0$. In other words, we can only directly observe the effective central charge
$c_{\text{eff}}=c-24h_0$. This fact allows us to write the critical exponents formally as those of a parafermionic CFT---where ``formally'' means
extending the parameter $k$ from integer to real values---with central charge $c_{\rm PF}$ given by (\ref{cpf2}). Equating
\beq\label{cpfcbh}
c_{\rm PF}-24\Delta_{\rm PF}=c_{\rm BH}-24h_{\rm BH} \,,
\eeq
where $\Delta_{\rm PF}$ is a critical exponent in the parafermion theory with available values \cite{GepnerQiu}
\begin{equation}\label{paraflm}
\Delta^{m}_l={l(l+2)\over 4k}-{m^2\over 4(k-2)}
\end{equation}
and $c_{\rm PF}$ is given in (\ref{cpf2}), and combining equations (\ref{confw2})--(\ref{cpfcbh}), we get:
\beq\label{pfcont}
\Delta_{\rm PF}=\frac{(n \pm wk - 1)(n \pm wk + 1)}{4k}+\frac{s^2}{k-2} \equiv \frac{l(l+2)}{4k}+\frac{s^2}{k-2} \,,
\eeq
where
\beq\label{lbh}
l=n \pm wk - 1 \,.
\eeq

\begin{figure}[h]
	\centering
	\includegraphics[scale=0.75]{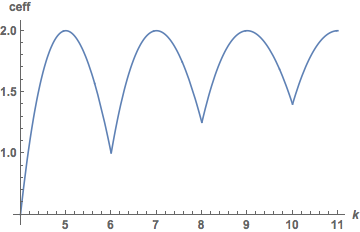}
\caption{The effective central charge as a function of $k$. For $k$ an odd integer we have $c_{\text{eff}}=2$, while for $k$ an even integer we have $c_{\text{eff}}=2-\frac{6}{k}$.}\label{ceff}
\end{figure}

Notice that (\ref{pfcont}) is consistent with our earlier observation that the loop model produces a continuum of critical exponents beginning at
(\ref{hparaf0})---and combining eqs.~(\ref{lcont}) and (\ref{lbh}) leads to $n=-1$ and $w=1$. It also follows that the effective central charge
$c_{\text{eff}}=c-24h$ is given by $c_{\text{eff}}=c=2-\frac{6}{k}$ for $k$ an even integer and $c_{\text{eff}}=2$ for $k$ an odd integer. The
general case is plotted in Figure \ref{ceff}, and is given by
\beq\label{ceffeq}
c_{\text{eff}} = 2-24\frac{(\text{frac}\left(\frac{k}{2}\right) -\frac{1}{2})^2}{k} \,,
\eeq
where the notation $\text{frac}\left(\frac{k}{2}\right)$ refers to the fractional part of $\frac{k}{2}$. The expression (\ref{ceffeq}) is consistent
with (\ref{hparaf0})--(\ref{lcont}). 

To conclude this section, we have 
\begin{equation}
\mathcal{Z}^{\text{cont.}}_{2j}(\text{loop})=q^{l(l+2)\over 4k} q^{-{c_{\text{PF}}\over 24}} \int_0^\infty {\rm d}s \, \rho_l(s) q^{s^2\over k-2} \,, \quad \mbox{with } l\equiv k-2j-2 \,,
\end{equation}
where $\rho_j$ is a density of states which we will not try to determine in this paper.

\subsubsection{Back to the Bethe Ansatz}

\begin{figure}[h]
	\centering
	\includegraphics[scale=0.75]{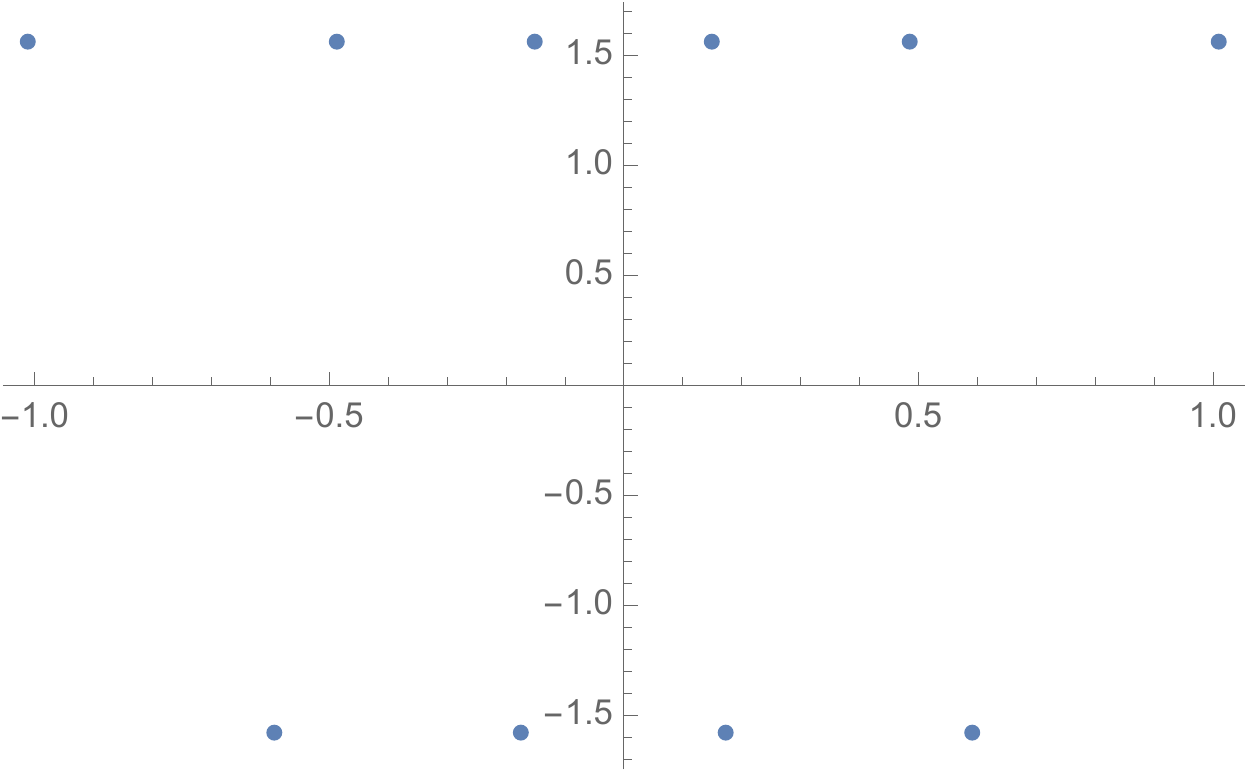}
\caption{An example of a solution to the bulk BAE that leads to a continuum state. This example corresponds to $L=12$,  with $n^+=0$ and $n^-=2$.}\label{periodic}
\end{figure}

\no It is worthwhile discussing in more detail the states on the lattice that converge logarithmically as in Figure \ref{noncompact2} and hence lead to the continuum in the field theory, particularly in the context of the Bethe Ansatz solution from section \ref{newsolution}. The appearance of a continuous spectrum in the periodic model, studied in \cite{IJS2008}, could be understood in terms of a particular class of solutions to the BAE
\beq\label{periodicbae}
\left(\frac{\sinh(u_j+\eta)}{\sinh(u_j-\eta)}\right)^{L}=-\prod\limits_{k\neq j}^{m}\frac{\sinh(\frac{1}{2}(u_j-u_k)+\eta)}{\sinh(\frac{1}{2}(u_j-u_k)-\eta)} \,.
\eeq
A subset of the solutions to these equations have the form
\begin{subequations}
\label{contsolbae}
\begin{eqnarray}
u_j^0 &=& \alpha^+_j+i\frac{\pi}{2} \,, \\
u_j^1 &=& \alpha^-_j-i\frac{\pi}{2} \,,
\end{eqnarray}
\end{subequations}
with $\alpha^{\pm}_j$ real.  States in the continuum arise from solutions of the form (\ref{contsolbae}) with different numbers of $u_j^0$-roots and $u_j^1$-roots. Denote the number of $u_j^0$-roots as $m^+$ and the number of $u_j^1$-roots as $m^-$. Then define
\begin{subequations}
\label{nplusminus}
\begin{eqnarray}
m^+ &=& \frac{L}{2}-n^+ \,, \\
m^- &=& \frac{L}{2}-n^- \,.
\end{eqnarray}
\end{subequations}
In the bulk, the continuum states converge logarithmically to the ``floor states'' $h_0$ as
\beq\label{hcont}
h=h_0+K(\gamma, L)(n^+-n^-)^2 \,,
\eeq 
where $K(\gamma, L)\rightarrow 0$ as $L\rightarrow \infty$. Figure \ref{periodic} shows an example of a configuration of Bethe roots with $n^+ \neq n^-$, therefore corresponding to a continuum state.

\begin{figure}[h]
	\centering
	\includegraphics[scale=0.75]{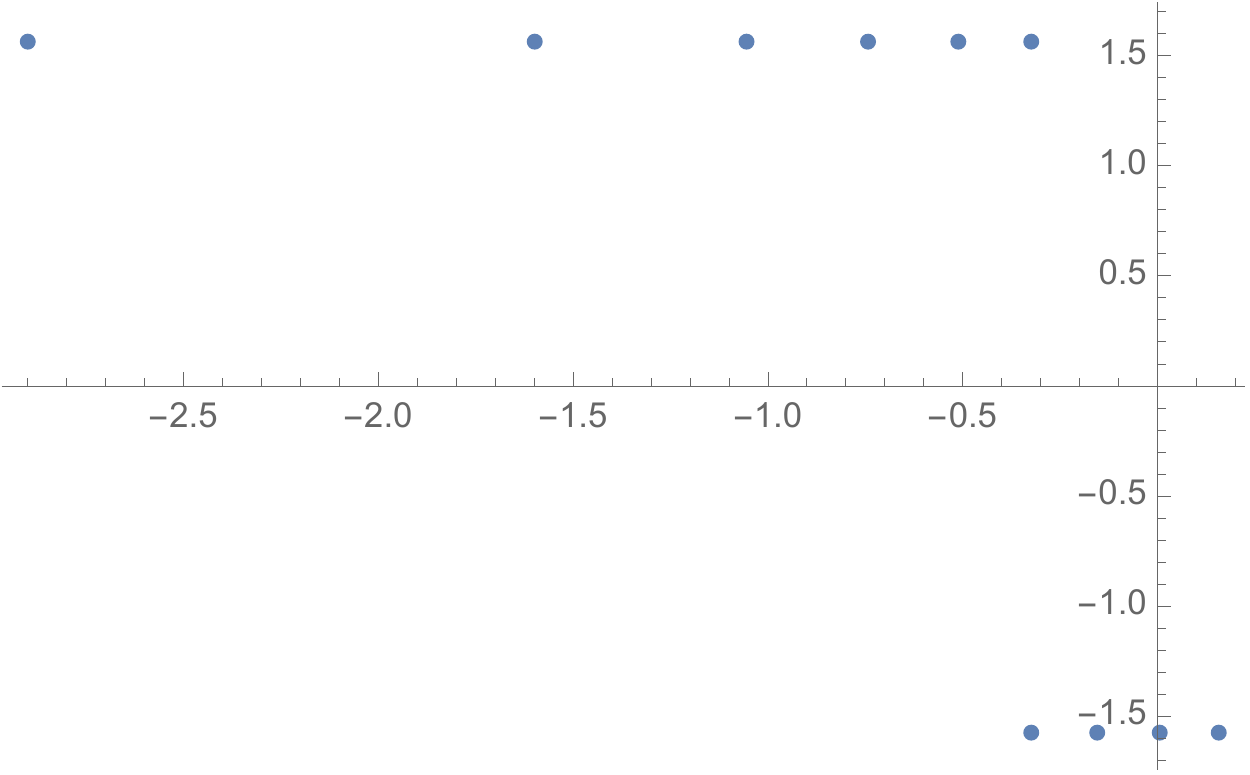}
\caption{An example of a solution to the boundary BAE (\ref{baechap5}) that \textit{do not} lead to a continuous spectrum, unlike the bulk case. This example corresponds to $L=12$, with $n^+=0$ and $n^-=2$.}\label{bndry}
\end{figure}
The \dtt model boundary conditions studied in \cite{Robertson2020} led to a discrete spectrum in the continuum limit. We can understand this in the following way. The BAE for the model with those boundary conditions are \cite{Robertson2020}
\beq\label{baechap5}
\left(\frac{\sinh(u_j+\eta)}{\sinh(u_j-\eta)}\right)^{2L}=\prod\limits_{k\neq j}^{m}\frac{\sinh(\frac{1}{2}(u_j-u_k)+\eta)}{\sinh(\frac{1}{2}(u_j-u_k)-\eta)}\frac{\sinh(\frac{1}{2}(u_j+u_k)+\eta)}{\sinh(\frac{1}{2}(u_j+u_k)-\eta)} \,.
\eeq
These equations permit solutions of the form (\ref{contsolbae}); one such solution is shown in Figure \ref{bndry}. This solution satisfies $n^+ \neq n^-$ but does not correspond to a continuum state, in contradistinction with the bulk case. Namely, if we multiply any number of Bethe roots $u_k$ by a sign, eqs.~(\ref{baechap5}) will still be satisfied and the energy is unchanged. By using this symmetry under $u_k\rightarrow -u_k$, we can therefore transform any solution to (\ref{baechap5}) with $n^+ \neq n^-$ into another one with $n^+ = n^-$, and hence the $K(\gamma, L)$ term in (\ref{hcont}) will not play any role. More precisely, the symmetry $u_k\rightarrow -u_k$ allows us to transform any $u_j^0$-root into a $u_j^1$-root, meaning that only $n=n^+ + n^-$ is a meaningful quantity for solutions to (\ref{baechap5}), whereas $n^+$ and $n^-$ have no physical meaning individually.  Compare this to the BAE (\ref{baecorrect}), where the symmetry $u_k\rightarrow -u_k$ is replaced by $u_k\rightarrow -u_k+i\pi$. The symmetry $u_k\rightarrow -u_k+i\pi$ does not affect $n^+$ or $n^-$, and therefore (\ref{hcont}) can be applied to the BAE (\ref{baecorrect}). States with $n^+ \neq n^-$ therefore lead to a continuum of exponents in this case.

\subsubsection{Discrete states}\label{secdiscretestates}

The full partition function is given by (\ref{contdis}), but we have so far only considered the contribution to the continuous part. In concrete terms, this means that there are states in the spectrum which do not correspond to the continuum states (\ref{hparaf0})--(\ref{lcont}), but rather to so-called discrete states that do not have the logarithmic corrections exemplified in Figure \ref{noncompact2}. These discrete states contribute to the $\mathcal{Z}^{\text{disc.}}$ term in (\ref{contdis}) and, using (\ref{flscaling}), are found to be given by the parafermion exponents of equation (\ref{paraflm}).

The generating function of scaling levels in the sector with $2j=l$ defect lines is
\beq\label{discreteloop}
\mathcal{Z}^{\text{disc.}}_{2j}(\text{loop})=\sum_{\substack{m \in \mathbb{Z} \\ -k+3 \leq m \leq k-2}}\Tr_{\tilde{F}_{lm}}q^{L_0-\frac{c}{24}} \,, \quad \mbox{with } l\equiv 2j \,,
\eeq
where
\begin{eqnarray}\label{trflm}
\hbox{Tr }_{\!\!\tilde{F}_{lm}}q^{L_0-c/24}={q^{{(l+1)^2\over 4k}-{m^2\over 4(k-2)}}\over \eta(q)^2}\left[\sum_{n=0}^\infty (-1)^n q^{{n^2\over 2}+{n(l+1-m)\over 2}} +\sum_{n=1}^\infty (-1)^n q^{{n^2\over 2}+ {n(l+1+m)\over 2}}\right]
\end{eqnarray}
for $l$ and $m$ having the same parity, and we define $\hbox{Tr }_{\!\!\tilde{F}_{lm}}q^{L_0-c/24}=0$ for $l$ and $m$ having opposite parity.
Then the term $\mathcal{Z}^{\text{disc.}}$ in (\ref{contdis}) is given by the sum over all of these $\mathcal{Z}^{\rm disc.}_l$. Furthermore, note from
(\ref{discreteloop}) that the bounds on $m$ change discontinuously when $k$ passes through an integer. Some care must be taken when applying
(\ref{trflm}) to extract the critical exponent $h$. When one expands (\ref{trflm}) for $m>l$, the first term is obtained from $n=m-l$.
This corresponds to the conformal weight
\beq\label{hml}
h = {l(l+2)\over 4k}-{m^2\over 4(k-2)}+\frac{1}{2}(m-l)\,,
\eeq
which is equivalent to the exponent $\Delta^m_l$ in (\ref{paraflm}), but with the replacements
\begin{subequations}
\begin{eqnarray}
 l &\rightarrow& k-2-l \,, \\
 m &\rightarrow& k-2-m \,.
\end{eqnarray}
\end{subequations}

\subsubsection{XXZ subset}\label{secxxz}

In our article \cite{Robertson2020}, it was shown that a subset of the solutions to the BAE for the boundary conditions considered there
corresponds to solutions of the BAE for the XXZ spin chain with particular boundary conditions. The open XXZ Hamiltonian with boundary
fields $H$ and $H'$ is given by
\beq
\mathcal{H}_{\rm XXZ}=-\frac{1}{2}\left[\sum\limits_{i=1}^{L-1}(\sigma_i^x\sigma_{i+1}^x+\sigma_i^y\sigma_{i+1}^y-\cos\gamma_0 \, \sigma_i^z\sigma_{i+1}^z)+H\sigma_1^z+H'\sigma_L^z\right] \,,
\eeq
and its eigenvalues can be written in terms of Bethe roots $\mu_k$ as \cite{SALEURBauer}
\beq\label{xxzeigs}
E=-\sum\limits_{k=1}^{m'}\frac{2\sin^2\gamma_0}{\cosh2\mu_k-\cos\gamma_0}+\frac{1}{2}(N-1)\cos\gamma_0+\text{boundary terms} \,.
\eeq
The CFT properties reside in the terms of $E$ proportional to $\frac{1}{N}$, so the second term and the boundary terms in (\ref{xxzeigs})
can be neglected. The $m'$ Bethe roots $\mu_k$ in (\ref{xxzeigs}) are given by the solutions to the BAE of the boundary XXZ chain
	\beq\label{xxzbae}
	\left(\frac{\sinh(\mu_j+i\frac{\gamma_0}{2})}{\sinh(\mu_j-i\frac{\gamma_0}{2})}\right)^{2L}\frac{\sinh(\mu_j+i\Lambda)}{\sinh(\mu_j-i\Lambda)}\frac{\sinh(\mu_j+i\Lambda')}{\sinh(\mu_j-i\Lambda')}=\prod\limits_{k\neq j}^{m'}\frac{\sinh(\mu_j-\mu_k+i\gamma_0)}{\sinh(\mu_j-\mu_k-i\gamma_0)}\frac{\sinh(\mu_j+\mu_k+i\gamma_0)}{\sinh(\mu_j+\mu_k-i\gamma_0)} \,,
	\eeq
where the parameters $\Lambda,\Lambda'$ are defined in terms of $H,H'$ as
\beq
e^{2i\Lambda}=\frac{H-\Delta-e^{i\gamma_0}}{(H-\Delta)e^{i\gamma_0}-1} \,,
\eeq
and similarly for $\Lambda'$.

We now turn to the BAE (\ref{baetype3subset}), corresponding to the BAE in (\ref{baecorrect}) when the solutions come in pairs $\{u_j,u_j+i\pi\}$.
In order to compare the \dtt BAE in (\ref{baetype3subset}) with the XXZ boundary BAE in (\ref{xxzbae}) we set $\gamma_0 = \pi - 2 \gamma$
(where $\eta = i\gamma$), and we consider solutions of the form $u_j = \alpha_j-i\frac{\pi}{2}$. The BAE (\ref{baetype3subset}) then become
	\beq\label{xxzsubset2}
	\left(\frac{\sinh(\alpha_j+i\frac{\gamma_0}{2})}{\sinh(\alpha_j-i\frac{\gamma_0}{2})}\right)^{2L}\frac{\sinh(\alpha_j-i\frac{\gamma_0}{2})}{\sinh(\alpha_j+i\frac{\gamma_0}{2})}=-\prod\limits_{k\neq j}^{m}\frac{\sinh(\alpha_j-\alpha_k+i\gamma_0)}{\sinh(\alpha_j-\alpha_k-i\gamma_0)}\frac{\sinh(\alpha_j+\alpha_k+i\gamma_0)}{\sinh(\alpha_j+\alpha_k-i\gamma_0)} \,.
	\eeq
Setting $\Lambda = -\frac{\gamma_0}{2}$ and $\Lambda' = \frac{3\pi}{2}$ we have that the solutions $\alpha_j$ to (\ref{xxzsubset2}) are also solutions to (\ref{xxzbae}), i.e., we have $\lambda_j = \alpha_j$. We write the energy (\ref{d22eigs3}) in terms of $\alpha_j$ and $\gamma_0$ to find
\beq
E_{D_2^2}=-\sum\limits_{j=1}^{m}\frac{2\sin^2 \gamma_0}{\cosh2\alpha_j-\cos\gamma_0} \,.
\eeq
This coincides with the energy of the XXZ chain,
\beq\label{enxxz}
E_{\text{XXZ}}=\sum\limits_{j=1}^{m}\frac{-2\sin^2 \gamma_0}{\cosh2\lambda_j-\cos\gamma_0} \,,
\eeq
when $\alpha_j=\lambda_j$. Since the solutions to the \dtt BAE admit twice as many roots as the XXZ chain, the final result is in fact
\beq
E_{D_2^2} = 2 E_{XXZ} \,.
\eeq 

The effective central charge of the open XXZ chain for generic $\Lambda$, $\Lambda'$ is \cite{SALEURBauer}
\beq\label{ceff2}
c_{\text{eff}}=1-\frac{6}{1-\frac{\gamma_0}{\pi}}\left(1-\frac{\gamma_0+\Lambda+\Lambda'-2\pi S (1-\frac{\gamma_0}{\pi})}{\pi}\right)^2 \,,
\eeq
where $S$ is the total magnetisation of the chain in a given sector. We have $\gamma_0=\pi-2\gamma$, $\gamma=\frac{\pi}{k}$, $\Lambda = -\frac{\gamma_0}{2}$ and $\Lambda' = \frac{3\pi}{2}$, yielding
\beq\label{ceffxxz}
c_{\text{eff}}=1-\frac{3}{k}(1-k+4S)^2 \,.
\eeq
When considering the \dtt chain, however, we must multiply the effective central charge (\ref{ceffxxz}) by two to get finally for the \dtt chain
\beq\label{ceffd22}
c_{\text{eff}}(D^2_2) =2-\frac{6}{k}(1-k+4S)^2 \,.
\eeq

We must however take considerable care when applying (\ref{ceffd22}). Since the \dtt chain contains many more states than the XXZ chain,
there is no guarantee that the state scaling with $c_{\text{eff}}$ in (\ref{ceffd22}) is the ground state of the \dtt system. Consider for example the cases $k=5$ and $k=7$. For $k=5$ the ground state of the system occurs in the $S=1$ sector and we recover the numerical result $c_{\text{eff}}=2$, which in this case does indeed correspond to (\ref{ceffd22}). For $k=7$, however, the ground state of the \dtt system does not correspond to a solution to the XXZ equations in (\ref{xxzsubset2}). In this case, then, the lowest-energy state that scales with (\ref{ceffd22}) will be an excited state of the \dtt chain. Note that as we increase $k$, the lowest energy state that scales like (\ref{ceffd22}) appears in higher and higher magnetisation sectors $S$. 

\subsubsection{Loop, Potts and vertex-model spectra}

The spectrum of the loop model is a central object from which the spectra of the vertex and Potts models can  be inferred using well known correspondence rules discussed in \cite{S-AF}. In a nutshell, the key point is that the vertex model and associated staggered XXZ spin chains enjoy $U_\q sl(2)$ symmetry, with $\q=e^{i\gamma}$. Accordingly, the spectrum of the vertex model can be split into sectors of total (integer) quantum spin $j=0,\ldots, L$. The loop model with a fixed number of through-lines gives the spectrum of the vertex model for a fixed value of $j$. Taking the traces in the vertex instead of loop Hilbert space gives 
\begin{equation}
\mathcal{Z}_{2j} (\text{vertex})=(2j+1) \mathcal{Z}_{2j}(\text{loop}) \,,
\end{equation}
both for the discrete and continuous parts of the spectrum, where $2j+1$ is simply the dimension of the spin-$j$ $U_\q sl(2)$ representation. 

The correspondence with the Potts model is very similar, but requires giving non-contractible loops 
wrapping around the annulus a weight $\sqrt{Q}$. This is obtained by introducing a ``modified'' trace, leading to the expression \cite{SALEURBauer}
\begin{equation}
\mathcal{Z}(\text{Potts})=\sum_{j=0}^\infty [2j+1]_\q \left\{\mathcal{Z}_{2j}^{\text{disc.}}(\text{loop})
+\mathcal{Z}_{2j}^{\text{cont.}}(\text{loop})\right\} \,,
\end{equation}
where $\q$-deformed numbers are defined as
\begin{equation}
 [2j+1]_\q={\q^{2j+1}-\q^{-2j-1}\over \q-\q^{-1}} \,.
\end{equation}
In all these expressions, the objects $\mathcal{Z}$ are defined as traces over the relevant ``Hilbert spaces'' (which in the case of $Q$ non-integer is of course only a formal object). 

\subsection{RSOS model}\label{secrsostype3}

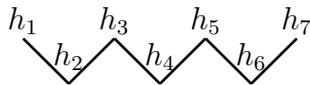
\begin{figure}[ht]
	\centering
\begin{tikzpicture}[scale=1.2]
	
\draw[black,line width = 1pt](1,1)--(1.5,1.5);
\draw[black,line width = 1pt](2,1)--(2.5,1.5);
\draw[black,line width = 1pt](3,1)--(3.5,1.5);
\draw[black,line width = 1pt](0.5,1.5)--(1,1);
\draw[black,line width = 1pt](1.5,1.5)--(2,1);
\draw[black,line width = 1pt](2.5,1.5)--(3,1);

\node at (0.5,1.7) {$h_1$};
\node at (1,1.3) {$h_2$};
\node at (1.5,1.7) {$h_3$};
\node at (2,1.3) {$h_4$};
\node at (2.5,1.7) {$h_5$};
\node at (3,1.3) {$h_6$};
\node at (3.5,1.7) {$h_7$};

\end{tikzpicture}

\caption{The state $\left| h_1h_2 h_3 h_4 h_5 h_6 h_7 \right\rangle$.}\label{RSOSrow}
\end{figure}

When $k$ (recall $\gamma={\pi\over k}$) is an integer, it is possible to consistently truncate the Hilbert space of the vertex model or spin chain by restricting the action of the Temperley-Lieb generators to a certain type of $U_\q sl(2)$ representation. The resulting model can be interpreted as a ``restricted solid-on-solid model'', and the restriction of the Temperley-Lieb action can be expressed by a change of basis from ``spins'' to ``heights''. We will restrict in what follows to the case $k$ integer. In the corresponding RSOS representation, the $e_i$ act on states of neighbouring heights $h_i=1,2,\ldots,k$ which are subject to the constraint $|h_{i+1}-h_i| = 1$. An example of an RSOS state is shown in Figure \ref{RSOSrow}. The explicit form of the $e_i$ is given by \cite{PasquierRSOS}
\begin{eqnarray} \label{tlrsos}
e_i \left| h_1,\ldots,h_{i-1},h_i,h_{i+1},\ldots,h_{N+1} \right\rangle &=& \nonumber \\
 \delta(h_{i-1},h_{i+1}) & & \!\!\!\!\!\!\!\!\!\!\!\! \sum\limits_{h_i'}\frac{\sqrt{S_{h_i}S_{h_i'}}}{S_{h_{i-1}}} \left| h_1,\ldots,h_{i-1},h_i',h_{i+1}, \ldots,h_{N+1} \right\rangle \,,
\end{eqnarray}
with the heights $\left| h_1,\ldots,h_{i-1},h_i,h_{i+1},\ldots,h_{N+1} \right\rangle$ being situated along a zig-zag cross-section of the lattice (see Figure \ref{RSOSrow}) and the $S_{h_i}$ being defined as
\beq\label{Sadef}
S_a=\frac{\sin(\frac{a\pi}{k})}{\sin(\frac{\pi}{k})} \,,
\eeq
where we recall that $Q$ is parametrised as $\sqrt{Q}=2\cos(\frac{\pi}{k})$.

It was discovered in our earlier papers  \cite{JS-AF} that the model obtained by choosing this RSOS version of the TL algebra in the AF Potts Hamiltonian gives rise, in the continuum limit, to the $Z_{k-2}$ parafermionic  (diagonal) theory \cite{FateevZamolodchikov,GepnerQiu}.%
\footnote{How this can arise starting from the black hole sigma model remains, however, a mystery.}
A particularly important set of objects in this context  are the $su(2)/u(1)$ string functions $c_l^m$, defined as
\beq\label{stringfuncdef}
 c_l^m = \frac{1}{\eta(q)^2}\sum_{\substack{n_1, n_2 \in \mathbb{Z}/2  \\ n_1-n_2\in \mathbb{Z} \\ n_1 \geq |n_2|, -n_1 > |n_2| }}(-1)^{2n_1}\text{sign}(n_1)q^{\frac{(l+1+2n_1k)^2}{4k}-\frac{(m+2n_2(k-2))^2}{4(k-2)}} \,.
\eeq
It was shown in \cite{S-AF,Robertson2020} that the generating function of scaling levels of the AF Potts model with free boundary conditions is
given by the string function $c^0_l$. It was furthermore shown in \cite{Robertson2019} that there exist other boundary conditions in the same
model for which the generating functions are given by the other string functions with $m\neq 0$.

We shall consider the model obtained by inserting the representation (\ref{tlrsos}) for the $e_i$ into the Hamiltonian (\ref{hamtype3TL}),
while fixing the leftmost RSOS height to $1$ and the rightmost height to $l+1$. We will write this boundary condition as
\beq\label{type3rsosbc}
1,\ldots,l+1 \,, \quad \mbox{or more concisely: } 1/l+1 \,.
\eeq
By directly diagonalising this Hamiltonian, we once again observe the parafermion central charge (\ref{cpf2}) and the parafermion exponents (\ref{paraflm}),
upon using the finite-size scaling result (\ref{flscaling}) for the energy eigenvalues. Furthermore, we find that the full generating function is given
by particular combinations of the string functions. For $k$ an odd integer the boundary condition (\ref{type3rsosbc}) produces the generating function
\beq\label{stringodd}
\mathcal{Z}_{1/l+1}(\text{RSOS})=c^{m=0}_l+2\sum\limits_{\substack{n=2\\ n \ {\rm even}}}^{k-3}c_l^{m=n} \,,
\eeq
while for $k$ an even integer we obtain
\beq\label{stringeven}
\mathcal{Z}_{1/l+1}(\text{RSOS})=c^{m=0}_l+2\sum_{\substack{n=2\\ n \ {\rm even}}}^{k-4}c_l^{m=n} + c_l^{m = k-2} \,.
\eeq

Table \ref{rsostab} presents explicit examples of this correspondence for $k=4,5,6,7$. We can gain some insight into these results by comparing with
the two well-studied cases $k=4$ (Ising model) and $k=5$ (three-state Potts model).
For $k=4$, the combination of string functions $c^{m=0}_{l=0}+c^{m=2}_{l=0}$ can be written in the suggestive form 
$\chi_{1,1}+\chi_{1,2}$, where $\chi_{r,s}$ denotes the usual minimal model generating function. This expression agrees
with the continuum limit for an open Ising chain with free boundary conditions on both boundaries.
Similarly, for $k=5$ the combination of string functions $c^{m=0}_{l=0}+2c^{m=2}_{l=0}$ can be written as
$\chi_{1,1}+\chi_{4,1}+2\chi_{4,3}$. Once again, this agrees with the continuum limit of the three-state Potts
model with free-free boundary conditions \cite{AOS}.

The boundary conditions identified in our previous papers \cite{Robertson2019,Robertson2020} can  be labelled by the pair $(l,m)$ and correspond, in the language of $Z_{k-2}$ boundary parafermionic CFT, to ``untwisted branes''. In contrast, the ones we find here come with a single label $l$, and must be interpreted as ``twisted branes''. In the notations of \cite{Fredenhagen2003} (see also \cite{Maldacena}), the partition functions with a fixed value of $l$ correspond to partition functions in the CFT with boundary condition $[0,+]$ on one side of the annulus, and $[l,(-1)^l]$ on the other side. For  the three-state Potts model, $[0,+]$ and $[1,-]$ are known to correspond to ``free'' and ``new'' boundary conditions, as introduced in \cite{AOS}.

By definition, $c^m_l=0$ when $m$ and $l$ do not have the same parity. Using this and the identity $c^{-m}_l=c^m_l$,
eqs.~(\ref{stringodd})--(\ref{stringeven}) can be seen to reduce to
\beq\label{stringcombo}
\mathcal{Z}_{1/l+1}(\text{RSOS})=\sum\limits_{m=-k+3}^{k-2}c^m_l \,.
\eeq
Further recall the well-studied connection (see, e.g., \cite{Robertson2019}) between the generating function of scaling levels in the loop model and in the RSOS model. For the AF Potts model with free boundary conditions, we have
\begin{equation} \label{strfct3}
c_l^0 = \sum\limits_{n=0}^{\infty} \left( K_{l+2nk}-K_{2(n+1)k-l-2} \right) = K_l-K_{2k-l-2}+K_{l+2k}-K_{4k-l-2}+\ldots \,,
\end{equation}
where $K_l$ is the generating function of the model with free boundary conditions in the loop representation, defined as
\begin{equation}
K_l=\hbox{Tr}_{\StTL{j=l/2}}q^{L_0-c/24}={q^{(l+1)^2/4k}\over \eta(q)^2}\left[1+2\sum_{n=1}^\infty (-1)^n q^{n(n+l+1)/2}\right]\label{Kfct} \,,
\end{equation}
and where the correspondence with the loop model is $l=2j$. For the states that correspond to the continuum in the loop model to disappear in the RSOS
model, we need them to appear in different sectors of $l$, so that these states cancel in (\ref{strfct3}). From (\ref{strfct3}) we see that a continuum state in
the sector with $2j$ through-lines in (\ref{hparaf0}) must also appear in the sector with $2k-2j-2$ through-lines. This is consistent with the discussion of the
critical exponents in section \ref{secloopreptype3}: From the right-hand side of (\ref{lcont}), we can see that the exponent $h$ in (\ref{hparaf0}) is invariant under $2j\rightarrow 2k-2j-2$, and we indeed observe that these cancellations do in fact appear as degeneracies in the loop model spectrum. The string functions, defined in (\ref{stringfuncdef}), can in fact be written as
\beq\label{strfctredef}
c^m_l=\sum\limits_{n=0}^{\infty}\Tr_{\tilde{F}_{l+2nk,m}}q^{L_0-\frac{c}{24}}-\Tr_{\tilde{F}_{2k-l-2+2nk,m}}q^{L_0-\frac{c}{24}} \,,
\eeq
where $\Tr_{\tilde{F}_{l+2nk,m}}$ was defined in (\ref{trflm}). We must be able to relate the generating function of \textit{discrete} scaling levels in (\ref{discreteloop}) to the combination of string functions in (\ref{stringcombo}) observed in the RSOS model, by summing the $\mathcal{Z}_m$ in the same way that we sum the $K_l$ in (\ref{strfct3}). Indeed, from (\ref{strfctredef}) we have
\beq
\sum\limits_{m=-k+3}^{k-2}c^m_l=\sum\limits_{n=0}^{\infty} \left[\mathcal{Z}^{\text{disc.}}_{l+2nk}(\text{loop})-\mathcal{Z}_{2(n+1)k-l-2}^{\text{disc.}}(\text{loop}) \right] \,,
\eeq
which is precisely the relation one would expect to hold from considering the correspondence between the loop and RSOS models, using the $U_{\q}sl(2)$ symmetry \cite{S-AF} with $\q=e^{i\gamma}$. 
\begin{table}[h]
\begin{center}
		\renewcommand{\arraystretch}{1.5} % Default value: 1
\begin{tabular}{c|c|c}
Boundary condition & $k$ & Generating function  \\
\hline
1,...,1  & 4  & $c^{m=0}_{l=0}+c^{m=2}_{l=0}$\\
\hline
1,...,1  & 5  & $c^{m=0}_{l=0}+2c^{m=2}_{l=0}$\\
\hline
1,...,1  & 6  & $c^{m=0}_{l=0}+2c^{m=2}_{l=0}+c^{m=4}_{l=0}$\\
\hline
1,...,1  & 7  & $c^{m=0}_{l=0}+2c^{m=2}_{l=0}+2c^{m=4}_{l=0}$\\
\hline
1,...,3  & 4  & $c^{m=0}_{l=2}+c^{m=2}_{l=2}$\\
\hline
1,...,3  & 5  & $c^{m=0}_{l=2}+2c^{m=2}_{l=2}$\\
\hline
1,...,3  & 6  & $c^{m=0}_{l=2}+2c^{m=2}_{l=2}+c^{m=4}_{l=2}$\\
\hline
1,...,3  & 7  & $c^{m=0}_{l=2}+2c^{m=2}_{l=2}+2c^{m=4}_{l=2}$\\
\hline
1,...,5  & 6  & $c^{m=0}_{l=4}+2c^{m=2}_{l=4}+c^{m=4}_{l=4}$\\
\hline
1,...,5  & 7  & $c^{m=0}_{l=4}+2c^{m=2}_{l=4}+2c^{m=4}_{l=4}$\\

\end{tabular}
\caption{The continuum limit of the RSOS model.}\label{rsostab}
\end{center}
\end{table}

\section{A boundary RG flow}\label{secboundaryrgflow}

Let us now consider the following Hamiltonian
\beq\label{hgeneral}
\mathcal{H}=\alpha(e_1+e_{2L-1})+2\cos\gamma\sum\limits_{m=1}^{2L-1}e_m-\sum\limits_{m=1}^{2L-2}(e_me_{m+1}+e_{m+1}e_m) \,,
\eeq
where $\alpha$ is a free boundary parameter. Setting $\alpha=0$ we get back the Hamiltonian (\ref{hamtype3TL}), the main object of
study in the present paper, while for $\alpha = -\frac{1}{\cos\gamma}$ we obtain the Hamiltonian (\ref{hamtype2TL}), which was studied in
our paper \cite{Robertson2020}. These two Hamiltonians admit a Bethe Ansatz solution but are described by different boundary conformal
field theories in the continuum limit. As we have seen, the most striking difference between the two field theories is that, in the loop or vertex representations,
the latter is compact, while the former is non-compact. It is thus interesting to consider the continuum limit when $\alpha$ varies continuously
between the two exactly solvable points, $\alpha=0$ and $\alpha=-\frac{1}{\cos\gamma}$.

We first discuss how this works in the loop model; the results are the same for the vertex model and the Potts model, up to some discrete degeneracies.  For concreteness, we will focus on the case $\gamma = \frac{\pi}{4}$ in the sector with $2j=2$
through-lines. When $\alpha=0$, the continuum will begin at $h=0$ according to (\ref{hparaf0})--(\ref{lcont}). Using (\ref{flscaling}), we therefore expect that the first excited state of the Hamiltonian will produce a gap $h_1-h_0$ that converges logarithmically to $0$ as in Figure \ref{noncompact2}. When $\alpha=-\frac{1}{\cos\gamma}$, the generating function of scaling levels is instead given by (\ref{Kfct}) \cite{Robertson2020}, and hence the first gap $h_1-h_0$ should converge to $1$ with no logarithmic corrections. If we perturb $\alpha$ slightly away from the non-compact point, $\alpha = 0$, to $\alpha = -\epsilon$ for $\epsilon$ small and positive, it turns out that the first gap converges to $h_1-h_0=1$. This is suggestive of a Renormalisation Group (RG) flow away from the non-compact BCFT at $\alpha = 0$ towards the compact BCFT at $\alpha = -\frac{1}{\cos\gamma}$.

\begin{figure}[h]
	\centering
	\includegraphics[scale=0.75]{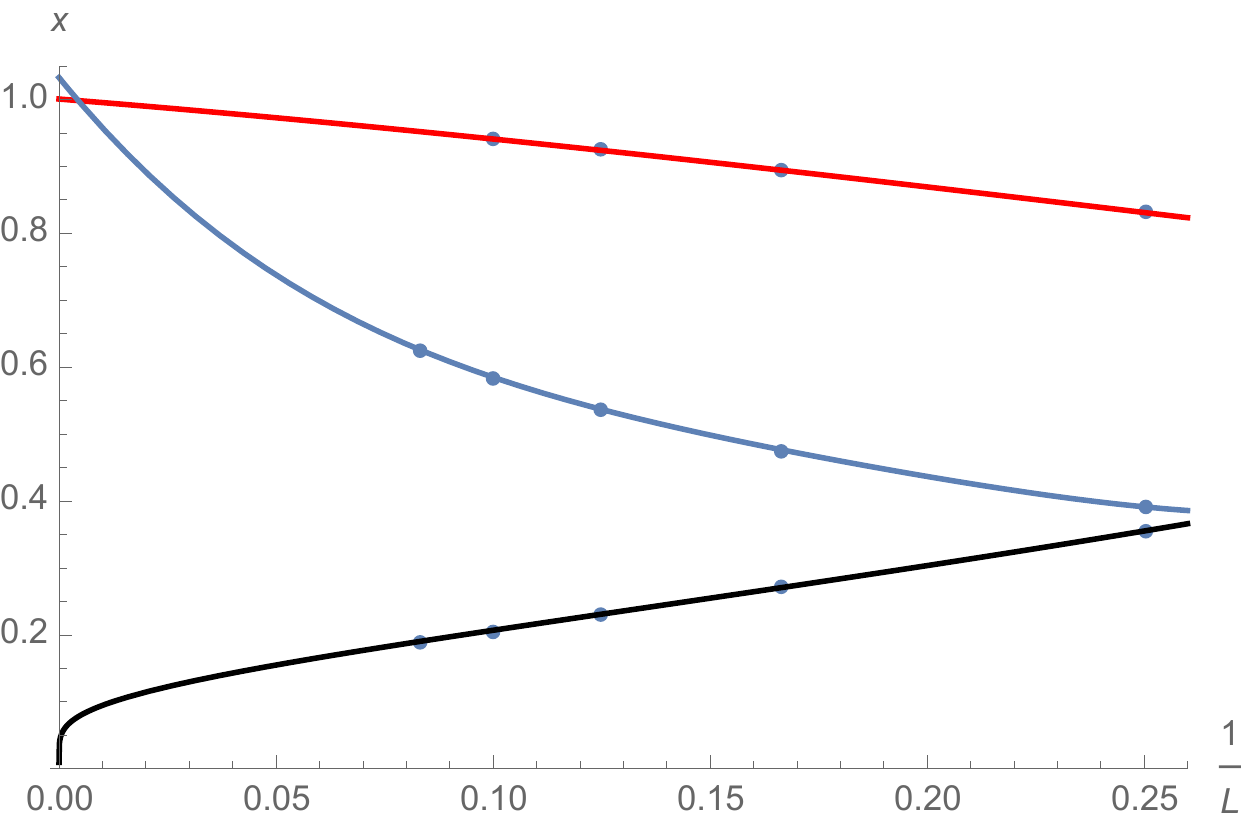}
\caption{The convergence of the first gap in the loop model in the sector with $2j=2$ through lines, with $\gamma=\frac{\pi}{4}$. The black line corresponds to $\alpha=0$ in (\ref{hgeneral}), the red line corresponds to $\alpha=-\frac{1}{\cos\gamma} \simeq -1.41$, and the blue line corresponds to $\alpha=-0.46$. We expect the first gap for $\alpha=-\frac{1}{\cos\gamma}$ to converge to $1$ and  the first gap for $\alpha=0$ to converge logarithmically to $0$, in agreement with the numerical results. Finally, we observe that when we perturb $\alpha$ slightly away from $0$ the model flows under RG to $\alpha=-\frac{1}{\cos\gamma}$. }\label{loopRG}
\end{figure}

\begin{figure}[h]
	\centering
	\includegraphics[scale=0.75]{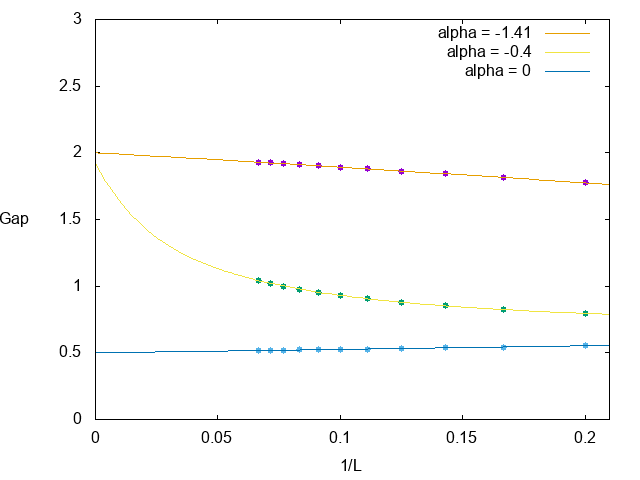}
\caption{The blue line corresponds to the case $\alpha = 0$. As expected, we see that the first gap over the ground state converges to $h=\frac{1}{2}$ since in this case the full generating function is given by the sum of the two string functions $c^{m=0}_{l=0}+c^{m=2}_{l=0}$. The orange line corresponds to $\alpha = -\frac{1}{\cos\gamma}$; in this case,  we see that the first gap  converges to $h=2$ since the generating function in this case is given only by the string function $c^{m=0}_{l=0}$. The yellow line corresponds to a perturbation of $\alpha$ away from zero, (in the figure we take $\alpha = -0.364$). We see that in this case the gap also converges to $2$, but it converges much more slowly than the case $\alpha = -\frac{1}{\cos\gamma}$: this is compatible with  an RG flow away from the non-compact theory with $\alpha = 0$, towards the compact theory with $\alpha = -\frac{1}{\cos\gamma}$.  }\label{RSOSRG}
\end{figure}

Our numerical results show this RG flow explicitly, for both the loop model (Figure \ref{loopRG}) and the RSOS model (Figure \ref{RSOSRG}).
In both figures we plot the first gap, $h_1-h_0$ against $\frac{1}{L}$, where $h_1$ and $h_0$ are the numerical approximations to the conformal dimensions appearing on the right-hand side of (\ref{flscaling}) for the first excited state and the ground state, respectively, of the Hamiltonian.

Figure \ref{loopRG} contains the results for the loop model. The black curve corresponds to $\alpha=0$ and, as expected from the discussion in section \ref{secloopreptype3}, the gap converges logarithmically to zero. The red curve corresponds to $\alpha=-\frac{1}{\cos\gamma}$, and, as expected from (\ref{Kfct}), this gap converges nicely to $1$. Finally, the blue curve corresponds to a slight perturbation away from $\alpha=0$. We observe that, for low sizes, the numerical approximation to $h_1-h_0$ is very close to the result for $\alpha=0$, but that in the limit $\frac{1}{L}\rightarrow 0$ the curve converges towards $1$. This strongly suggests that $\alpha=0$ is a repulsive fixed point and that $\alpha=-\frac{1}{\cos\gamma}$ is an attractive fixed point. In other words, \eqref{hgeneral} provides a lattice realisation of an RG flow away from a non-compact BCFT
towards a compact one.

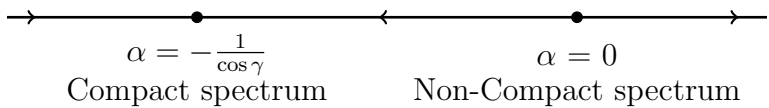
\begin{figure}[h]
	\centering
\begin{tikzpicture}
	
	\begin{scope}[thick,decoration={
	    markings,
	    mark=at position 0.5 with {\arrow{<}}}
	    ] 
	\draw[black,line width = 1pt,postaction={decorate}] (2.5,1)--(7.5,1);
	\end{scope}
	
	\begin{scope}[thick,decoration={
	    markings,
	    mark=at position 0.85 with {\arrow{>}}}
		]
	\draw[black,line width = 1pt,postaction={decorate}] (7.5,1)--(10,1);
	    ] 
	\end{scope}
	
	\begin{scope}[thick,decoration={
	    markings,
	    mark=at position 0.15 with {\arrow{>}}}
		]
	\draw[black,line width = 1pt,postaction={decorate}] (0,1)--(2.5,1);
	    ] 
	\end{scope}
	
	\filldraw[black] (7.5,1) circle (2pt);
	\node at (7.5,0.5) {$\alpha$ = 0};
	\node at (7.5,0) {Non-Compact spectrum};
	
	\filldraw[black] (2.5,1) circle (2pt);
	\node at (2.5,0.5) {$\alpha =- \frac{1}{\cos\gamma}$};
	\node at (2.5,0) {Compact spectrum};

\end{tikzpicture}
\caption{RG flow from the Hamiltonian (\ref{hamtype3TL}) to the Hamiltonian of \cite{Robertson2020}, which correspond to a compact and a non-compact continuum limit respectively.}\label{rgflow}
\end{figure}

The same analysis for the RSOS model is presented in Figure \ref{RSOSRG}. In this case, the generating function obtained from the boundary condition (\ref{type3rsosbc}) with $\alpha=0$ is given by (\ref{stringodd})--(\ref{stringeven}), or equivalently by (\ref{stringcombo}), whereas the generating function for the case $\alpha = -\frac{1}{\cos\gamma}$ is given by $c^{0}_{l}$ \cite{Robertson2020}. This results in the following RG flow
\beq\label{stringcomboflowl}
\sum\limits_{m=-k+3}^{k-2}c^m_l \rightarrow c^0_l \,.
\eeq
where it is convenient to express  the effect of the flow on  the generating functions of levels. The observed RG flow is summarised in Figure \ref{rgflow}. The points $\alpha=0$ and $\alpha=-\frac{1}{\cos\gamma}$ are again found to be fixed points
under the boundary RG, with $\alpha=0$ a repulsive fixed point and $\alpha=-\frac{1}{\cos\gamma}$ an attractive fixed point.

It is enlightening to study this flow in the context of the known boundary RG flows in the Ising and the three-state Potts model. Consider the RSOS model with $k=4$, which in the bulk corresponds to the  $Z_2$ parafermions, i.e., the Ising CFT. The flow in (\ref{stringcomboflowl}) is, in this case,
\beq\label{flowk4m0}
c^{0}_{0}+c^{2}_{0}\rightarrow c^{0}_{0} \,.
\eeq
As discussed in section \ref{secrsostype3}, the left-hand side of (\ref{flowk4m0}) is the generating function obtained by imposing free boundary conditions on both sides in the Ising chain. These boundary conditions flow under RG to fixed boundary conditions on both sides, where the generating function is given by the right-hand side of (\ref{flowk4m0}).

Similarly, for $k=5$ (corresponding now to $Z_3$ parafermions, i.e., the ordinary critical three-state Potts model) the flow from $\alpha = 0$ to $\alpha = -\frac{1}{\cos\gamma}$ corresponds to
\beq
c^{0}_{0}+2c^{2}_{0}\rightarrow c^{0}_{0} \,.
\eeq
This is in fact the same flow from free to fixed boundary conditions in the three-state Potts chain that was observed in \cite{AOS}. Note that this discussion requires some clarification regarding the term ``free boundary conditions". The free boundary conditions referred to in the context of the Potts (or $Z_3$ parafermionic) chain of  \cite{AOS}  do not correspond to free boundary conditions in the generic \textit{antiferromagnetic} Potts model in Hamiltonian (\ref{hgeneral}) when $Q$ is set equal to $3$.%
\footnote{In general, the correspondence between boundary conditions is quite complex indeed. Starting with the AF Potts model with wired boundary conditions, we have seen relations with RSOS models with fixed  boundary conditions (and a special boundary interaction), and, in turn,  $Z_{k-2}$ parafermions with ``twisted'' boundary conditions!}

Some of the boundary RG flows observed here in the RSOS models for general $k$ have been previously classified \cite{Fredenhagen2003} in the context of boundary RG flows in parafermion theories. In particular, when we specialise to the case $l=0$, the flow from $\alpha=0$ to $\alpha = -\frac{1}{\cos\gamma}$ corresponds to
\beq\label{stringcombol0}
\sum\limits_{m=-k+3}^{k-2}c^m_0 \rightarrow c^0_0 \,.
\eeq
The first relevant operator appearing in (\ref{stringcombol0}) has conformal dimension given by (\ref{hml}) with $m=2$ and $l=0$, resulting in
\beq
h=\frac{k-3}{k-2} \,.
\eeq
This flow is known to correspond to the flow from free to fixed boundary conditions in the $Z_{k-2}$ parafermionic theories \cite{Fredenhagen2003}. Note that in the Ising case, $k=4$, we get $h=\frac{1}{2}$, corresponding to the energy operator. For $l\neq 0$, the  partition functions must be interpreted in terms of a BCFT with  the insertion of a boundary-condition changing operator at the origin.  Accordingly, the flow from (\ref{stringcombo}) to $c^0_l$ has a more complicated, and not particularly illuminating,  interpretation.  Standard  calculations \cite{affleck1991} show that, corresponding to the flow (\ref{stringcombol0}), we have the ratio of boundary degeneracies
\begin{equation}
{g_{\text{UV}}\over g_{\text{IR}}}=\sqrt{k-2} \,,
\end{equation}
as expected for the flow from free to fixed boundary conditions in $Z_{k-2}$ parafermionic theories, where ${g_{\text{free}}\over g_{\text{fixed}}}=\sqrt{k-2}$ \cite{Lukyanov06}.

\section{Conclusion}

The analysis of conformal boundary conditions is an essential part of the solution of a CFT \cite{PetkovaZuber}. We hope that, now that lattice boundary conditions for the antiferromagnetic Potts model and the staggered XXZ spin chain giving rise to discrete and continuous $sl(2,\mathbb{R})/u(1)$ characters have been identified, a more precise study of the relationship between the lattice formulation and the corresponding continuum limit can be carried out. This should involve, among other issues, understanding  the emergence of a non-compact direction from the compact lattice variables. We emphasize  that, while  none of our results seem to really match those obtained in \cite{RibaultSchomerus}, it has recently been suggested in \cite{BKKL20a,BKKL20b} that the continuum limit of our lattice model is not the Euclidian but the {\sl Lorentzian} version of the sigma model. Hopefully, an analysis of the conformal boundary conditions in the latter model could help settle this question. 

From a maybe more basic integrable point of view, we have found several intriguing aspects. One of them is that the ordinary AF Potts model with free boundary conditions appears to be, in fact, integrable.%
\footnote{We showed in section \ref{sectmat} that when the geometrical construction outlined in section \ref{geomchange} is applied to the integrable transfer matrix in \eqref{tmatrixopendefects}, and when the spectral parameter $u$ is set to $\frac{u_0'}{2}$ where $u_0'$ is defined in \eqref{u0primedef}, and where the parameters $\omega_i$ are defined as in \eqref{defectchoice}, one obtains a transfer matrix describing the free AF Potts model; see equation \eqref{tmataf} and the surrounding discussion.}
This is a bit unexpected, and might prompt a new discussion of the results in \cite{S-AF}, even though we do not expect any of the qualitative results presented there to change. An interesting question we leave for further study is  what the corresponding Hamiltonian might be.

\subsection*{Acknowledgments}

This work was supported by the ERC Advanced Grant NuQFT. We thank R. Nepomechie and A.L. Retore for their patience and useful comments and for sharing their upcoming work with us. We thank S.\ Lukyanov and S.\ Ribault for useful discussions.

\end{document}